\documentclass{aa}

\usepackage{graphicx}
\usepackage{color}

\begin{document}

\title{On the difference between Nuclear and Contraction ages
  \thanks{Based on observations collected at Cerro Tololo
    Interamerican Observatory}
  \thanks{Tables 6, 7, 8, and 9 are
    only available in electronic form at the CDS via anonymous ftp to
    {\tt cdsarc.u-strasbg.fr (130.79.128.5)} or via {\tt
    http://cdsweb.u-strasbg.fr/cgi-bin/qcat?J/A+A//}}
}

\author{W. Lyra\inst{1,2,3},~
  A. Moitinho\inst{4},~
  N. S. van der Bliek\inst{1} and
  J. Alves\inst{5}}

\offprints{wlyra@astro.uu.se}

\institute{Cerro Tololo Interamerican Observatory, Casilla 603 La Serena, Chile
\and Observat\'{o}rio do Valongo/UFRJ, Ladeira do Pedro Ant\^{o}nio 43, 20080-090 Rio de Janeiro, Brazil
\and Department of Astronomy and Space Physics, Uppsala Astronomical Observatory, Box 515, 751\,20 Uppsala, Sweden
\and Observat\'{o}rio Astron\'{o}mico de Lisboa, Tapada da Ajuda, 1349-018 Lisbon, Portugal
\and European Southern Observatory, Karl-Schwarzschild 2, D-85748 Garching, Germany 
}

\date{Received; accepted}
\authorrunning{Lyra et al.}
\titlerunning{Young Open Clusters}

\abstract
{Ages derived from the low mass stars still contracting onto the main sequence often differ from ages derived from the high mass ones that have already evolved away from it.}
{We investigate the general claim of disagreement between these two independent age determinations by presenting $UBVRI$ photometry of the young galactic open clusters NGC~2232, NGC~2516, NGC~2547 and NGC~4755, spanning the age range $\sim$10-150 Myr}
{We derive reddenings, distances and nuclear ages by fitting ZAMS and isochrones to color-magnitudes and color-color diagrams. To derive contraction ages, we use four different pre-main sequence models, with an empirically calibrated color-temperature relation to match the Pleiades cluster sequence.}
{When using exclusively the $V$ vs. $V-I$ color-magnitude diagram and empirically calibrated isochrones, there is consistency between nuclear and contraction ages for the studied clusters. Although the contraction ages seem systematically underestimated, in none of cases they deviate by more than one standard deviation from the nuclear ages.}
{}

\maketitle

\section{Introduction}

\begin{figure*}
\begin{center}
\resizebox{4.4cm}{4.4cm}{\includegraphics{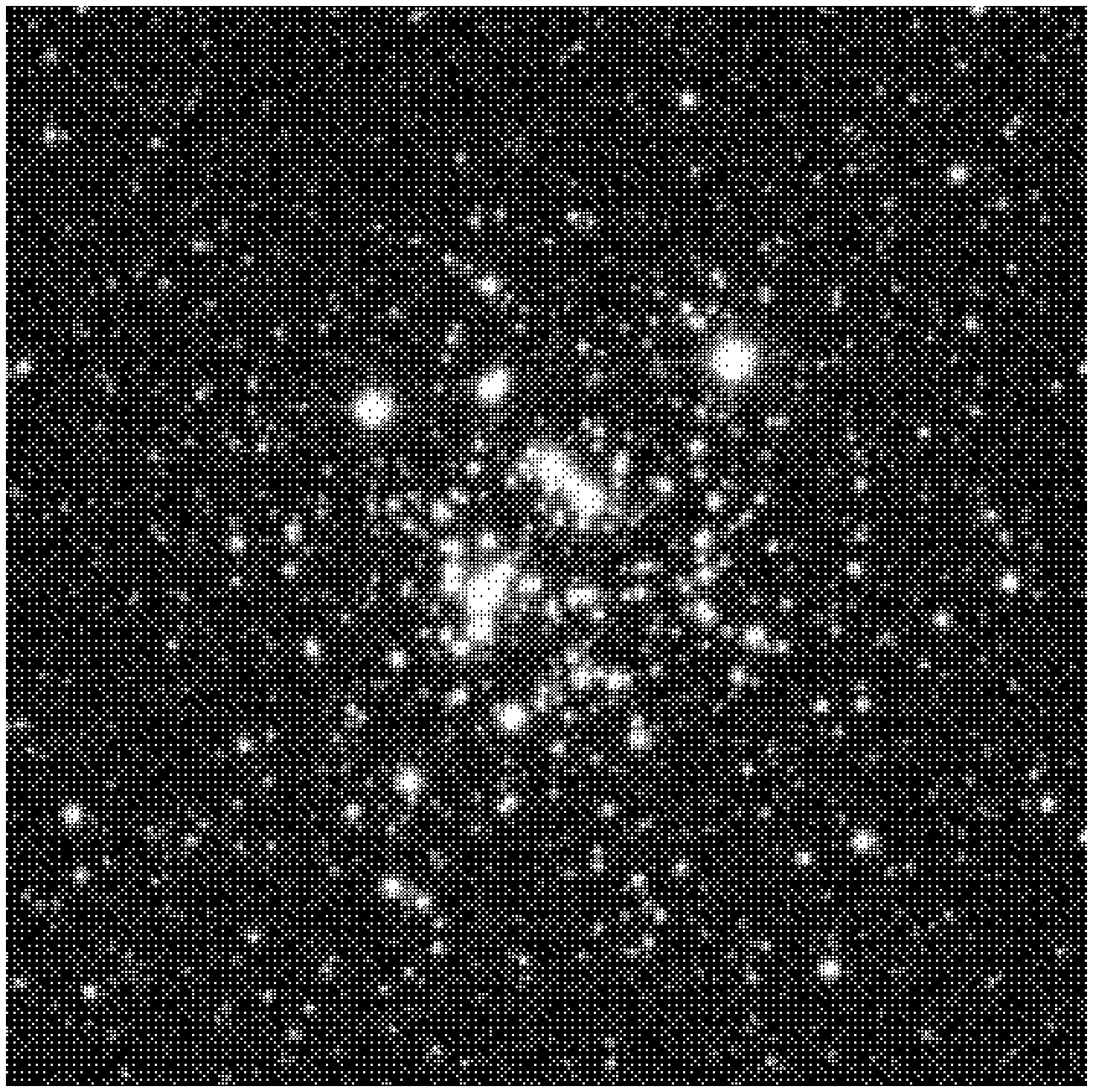}}
\resizebox{4.4cm}{4.4cm}{\includegraphics{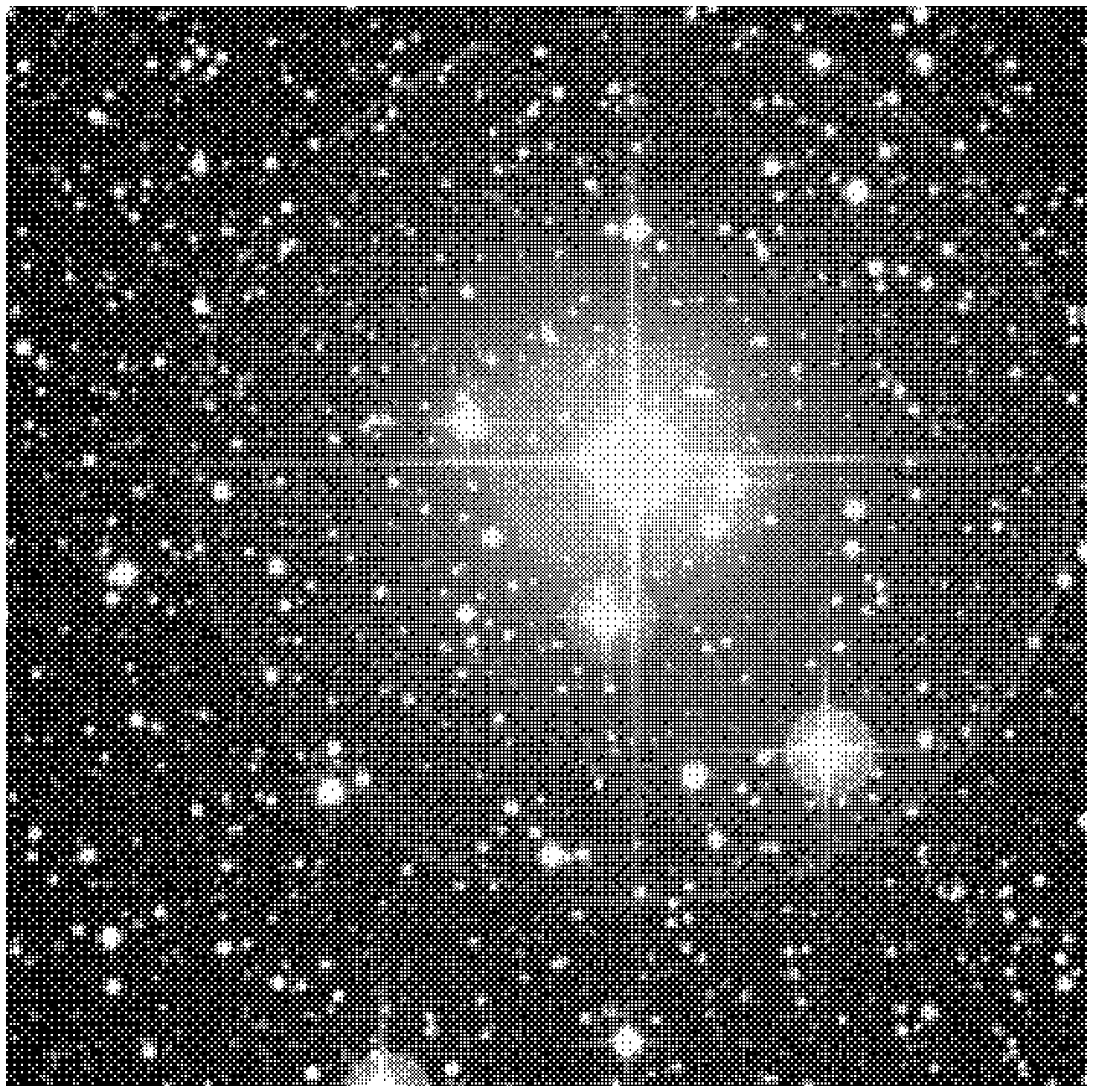}}
\resizebox{4.4cm}{4.4cm}{\includegraphics{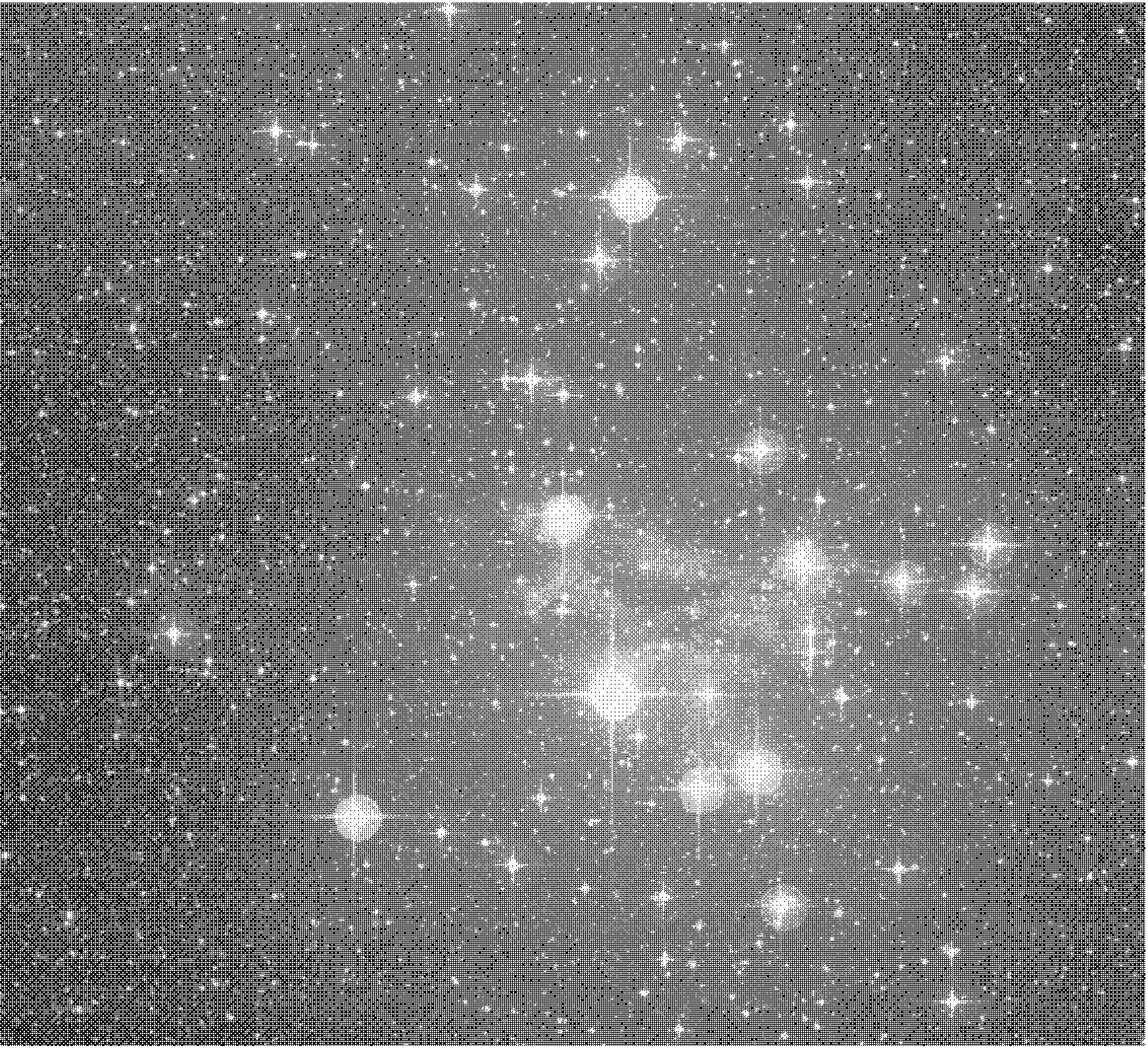}}
\resizebox{4.4cm}{4.4cm}{\includegraphics{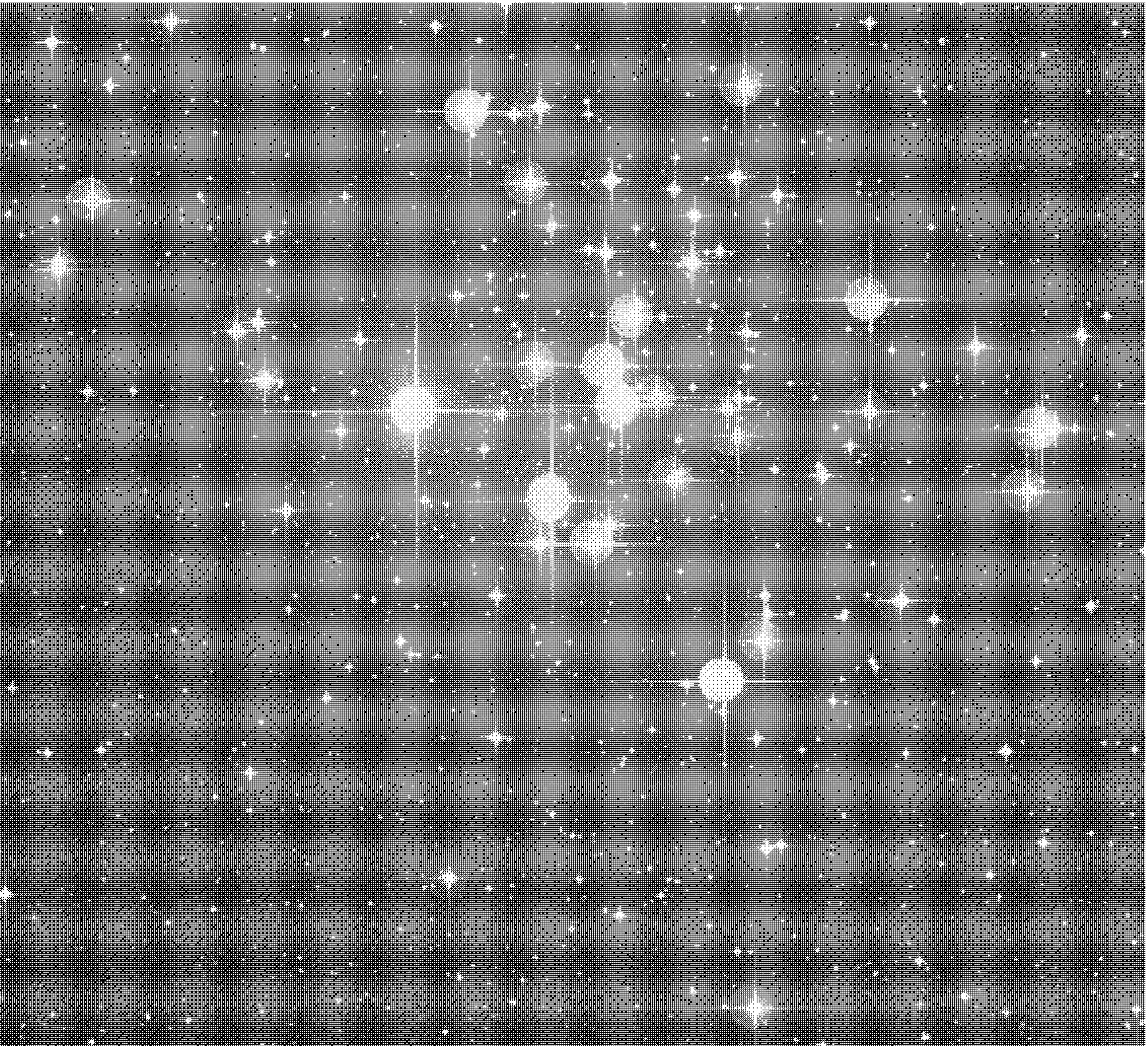}}
\end{center}
\caption[]{From left to right, the observed area of the young stellar clusters NGC~4755, NGC~2232, NGC~2547 and NGC~2516. Field of view is 14'~x~14 for NGC~2232 and NGC~4755. For the other two clusters, the field of view is 28'x28'. These images are from the Digital Sky Survey (DSS). The DSS charts correspond closely to the regions we have surveyed in each cluster with the CTIO 0.9m telescope.}
\label{images}
\end{figure*}

Stellar Clusters, being entities that contain statistically significant numbers of stars spanning a large range of mass and luminosity but sharing the same age and chemical composition, have long been regarded as crucial laboratories for research on stellar astrophysics and galactic structure. In particular, they provide a fundamental unit where to test theories of stellar evolution, from formation to death, spanning many evolutionary stages of the life of a star.

Although a theory of stellar evolution is firmly established for Main Sequence stars, our current understanding of the processes occurring during Pre Main Sequence (PMS) evolution is just marginal. As the PMS phase is relatively quick for high and intermediate mass stars, even most of the youngest optically revealed clusters have their F and G stars already on the main sequence. To make matters worse observationally, low mass stars undergoing PMS evolution are intrinsically faint and have cold atmospheres dominated by molecular lines, which considerably complicates the theoretical modeling of their colors.
 
As a result of these difficulties we lack today large samples of well determined parameters for low mass PMS stars, and the different PMS models predict discrepant stellar parameters. One particular well known problem is that ages derived from the low mass stars still contracting onto the main sequence ({\it contraction ages}, $\tau_c$) often differ from ages derived by the high mass ones that have already evolved away from it ({\it nuclear ages}, $\tau_n$).Quantitatively, the contraction ages are often younger than the nuclear ages for redder colors ($V-I$), and older for bluer colors ($B-V$), with the disagreement lying well beyond the intervals of confidence of both determinations.

The disagreement between these two independent age determinations has
long been a problem in the study of stellar open clusters, and the
matter of a lengthy debate. It was first noted by Herbig (1962) that in a 
 $V$ vs. $B-V$ color-magnitude diagram (CMD) the ordinary low mass pleiad 
was fainter than expected, lying below the isochrone that fit the
higher mass stars. He took this as an indication that they had spent more
time gravitationally contracting than the blue giants had spent on
burning their hydrogen, concluding that the star formation process is
not coeval when considering a range of mass. He further suggested a ``bimodal star
formation'' model, in which the formation of low and high mass stars
are independent. 

Jones (1972) argued against this strong claim, stating that coevality would not have to be discarded if the fainter luminosity of the low mass stars were due not to an older contraction age, but to a stronger, reddening-free, extinction. Such would be the case if contracting stars were surrounded by dust cocoons, he noted, and supported his hypothesis with the fact that the low mass pleiades were not only fainter than expected for the isochrone corresponding to the nuclear age, but also fainter than the main sequence itself. However, no evidence for these mysterious reddening-free cocoons were ever found.

A possible step towards resolving the age discrepancy was proposed by Mazzei \& Pigatto (1989) who obtained a revised nuclear age for the Pleiades. Using isochrones derived from stellar models with significant convective core overshoot, the nuclear age was shifted from 65 to 150 Myr, much nearer to Herbig's contraction age ($>$220 Myr). This determination was fine-tuned by Meynet et al. (1993), who used a smaller amount of convective core overshoot and improved opacities, deriving a nuclear age of 100 Myr. However, this would of course leave the quandary of why the low mass stars in the Pleaides fall below the Zero Age Main Sequence (ZAMS).

 Van Leeuwen et al. (1987) arrived at an interesting result as a byproduct of a photometric survey of the Pleiades in search for variability among its G and K stars. They compared Pleiades K dwarfs to field K dwarfs in color-color diagrams, showing that the Pleiades K dwarfs were too blue in the ultraviolet for their $B-V$ colors. Based on their photometry, they concluded that the ultraviolet excess of the Pleiades starts at the Balmer jump, and recalled the widely known spectroscopic result linking age, rotation and activity to suggest that this excess could be due to chromospheric emission in the Balmer continuum. 

The long lasting problem was recently studied by Stauffer et al. (2003), who obtained further evidence that the spectral energy distributions of low mass dwarfs in the Pleiades are unusual. In this work, the authors argue that if the observed ultraviolet excess extended into the B waveband, the $B-V$ colors of active stars would be bluer than for older, less active stars, thus providing an explanation to the problem posed by Jones (1972). Following the argument, this unaccounted shift might have led, since Herbig, to erroneous determinations of contraction ages.

Stauffer et al. (2003) show evidence for flux excess in the B waveband by comparing spectra of a couple Pleiades K dwarfs against K dwarfs belonging to the older Praesepe cluster ($\sim$700 Myr). In the ratios of spectra, one sees not only residuals of emission in the cores of saturated lines, but also a residual in the blue continuum that, integrated, dominates over the quasi-emission features. They suggest that this excess is not uniquely due to chromospheric emission as speculated by van Leeuwen et al.(1987), but dominantly due to a combination of cool spots and warm plages, or perhaps to reprocessing of coronal X-ray flux. 

If these results seriously jeopardize the reliability of the
color index $B-V$ for determinations of contraction ages, Stauffer et
al. (2003) show in the same work a $V$ vs. $V-I$ CMD for the Pleiades
in which no shift with respect to the Praesepe cluster sequence can be
seen.

Three years before, however, Jeffries \& Tolley (1998) had noticed
another curious effect. While dating the cluster NGC~2547, they had
found that the contraction age was younger (14 Myr) in comparison to
the cluster's nuclear age (55 Myr). As their main sequence was tight
enough to discard that the younger age could be due to binarity or
photometric errors, and as they had around 60 PMS stars and only 3
stars in the upper main sequence, they had concluded that the
contraction age was statistically more likely.

Adding to the debate, Piskunov et al. (2004) note that ages from
turn-on fitting to the luminosity function of open clusters were
several times lower than the nuclear ages, stressing their belief that
contraction ages were more accurate than nuclear ages. However, it is
interesting to note that the age they ascribe to NGC~4755 (16 Myr) is
in close agreement with previous nuclear age determinations (10-15
Myr), even slightly older.

In this paper we investigate the general claim of the disagreement
between $\tau_n$ when compared to $\tau_c$ by constructing an
homogeneous UBVRI photometric database for four galactic open cluster
spanning the age range of 10 - 100 Myr, and deriving contraction and
nuclear ages in several color combinations. In the next sections we
describe the target selection and observations, photometry and
calibration to the standard system, leading to the construction of
color-color and color-magnitude diagrams, and the derivation of
$\tau_n$ and $\tau_c$. 

\begin{table}	       
\label{coordinates}    
\caption{The observed clusters and their coordinates.}		       
\begin{center}
\begin{tabular}{l c c c c} \hline     
Cluster	 & $\alpha$(2000) & $\delta$(2000) & $l$ & $b$   \\  \hline		
NGC~4755 & 12$^{\rm h}$ 54$^{\rm m}$ &$-$60$^{\rm o}$ 22' & 303$^{\rm o}$.21 &$+$02$^{\rm o}$.50  \\
NGC~2232 & 06$^{\rm h}$ 28$^{\rm m}$ &$-$04$^{\rm o}$ 50' & 214$^{\rm o}$.60 &$-$07$^{\rm o}$.41  \\
NGC~2547 & 08$^{\rm h}$ 10$^{\rm m}$ &$-$49$^{\rm o}$ 10' & 264$^{\rm o}$.45 &$-$08$^{\rm o}$.53  \\ 
NGC~2516 & 07$^{\rm h}$ 58$^{\rm m}$ &$-$60$^{\rm o}$ 52' & 273$^{\rm o}$.93 &$-$15$^{\rm o}$.88 \\  \hline
\end{tabular}
\end{center}
\end{table}

\begin{figure*}
\begin{center}
\resizebox{3.5cm}{3.5cm}{\includegraphics{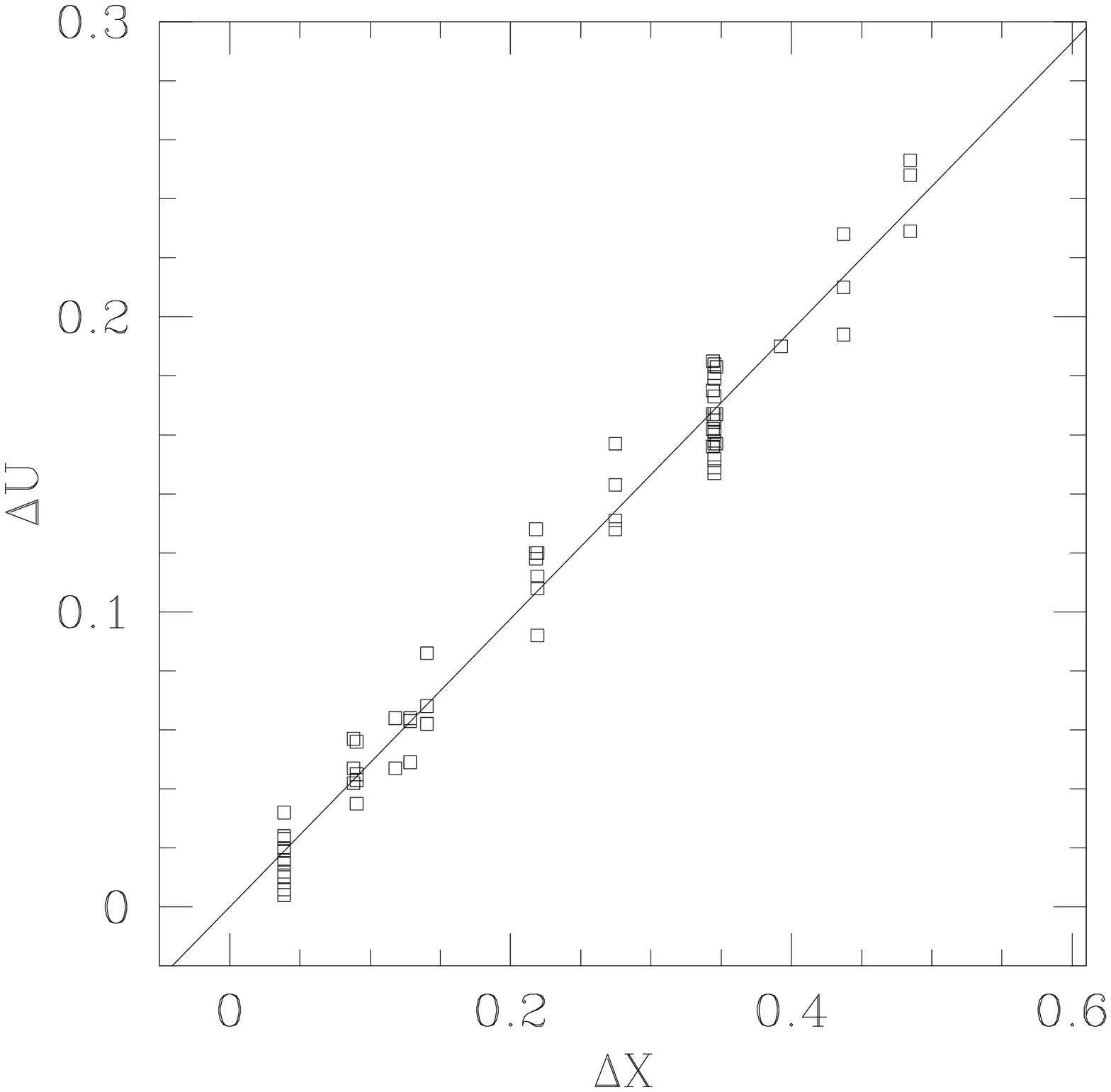}}
\resizebox{3.5cm}{3.5cm}{\includegraphics{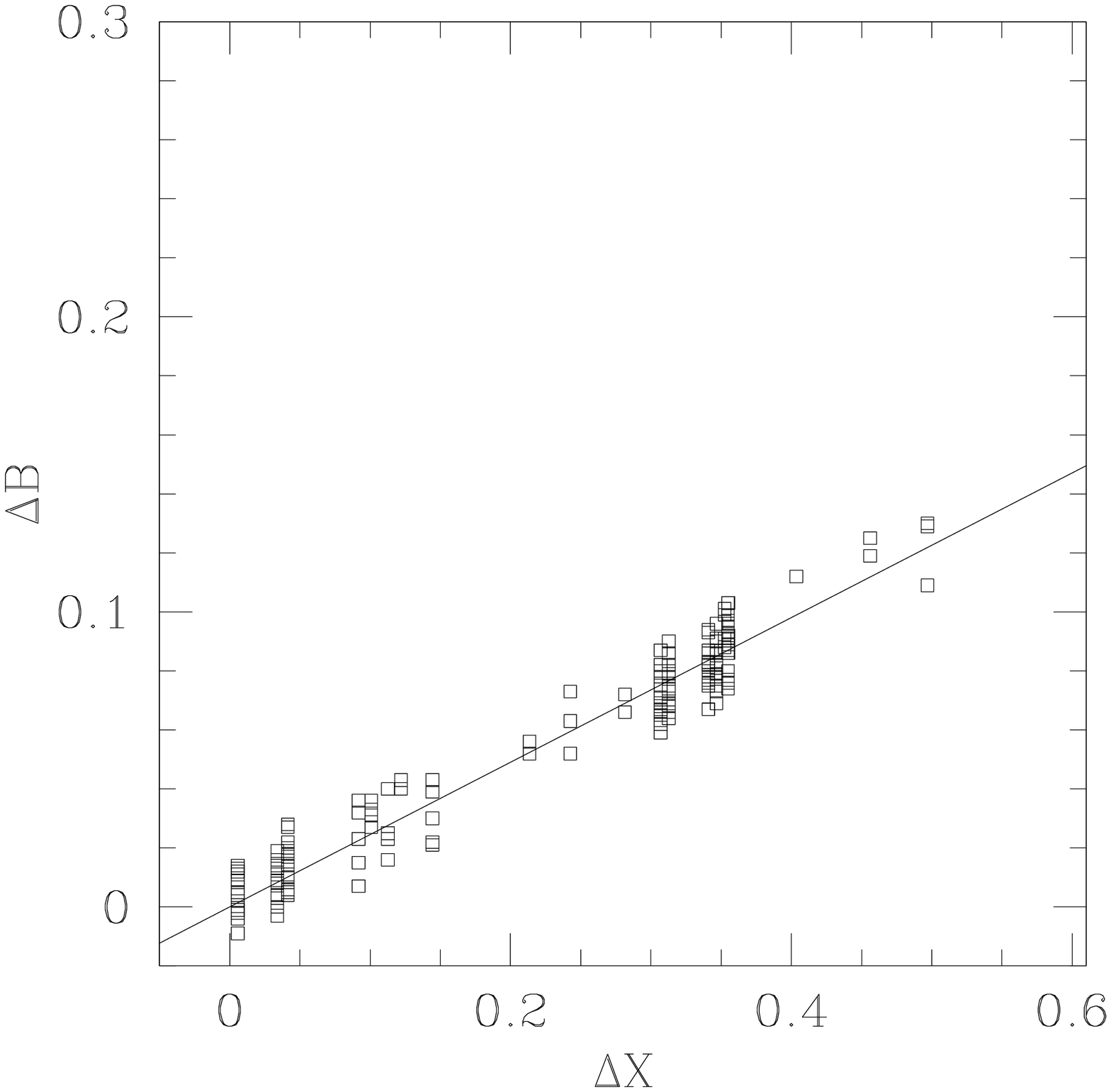}}
\resizebox{3.5cm}{3.5cm}{\includegraphics{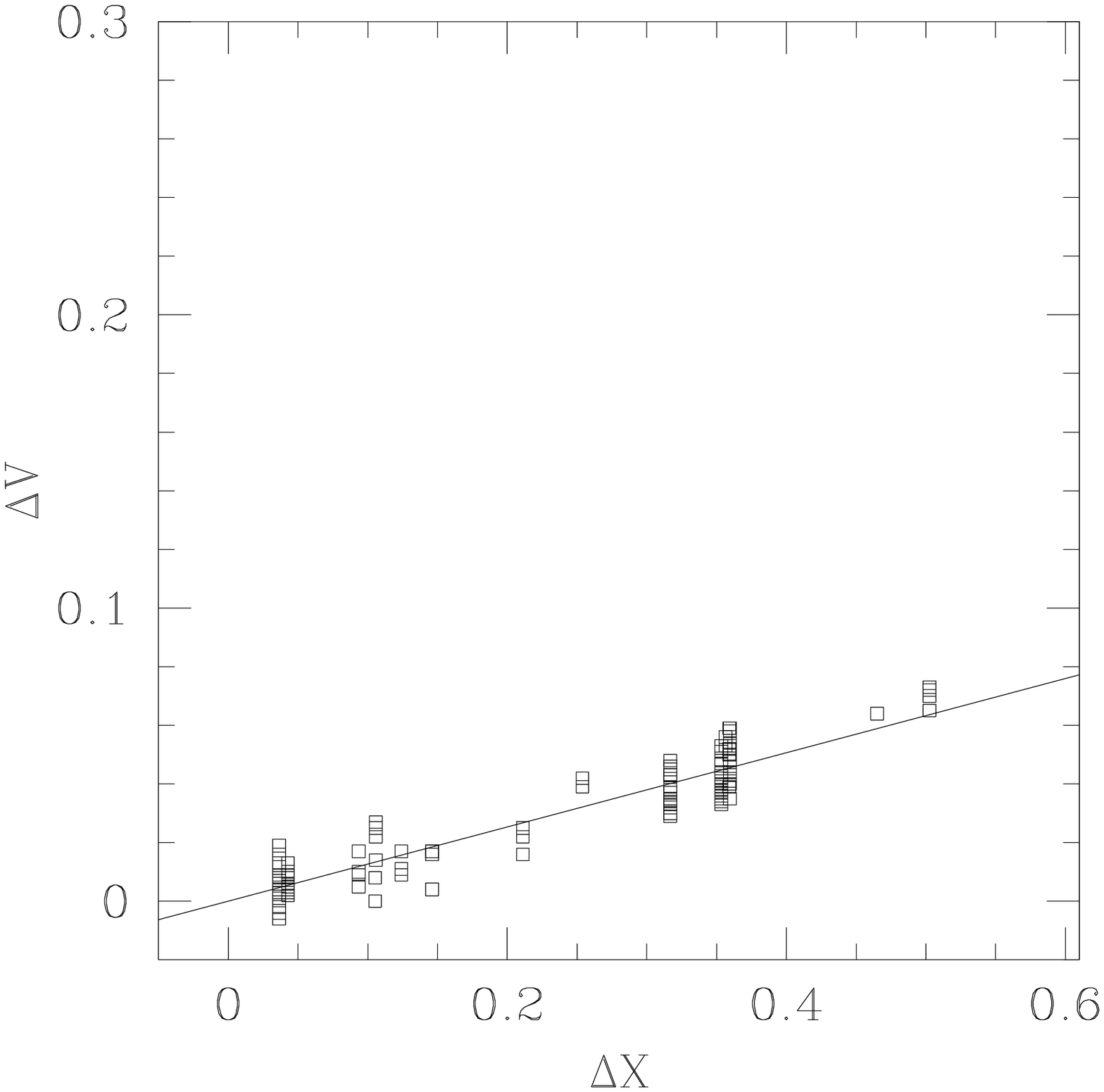}}
\resizebox{3.5cm}{3.5cm}{\includegraphics{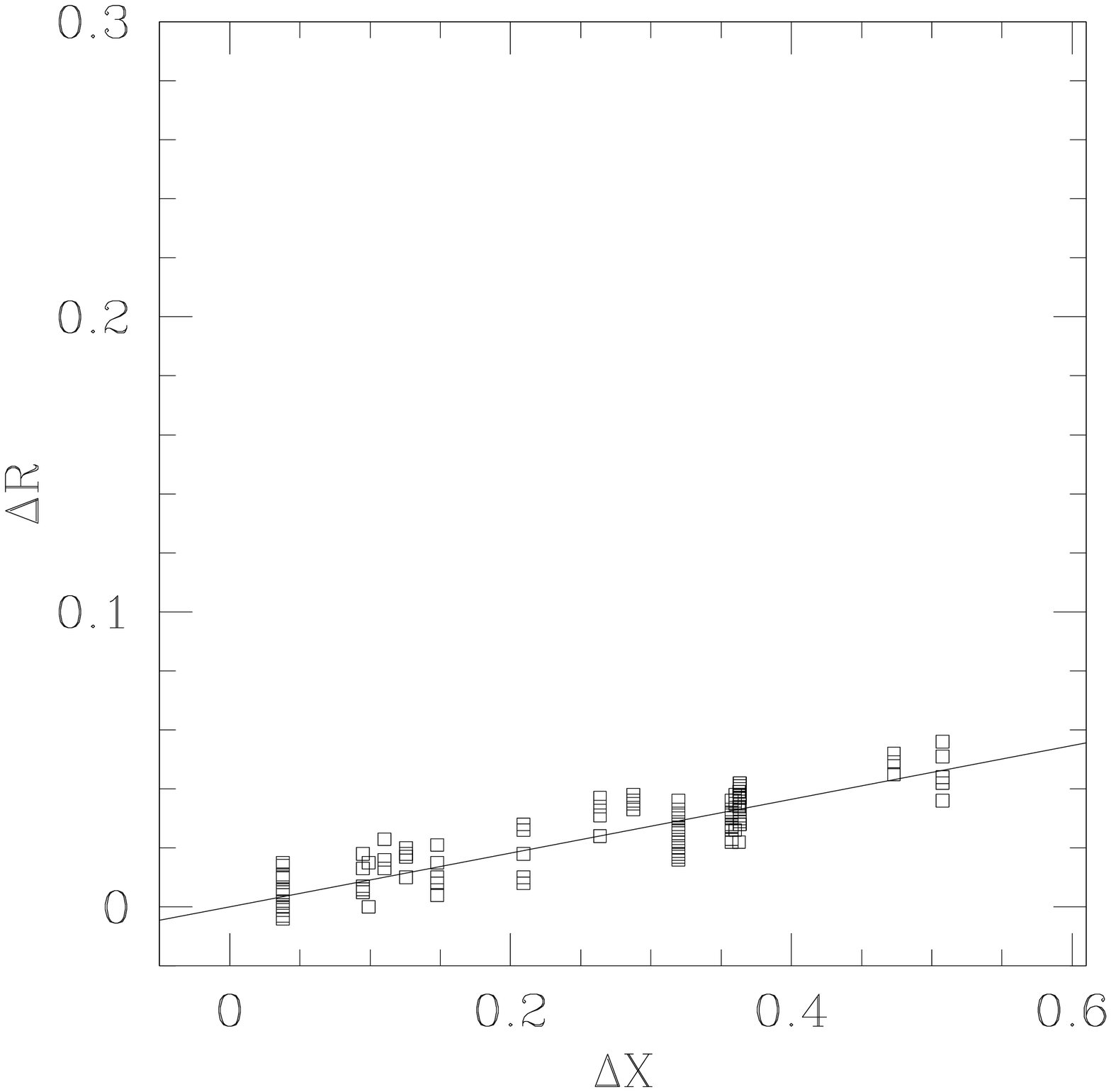}}
\resizebox{3.5cm}{3.5cm}{\includegraphics{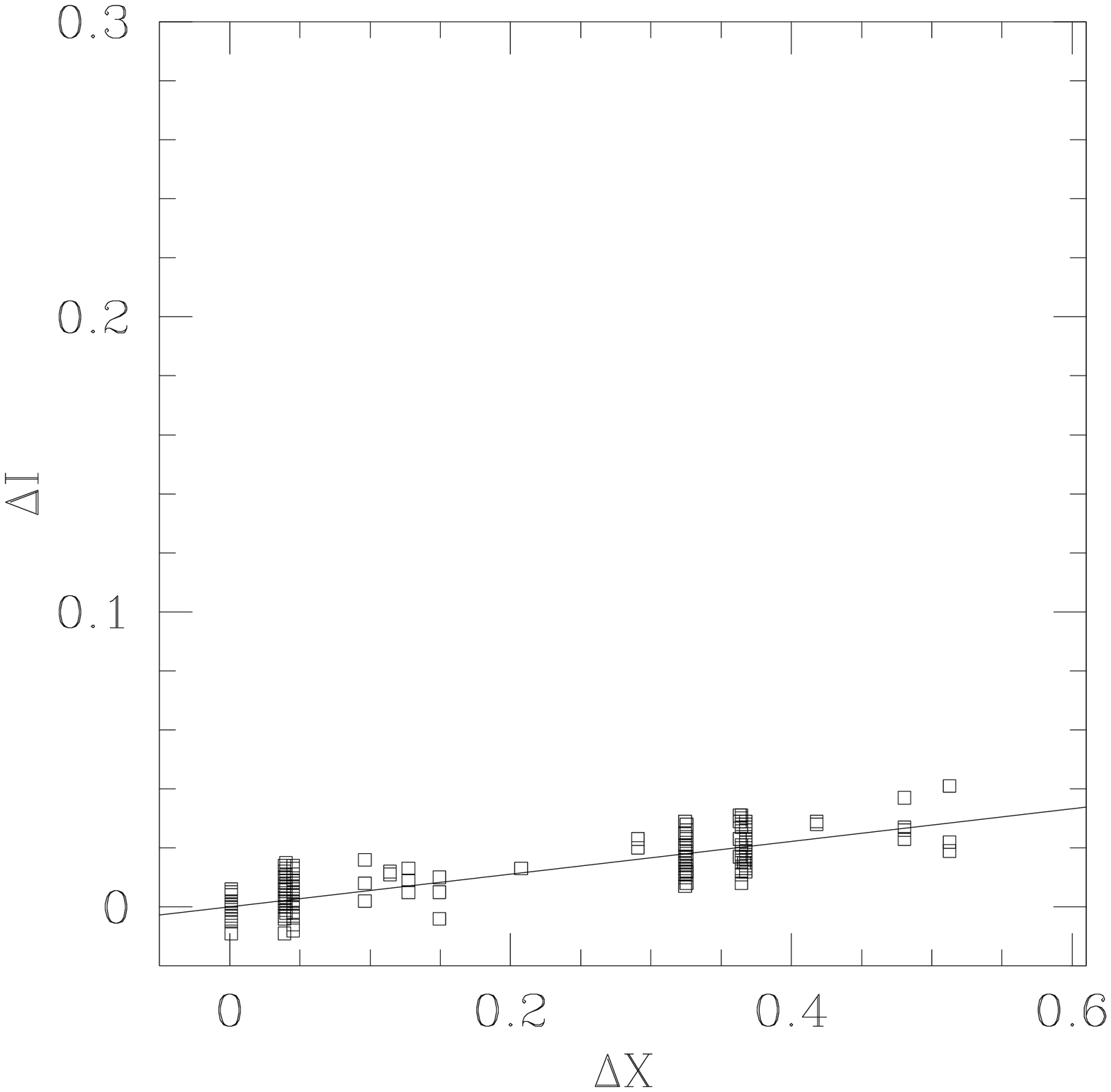}}
\end{center}
\caption[]{Best fits that determined the extinction correction for each filter. Data from all photometric nights are gathered together, resulting in average atmospheric extinction coefficients. In the plots, $\Delta M$ is the difference of a star's magnitude measured at different airmasses and $\Delta X$ is the difference of the airmasses (see Sec\,3.2 of Moitinho (2001)).}
\label{extinction}
\end{figure*}

\begin{table*}
\label{ext_table}
\caption[]{Night averaged atmospheric extinction coefficients and night-to-night photometric zeropoints referred to night 5. The linear fit rms for the extinction coefficients are  0.009 in U, 0.008 in B, 0.005 in V, 0.005 in R and 0.009 in I. }
\begin{center}
\begin{tabular}{c c c c c c c c c c c } \hline

Night  & $K_U$ & $ZP_U$ & $K_B$ & $ZP_B$ & $K_V$ & $ZP_V$ & $K_R$ & $ZP_R$ & $K_I$ & $ZP_I$ \\ \hline

1      & 0.496 & -0.040 & 0.273 & -0.051 & 0.144 & -0.040 & 0.111 & -0.034 & 0.051 & -0.037   \\ 
2      &       & ~0.012 &       & -0.025 &       & -0.010 &       & -0.012 &       & -0.018   \\
3      &       & ~0.005 &       & -0.008 &       & -0.003 &       & -0.005 &       & -0.009   \\
4      &       & ~0.002 &       & -0.004 &       & -0.007 &       & -0.007 &       & -0.003   \\ 
5      &       & ~0.000 &       & ~0.000 &       & ~0.000 &       & ~0.000 &       & ~0.000   \\
6      &       & ~0.018 &       & ~0.014 &       & ~0.011 &       & ~0.011 &       & ~0.001   \\ \hline
\end{tabular}
\end{center}
\end{table*} 

\section{Sample Selection and Observations}

	Data were collected at the CTIO 0.9m telescope during the second halves of six photometric nights in Feb. 2003. Therefore, the criteria for the sample selection were constrained to match the specifications for optical photometry under these conditions: the youngest but optically revealed clusters, not too distant and accessible on the second half of a February night at CTIO. The clusters thus selected were NGC~2232, NGC~2516, NGC~2547 and NGC~4755 (coordinates from Dias et al. (2002) in Table~\ref{coordinates}).

 	Images were acquired with a 2048 x 2048 Tek CCD at the 0.9m telescope and the standard Johnson UBV and Cousins RI filters available at CTIO. The field of view was 13'.7 x 13'.7, with a plate scale of 0".39 pixel$^{-1}$. Plates from the Digital Sky Survey (DSS) corresponding to the observed areas are shown in Fig.~\ref{images}. The gain was set at 3.1 e$^{-}$/adu and the readout noise was determined to be 4.9 e$^{-}$. We obtained a total of 206 images for the clusters (hereafter {\it on} fields) and 99 for control fields (hereafter {\it off} fields), being (10,10) in NGC~2232,(56,30) in NGC~2516, (40,45) in NGC~2547, and (61,5) in NGC~4755; where the notation means {\it (number of {\em on} images, number of {\em off} images)}. The exposure time for each filter is set to ($U$, $B$ , $V$, $R$, $I$ = 900, 390, 240, 300, 400s). 

 	The nearby ($\sim 30' away$) {\it off} fields images were taken with the same integration time, allowing for a photometry as deep as in the {\it on} fields. Also, for all clusters we acquired short 5s exposure images to get photometry for the bright stars still in the CCD's linear regime. For the bias level and flat field correction, zero second and twilight sky exposures in all filters were taken.  

	All the frames were processed within IRAF{\footnote {{\it Image Reduction and Analysis Facility} (IRAF) is distributed by NOAO, which is operated by AURA, Inc., under contract of NSF}}. Images were subjected to overscan correction and trimming using ARED/QUAD package, proper for the quadformat images obtained with CTIO ARCON. After this first processing, we proceeded to the usual bias subtraction and flatfielding within QUADPROC. 

\begin{table}
\label{transformation}
\caption[]{Coefficients for transformation to the standard system. The parameters refer to equations (1) to (7).}
\begin{center}
\begin{tabular}{c c c c c c} \hline
Parameter      	& Value  & error & Parameter     &  Value  & error  \\ \hline
$\alpha_0$      & ~2.968 & 0.004 & $\alpha_1$    & -0.015  & 0.005  \\
$\beta_0$       & ~0.190 & 0.003 & $\beta_1$     & ~1.113  & 0.003  \\
$\gamma_0$      & ~1.549 & 0.022 & $\gamma_1$    & ~0.014  & 0.020  \\
$\gamma_2$      & ~0.787 & 0.019 &               &         &        \\
$\delta_0$      & -0.883 & 0.030 & $\delta_1$    &  ~0.997 & 0.003  \\
$\epsilon_0$    & -0.044 & 0.002 & $\epsilon_1$  &  ~0.968 & 0.004  \\
$\phi_0$        & -0.048 & 0.003 & $\phi_1$      &  ~0.973 & 0.005  \\
$\zeta_0$       & ~2.969 & 0.004 & $\zeta_1$     &  -0.016 & 0.004  \\ \hline
\end{tabular}
\end{center}
\end{table}

\begin{figure*}
\begin{center}
\resizebox{16cm}{!}{\includegraphics{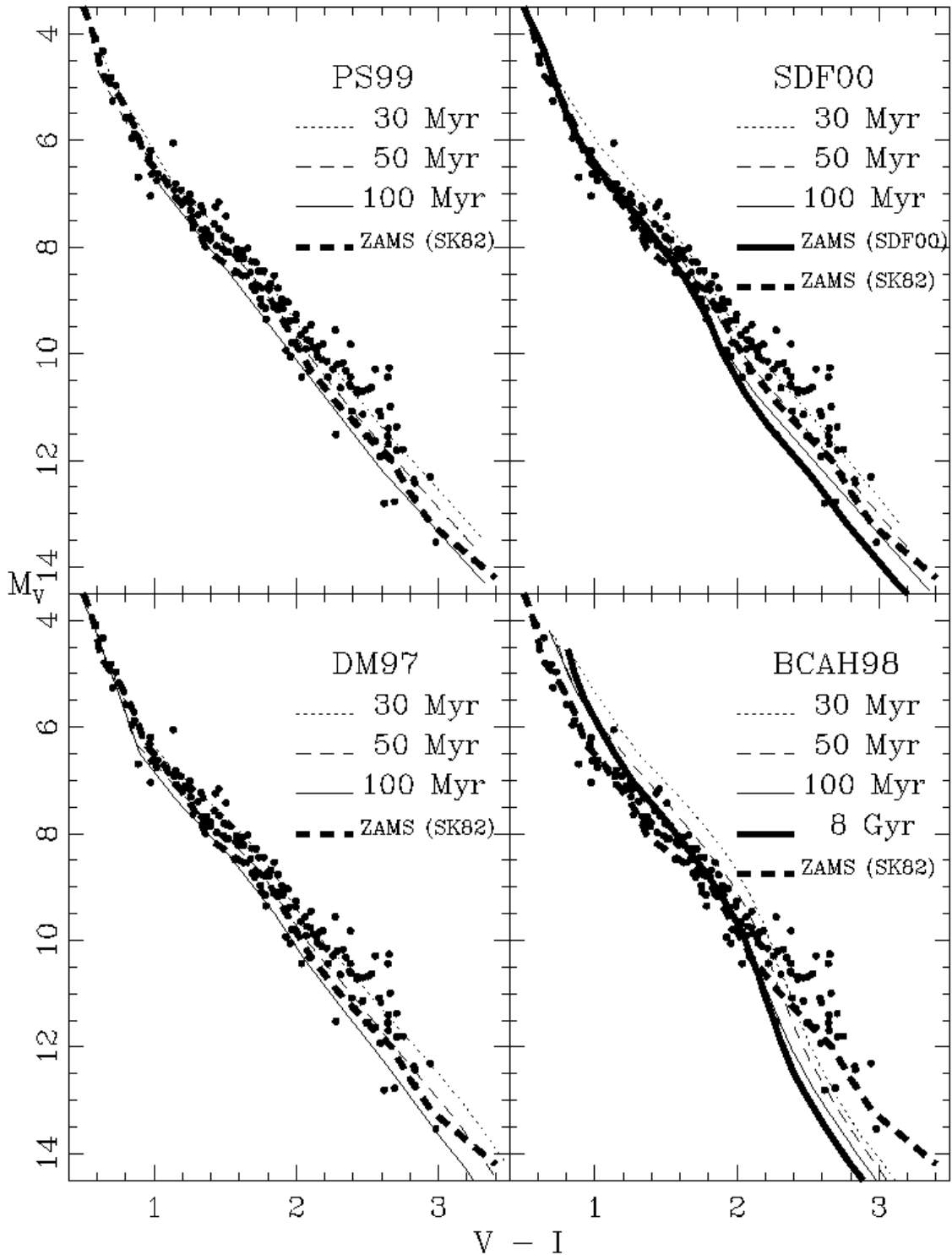}}
\end{center}
\caption[]{We use K and M dwarfs from the Gliese catalog to test the performance of the four PMS models we used to derive contraction ages. PS99, SDF00 and DM97 yield spuriously young ages. The model of BCAH98 provide isochrones that seem consistent with the presumably old age of the M field dwarfs in the range $1.6 < V-I < 2.0$. The empirical ZAMS of SK82 is shown as comparison.}
\label{gliese}
\end{figure*}

\begin{figure*}
\begin{center}
\resizebox{16cm}{!}{\includegraphics{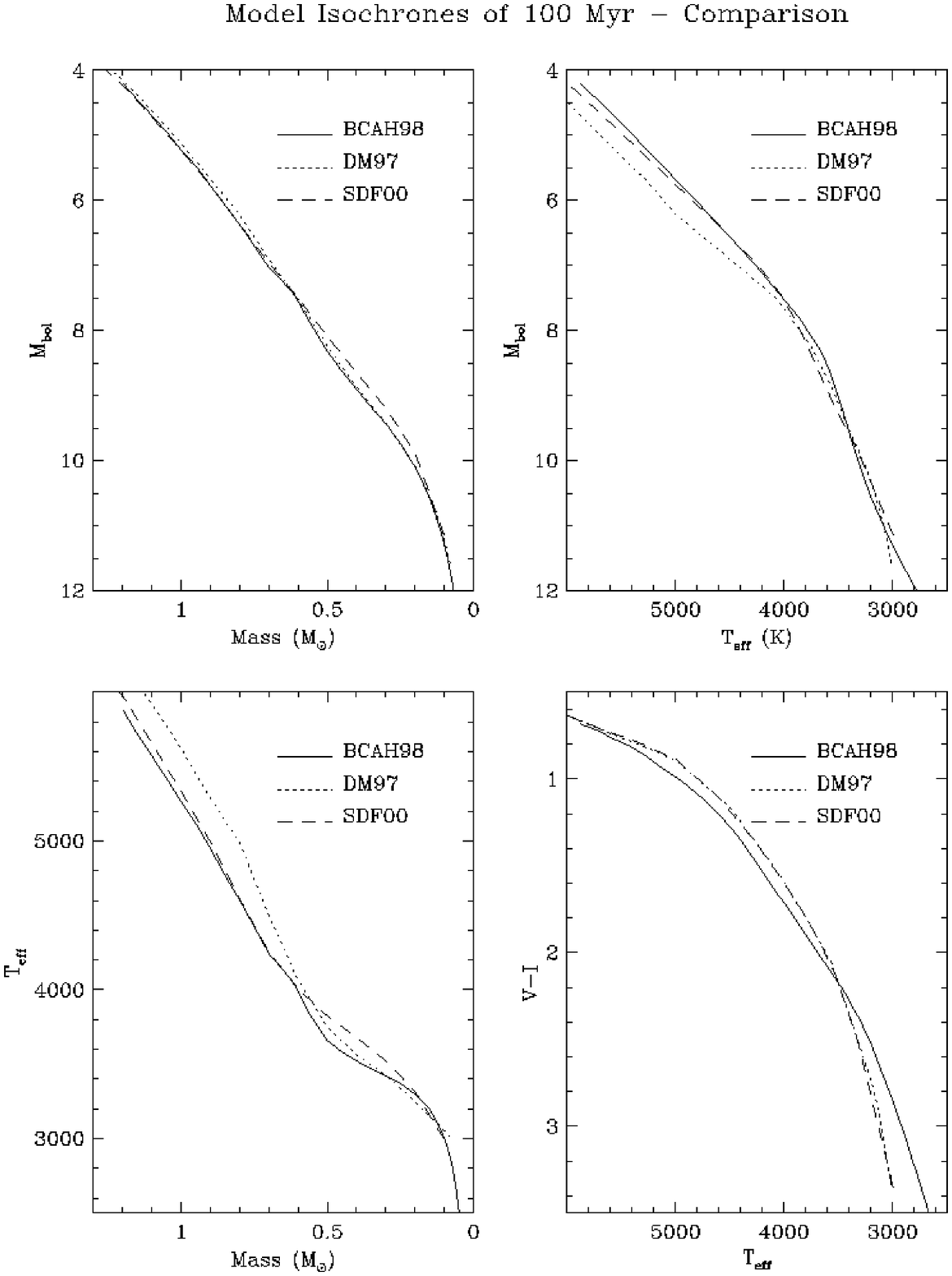}}
\end{center}
\caption[]{The PMS models plotted on theoretical planes. The models agree on Luminosity ($M{\rm _bol}$) and mass. DM97 yields hotter temperatures for the K dwarfs ($\sim$400~K hotter) but agrees with SDF00 and BCAH98 for the M dwarfs. The biggest difference between the models lies in the color-$T_{\rm eff}$ relation, crucial for an accurate age determination. The model of PS99 could not be used.}
\label{theo-plane}
\end{figure*}



\section{Atmospheric Extinction corrections and transformation to the Standard System}

	In order to correct for atmospheric extinction and to perform the transformation to the standard system of magnitudes, we observed two Landolt (1992) fields at several zenithal distances.  

Together, these two fields account for 30 standard stars observed across all nights in a broad range of airmasses, summing up to 135 frames taken. 

Photometry was performed with the IRAF/APPHOT package, using a 13.3" aperture (17 pixel radius; See Sect. 4 for details), which is approximately the same as the 14'' used by Landolt (1992) in his observations. Extinction is assumed to follow Bouguer's law, $M_i = M_{i0} + K_iX$, where $M_i$ is the observed magnitude in the band $i$, $M_{i0}$ is the magnitude the star would have in the band $i$ if observed outside the atmosphere, $K_i$ is the extinction coefficient in the band $i$ and $X$ stands for the airmass.   

	However, we used this equation in a slightly modified version, following a procedure described by Moitinho (2001). This version, being independent of the star used, allows to use several stars in the same linear regression. Plus, accuracy is favored because the linear regression to fit this equation has just one degree of freedom, instead of two as in the original version. 
        
	A per night analysis was not performed because of the limited number of standard stars and airmass ranges observed in some of the nights. Therefore, data collected across the nights were gathered and used together to determine average extinction coefficients for each filter. Linear regression was performed iteratively until no more points were rejected by a 2 sigma criterion. Fig.~\ref{extinction} shows the best fit achieved for each filter, and the extinction coefficients are tabulated in Table~2. 
      
	To correct for night to night variations on the instrumental magnitudes, one needs to choose one night to call the reference, and then compare the extinction corrected magnitudes of each night relatively to this standard.  This was chosen to be the 5$^{\rm th}$ night because it presented the biggest number of standard star measurements. Therefore, it is our best case where to perform extinction correction studies.

The zero points that must be added to each filter in each night to transform them to system of night 5 are also summarized in Table~2. The photometric relations between the instrumental system of the reference night and the standard system of magnitudes are assumed to be of the form: 	
\begin{eqnarray}
v-V   & = & \alpha_{0} 		+ \alpha_{1}      (B-V)                      \label{eq:calibv1}\\
(b-v) & = & \beta_{0}  		+ \beta_{1}       (B-V)    		     \label{eq:calibbv}\\ 
(u-b) & = & \gamma_{0} 		+ \gamma_{1}      (B-V)  + \gamma_{2} (U-B)  \label{eq:calibub}\\
(v-i) & = & \delta_{0} 		+ \delta_{1}      (V-I)                      \label{eq:calibvi}\\
(v-r) & = & \varepsilon_{0} 	+ \varepsilon_{1} (V-R) 		     \label{eq:calibvr}\\
(r-i) & = & \phi_{0} 		+ \phi_{1}        (R-I)                      \label{eq:calibri}\\
v-V   & = & \zeta_{0}           + \zeta_{1}       (V-I),                     \label{eq:calibv2}
\end{eqnarray} as thoroughly described in Moitinho (2001). Here, capital letters refer to the standard system, lower case to the instrumental system of the reference night.  

Performing least square fits, using the 30 observed standard stars, we determined the coefficients for equations (1) to (7), shown in Table~3.

\section{Photometry}

        Photometry was performed with the IRAF/DAOPHOT package (Stetson 1990). Since crowding was found not to be severe in most cluster fields, aperture measurements were extracted (except for NGC 4755; see below).

	A growth-curve analysis lead us to adopt a 17 pixel aperture radius ($\sim$13'' diameter) for the standard star measurements. For the cluster fields, magnitudes were obtained with a smaller aperture of radius equal to the image's typical full width at half maximum of the stellar profile. Aperture flux corrections, to be added to these small aperture measurements, were determined from the growth-curve analysis of $\sim$40 of the brightest isolated star in each image. By using small aperture measurements we minimize flux contamination from neighboring stars and substantially increase the signal-to-noise ratio of the fainter stars. 

	Following this procedure, magnitudes were measured and aperture corrected on the {\it on}/{\it off} fields for 1408/2074 stars in NGC~2232, 3461/2783 in NGC~2516 and 12813/5047 in NGC~2547. These photometric lists were then corrected for atmospheric extinction, transformed to the instrumental system by adding the zero points relative to the reference night and, finally, transformed to the standard system by the coefficients of Table~3. 

	Since in the case of NGC~4755 crowding was significant, photometry was obtained following the PSF fitting method (Stetson 1990). The PSF seemed to vary quadratically over the CCD, so we selected a great number of PSF stars in each frame, from 50 to 200. After selecting the stars, at least eight iterations were done to compute the quadratically variable PSF in each image. Some stars that had high {\it chi} values ($>$ 3) or that displayed odd looking residuals -- possibly due to unresolved blends, non-stellar nature or unidentified cosmetic defects -- after subtraction were eliminated from the PSF star list. PSF subtraction was performed for all the stars identified in the original image. The subtracted image was inspected for unidentified objects (mostly faint companions) which were used in contructing an improved list of stars. This process was repeated three times.

	The aperture correction in this case is the difference between the aperture photometry using the same radius as used in the standard stars and the PSF photometry just performed. In the images without neighbors, we measured the magnitudes of the PSF stars to compute this correction, excluding those with high photometric error ( $>$ 0.1mag). Finally, a second order multilinear fit in CCD X and Y positions was performed to compensate for the PSF's spatial variations.

\clearpage

\begin{figure*}
\begin{center}
\resizebox{16cm}{!}{\includegraphics{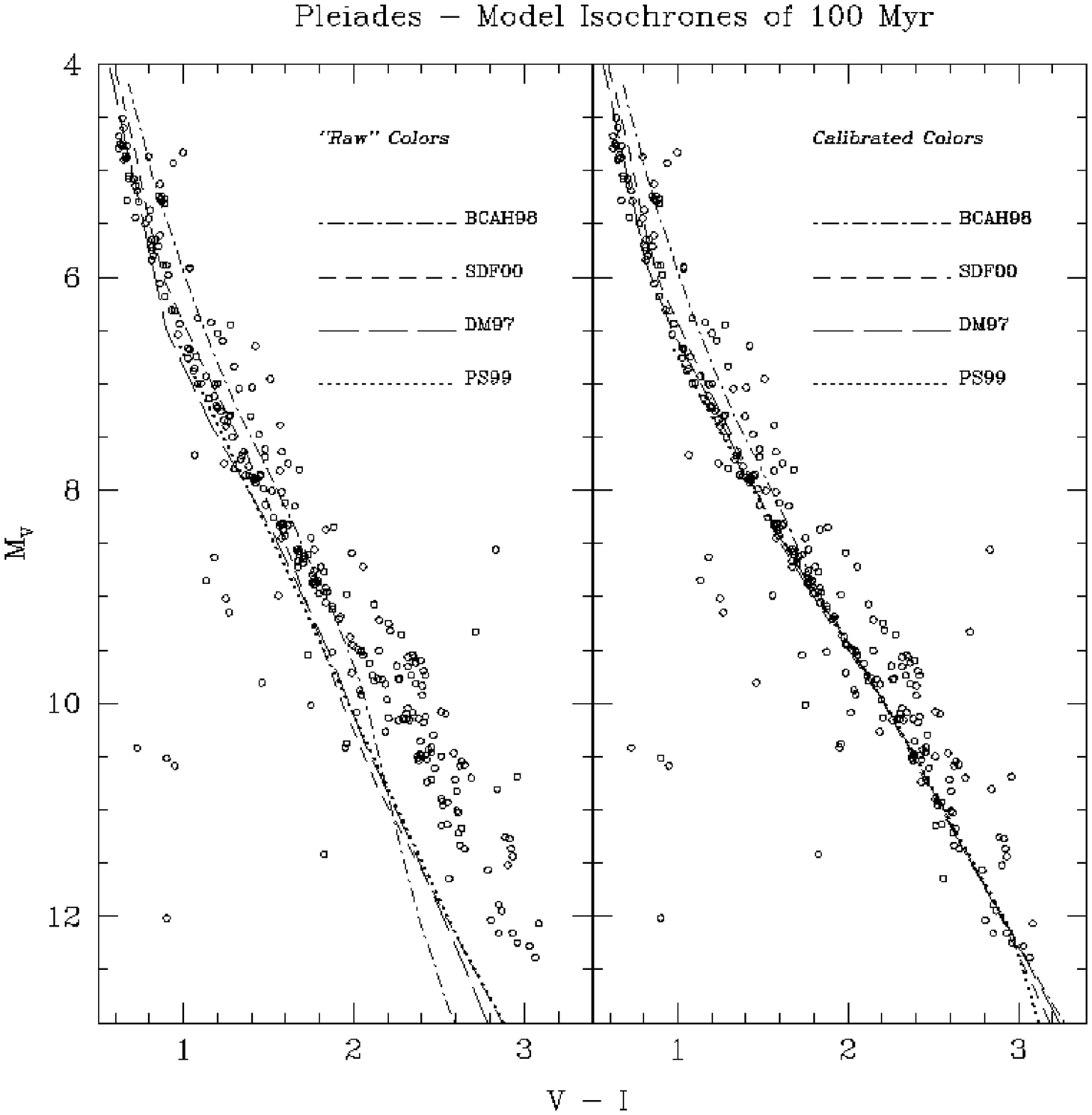}}
\end{center}
\caption[]{The construction of an empirical color-$T_{\rm eff}$ relation for young stars is done using the well-observed Pleiades cluster. For each PMS model, the isochrones of 100 Myr are forced to follow the photometry of low mass stars in the color range $1.6 < V-I < 3.0$ . The derived color-$T_{\rm eff}$ relation is then used on other clusters to derive consistent contraction ages.}
\label{calib-models}
\end{figure*}

\begin{table*}
\label{bright}
\caption[]{Comparison with previous studies.}
\begin{center}
\begin{tabular}{l c c r c c c r c c c r c c c r c c c r r l} \hline
\tiny Cluster      &\multicolumn{3}{c}{\tiny $V$}& &\multicolumn{3}{c}{\tiny $B-V$}& &\multicolumn{3}{c}{\tiny $U-B$}&&\multicolumn{3}{c}{\tiny $V-R$}& &\multicolumn{3}{c}{\tiny $V-I$}& \tiny Ref & \tiny Data \\ \cline{2-4}\cline{6-8}\cline{10-12}\cline{14-16}\cline{18-20}  
         &\tiny $\Delta$&\tiny $\sigma$&\tiny $N$ &&\tiny $\Delta$&\tiny $\sigma$&\tiny $N$ &&\tiny $\Delta$&\tiny $\sigma$&\tiny $N$&&\tiny $\Delta$&\tiny $\sigma$&\tiny $N$&&\tiny $\Delta$&\tiny $\sigma$&\tiny $N$&&       \\ \hline

\tiny  NGC~2232 &\tiny  0.014  &\tiny  --    &\tiny  1   &\tiny &\tiny  0.032  &\tiny  --    &\tiny  1   &\tiny &\tiny  0.023 &\tiny  --    &\tiny  1  &\tiny &\tiny    --  &\tiny  --    &\tiny  -- &\tiny &\tiny  --     &\tiny  --    &\tiny  -- &\tiny  1  &\tiny  pe  \\

\tiny&\tiny&\tiny&\tiny&\tiny&\tiny&\tiny&\tiny&\tiny&\tiny&\tiny&\tiny&\tiny&\tiny&\tiny&\tiny&\tiny&\tiny&\tiny&\tiny&\tiny&\tiny\\

\tiny NGC~2516&\tiny-0.010&\tiny0.025&\tiny2  &\tiny&\tiny 0.025&\tiny0.035&\tiny2  &\tiny&\tiny  -- &\tiny --  &\tiny-- &\tiny&\tiny   -- &\tiny   --&\tiny--&\tiny&\tiny   -- &\tiny --  &\tiny--  &\tiny 2&\tiny pe \\
\tiny        &\tiny-0.036&\tiny0.017&\tiny8  &\tiny&\tiny 0.018&\tiny0.024&\tiny17 &\tiny&\tiny0.038&\tiny0.022&\tiny11 &\tiny&\tiny   -- &\tiny   --&\tiny--&\tiny&\tiny   -- &\tiny --  &\tiny--  &\tiny 3&\tiny pe \\
\tiny        &\tiny-0.008&\tiny0.050&\tiny28 &\tiny&\tiny 0.005&\tiny0.033&\tiny29 &\tiny&\tiny0.049&\tiny0.046&\tiny20 &\tiny&\tiny   -- &\tiny   --&\tiny--&\tiny&\tiny   -- &\tiny --  &\tiny--  &\tiny 4&\tiny pe \\
\tiny        &\tiny-0.106&\tiny0.103&\tiny54 &\tiny&\tiny-0.011&\tiny0.061&\tiny53 &\tiny&\tiny0.069&\tiny0.037&\tiny35 &\tiny&\tiny   -- &\tiny   --&\tiny--&\tiny&\tiny   -- &\tiny --  &\tiny--  &\tiny 4&\tiny pg \\ 
\tiny        &\tiny-0.041&\tiny0.023&\tiny30 &\tiny&\tiny 0.016&\tiny0.029&\tiny41 &\tiny&\tiny0.045&\tiny0.017&\tiny30 &\tiny&\tiny   -- &\tiny   --&\tiny--&\tiny&\tiny   -- &\tiny --  &\tiny--  &\tiny 5&\tiny pe \\
\tiny        &\tiny 0.008&\tiny0.052&\tiny4  &\tiny&\tiny   -- &\tiny --  &\tiny-- &\tiny&\tiny  -- &\tiny --  &\tiny-- &\tiny&\tiny-0.049&\tiny0.038&\tiny3 &\tiny&\tiny   -- &\tiny --  &\tiny--  &\tiny 6&\tiny pe \\
\tiny        &\tiny 0.002&\tiny0.042&\tiny21 &\tiny&\tiny 0.000&\tiny0.030&\tiny30 &\tiny&\tiny0.059&\tiny0.050&\tiny22 &\tiny&\tiny   -- &\tiny   --&\tiny--&\tiny&\tiny   -- &\tiny --  &\tiny--  &\tiny 6&\tiny pe \\
\tiny        &\tiny 0.069&\tiny0.039&\tiny4  &\tiny&\tiny   -- &\tiny --  &\tiny-- &\tiny&\tiny  -- &\tiny --  &\tiny-- &\tiny&\tiny 0.202&\tiny0.014&\tiny3 &\tiny&\tiny 0.012&\tiny --  &\tiny 1  &\tiny 7&\tiny ccd\\
\tiny        &\tiny-0.008&\tiny0.016&\tiny185&\tiny&\tiny-0.015&\tiny0.015&\tiny214&\tiny&\tiny  -- &\tiny --  &\tiny-- &\tiny&\tiny   -- &\tiny   --&\tiny--&\tiny&\tiny-0.016&\tiny0.011&\tiny 178&\tiny 8&\tiny ccd\\
\tiny        &\tiny-0.019&\tiny0.036&\tiny11 &\tiny&\tiny-0.014&\tiny0.014&\tiny11 &\tiny&\tiny  -- &\tiny --  &\tiny-- &\tiny&\tiny   -- &\tiny   --&\tiny--&\tiny&\tiny-0.019&\tiny0.005&\tiny 8  &\tiny 9&\tiny ccd\\
\tiny        &\tiny 0.266&\tiny0.209&\tiny9  &\tiny&\tiny   -- &\tiny --  &\tiny-- &\tiny&\tiny  -- &\tiny --  &\tiny-- &\tiny&\tiny   -- &\tiny   --&\tiny--&\tiny&\tiny-0.148&\tiny0.145&\tiny 9  &\tiny10&\tiny ccd\\
\tiny        &\tiny-0.066&\tiny0.003&\tiny2  &\tiny&\tiny-0.013&\tiny0.017&\tiny3  &\tiny&\tiny0.102&\tiny0.018&\tiny 3 &\tiny&\tiny   -- &\tiny   --&\tiny--&\tiny&\tiny   -- &\tiny  -- &\tiny -- &\tiny11&\tiny pe \\
\tiny        &\tiny 0.016&\tiny0.051&\tiny252&\tiny&\tiny-0.001&\tiny0.020&\tiny239&\tiny&\tiny0.027&\tiny0.019&\tiny161&\tiny&\tiny   -- &\tiny   --&\tiny--&\tiny&\tiny-0.024&\tiny0.036&\tiny 212&\tiny12&\tiny ccd\\ 

&\tiny         &\tiny        &\tiny      &\tiny &\tiny         &\tiny        &\tiny      &\tiny &\tiny        &\tiny        &\tiny     &\tiny &\tiny        &\tiny        &\tiny     &\tiny &\tiny         &\tiny        &\tiny     &\tiny      &\tiny      \\

\tiny  NGC~2547 &\tiny  -0.037 &\tiny  0.021 &\tiny 34 	&\tiny &\tiny -0.017  &\tiny  0.023 &\tiny 40   &\tiny &\tiny 0.080  &\tiny  0.044 &\tiny 24  &\tiny &\tiny --     &\tiny --     &\tiny --  &\tiny &\tiny    --   &\tiny --     &\tiny --  &\tiny 13  &\tiny pe   \\
\tiny           &\tiny  -0.032 &\tiny  0.010 &\tiny 7 	&\tiny &\tiny  0.002  &\tiny  0.043 &\tiny 10 	&\tiny &\tiny   --   &\tiny --     &\tiny --  &\tiny &\tiny --     &\tiny --     &\tiny --  &\tiny &\tiny    --   &\tiny --     &\tiny --  &\tiny 2 &\tiny pe   \\
\tiny           &\tiny  -0.009 &\tiny  0.064 &\tiny 11 	&\tiny &\tiny -0.023  &\tiny  0.042 &\tiny 12 	&\tiny &\tiny   --   &\tiny --     &\tiny --  &\tiny &\tiny --     &\tiny --     &\tiny --  &\tiny &\tiny    --   &\tiny --     &\tiny --  &\tiny 14  &\tiny pe   \\
\tiny           &\tiny  -0.004 &\tiny  0.043 &\tiny 36 	&\tiny &\tiny -0.005  &\tiny  0.022 &\tiny 29 	&\tiny &\tiny   --   &\tiny --     &\tiny --  &\tiny &\tiny --     &\tiny --     &\tiny --  &\tiny &\tiny  -0.045 &\tiny  0.043 &\tiny 36  &\tiny 15  &\tiny ccd  \\
\tiny           &\tiny   0.013 &\tiny  0.020 &\tiny 336 	&\tiny &\tiny  0.010  &\tiny  0.011 &\tiny 303 	&\tiny &\tiny   --   &\tiny --     &\tiny --  &\tiny &\tiny --     &\tiny --     &\tiny --  &\tiny &\tiny  -0.029 &\tiny  0.013 &\tiny 308 &\tiny 16  &\tiny ccd  \\

\tiny           &\tiny         &\tiny        &\tiny      &\tiny &\tiny         &\tiny        &\tiny      &\tiny &\tiny        &\tiny        &\tiny     &\tiny &\tiny        &\tiny        &\tiny     &\tiny &\tiny         &\tiny        &\tiny     &\tiny      &\tiny      \\

\tiny  NGC~4755 &\tiny  -0.005 &\tiny  0.053 &\tiny  5   &\tiny &\tiny  -0.036 &\tiny  0.052 &\tiny  3   &\tiny &\tiny    --  &\tiny    --  &\tiny  -- &\tiny &\tiny    --  &\tiny  --    &\tiny  -- &\tiny &\tiny  --     &\tiny  --    &\tiny  -- &\tiny  17 &\tiny   pe \\
\tiny           &\tiny   0.178 &\tiny  0.101 &\tiny  110 &\tiny &\tiny  -0.091 &\tiny  0.013 &\tiny  40  &\tiny &\tiny    --  &\tiny    --  &\tiny  -- &\tiny &\tiny    --  &\tiny  --    &\tiny  -- &\tiny &\tiny  --     &\tiny  --    &\tiny  -- &\tiny   17 &\tiny   pg \\
\tiny           &\tiny  -0.005 &\tiny  0.031 &\tiny  35  &\tiny &\tiny  -0.040 &\tiny  0.027 &\tiny  29  &\tiny &\tiny  0.064 &\tiny  0.019 &\tiny  24 &\tiny &\tiny    --  &\tiny  --    &\tiny  -- &\tiny &\tiny  --     &\tiny  --    &\tiny  -- &\tiny  18 &\tiny   pe \\
\tiny           &\tiny  -0.029 &\tiny  0.045 &\tiny  84  &\tiny &\tiny  -0.030 &\tiny  0.064 &\tiny  82  &\tiny &\tiny  0.042 &\tiny  0.095 &\tiny  70 &\tiny &\tiny    --  &\tiny  --    &\tiny  -- &\tiny &\tiny  --     &\tiny  --    &\tiny  -- &\tiny  18 &\tiny   pg \\
\tiny           &\tiny  -0.017 &\tiny  0.011 &\tiny  18  &\tiny &\tiny  -0.009 &\tiny  0.015 &\tiny  20  &\tiny &\tiny    --  &\tiny    --  &\tiny  -- &\tiny &\tiny    --  &\tiny  --    &\tiny  -- &\tiny &\tiny  --     &\tiny  --    &\tiny  -- &\tiny   19 &\tiny  ccd \\
\tiny           &\tiny   0.021 &\tiny  0.021 &\tiny  23  &\tiny &\tiny  -0.023 &\tiny  0.009 &\tiny  17  &\tiny &\tiny  0.071 &\tiny  0.023 &\tiny  17 &\tiny &\tiny    --  &\tiny  --    &\tiny  -- &\tiny &\tiny  --     &\tiny  --    &\tiny  -- &\tiny  20 &\tiny   pe \\
\tiny           &\tiny   0.040 &\tiny  0.012 &\tiny  49  &\tiny &\tiny  -0.042 &\tiny  0.015 &\tiny  49  &\tiny &\tiny    --  &\tiny    --  &\tiny  -- &\tiny &\tiny  0.007 &\tiny  0.016 &\tiny  17 &\tiny &\tiny  -0.029 &\tiny  0.052 &\tiny  68 &\tiny   21 &\tiny  ccd \\
\tiny           &\tiny   0.013 &\tiny  0.012 &\tiny  151 &\tiny &\tiny  -0.017 &\tiny  0.016 &\tiny  145 &\tiny &\tiny    --  &\tiny    --  &\tiny  -- &\tiny &\tiny    --  &\tiny  --    &\tiny  -- &\tiny &\tiny  --     &\tiny  --    &\tiny  -- &\tiny  22 &\tiny ccd \\\hline

\end{tabular}		   	       
\end{center}

\begin{tabbing}
6  - Feinstein et al. (1973)kkkk\=9  - Jeffries et al. (1998)kk\=15 - Jeffries \& Tolley (1998)kkk\= \kill

1  - Clari\a'{a} (1972)\>         7  - Hawley et al. (1999)  \>   13 - Clari\a'{a} (1982)\>	 19 - Kjeldsen \& Frandsen (1991)\\    
2  - Cousins \& Stoy (1962)\>	  8  - Jeffries et al. (2001)\>   14 - Fernie (1959)\>		 20 - Perry et al. (1976)  \\   
3  - Dachs (1970)\>		  9  - Jeffries et al. (1998) \>  15 - Jeffries \& Tolley (1998)\>21 - Sagar \& Cannon (1995)\\	   
4  - Dachs \& Kabus (1989)\>	  10 - Micela et al. (2000) \>    16 - Naylor et al. (2002) \>	  22 - Sanner et al. (2001) \\        
6  - Feinstein et al. (1973)kkkk\=9  - Jeffries et al. (1998)kk\=15 - Jeffries \& Tolley (1998)kkk\= \kill
5  - Eggen (1972)	\>	  11 - Pedredos (1984)	   \>    17 - Arp \& van Sant (1958)\>\\	  
6  - Feinstein et al. (1973)\>	  12 - Sung et al. (2002)   \>    18 - Dachs \& Kaiser (1984)\>\\	    
\end{tabbing}

\begin{center}
\begin{tabular}{l c c c c c c c c c c c c c c c c c c c c c c c c c c c c c c c c c c c c c c c c l } \hline
                 & & & & & & & & & & & & & & & & & & & & & & & & & & & & & & & & & & & & & & & &\\ 
\end{tabular}
\end{center}

\end{table*}

	The full photometric list of NGC~4755 has 7188 stars in the cluster images, and 6298 in the control field. To build these lists, measurements were subjected to {\it chi} $<3$, {abs(\it sharp)} $<$ 1 and $\sigma_{\rm mag} <$ 0.1 cut-offs. Repeated measurements were weight-averaged, with the weight given by the errors output by ALLSTAR.

\section{Bright Star Photometry}

	A quick look at our images reveals that even for the shortest exposure times the brightest stars were saturated. Due to this, the blue luminous edge of our cluster sequences have an artificial cut-off. For some clusters, as NGC~2232, the cluster sequence starts as faint as 3.5 magnitudes below bright star data published in other studies.

	Because the brighter stars are crucial for deriving nuclear ages from isochrone fitting in color-magnitude diagrams, we have included sources from other studies, when available.

\begin{figure*}
\begin{center}
\resizebox{8cm}{!}{\includegraphics{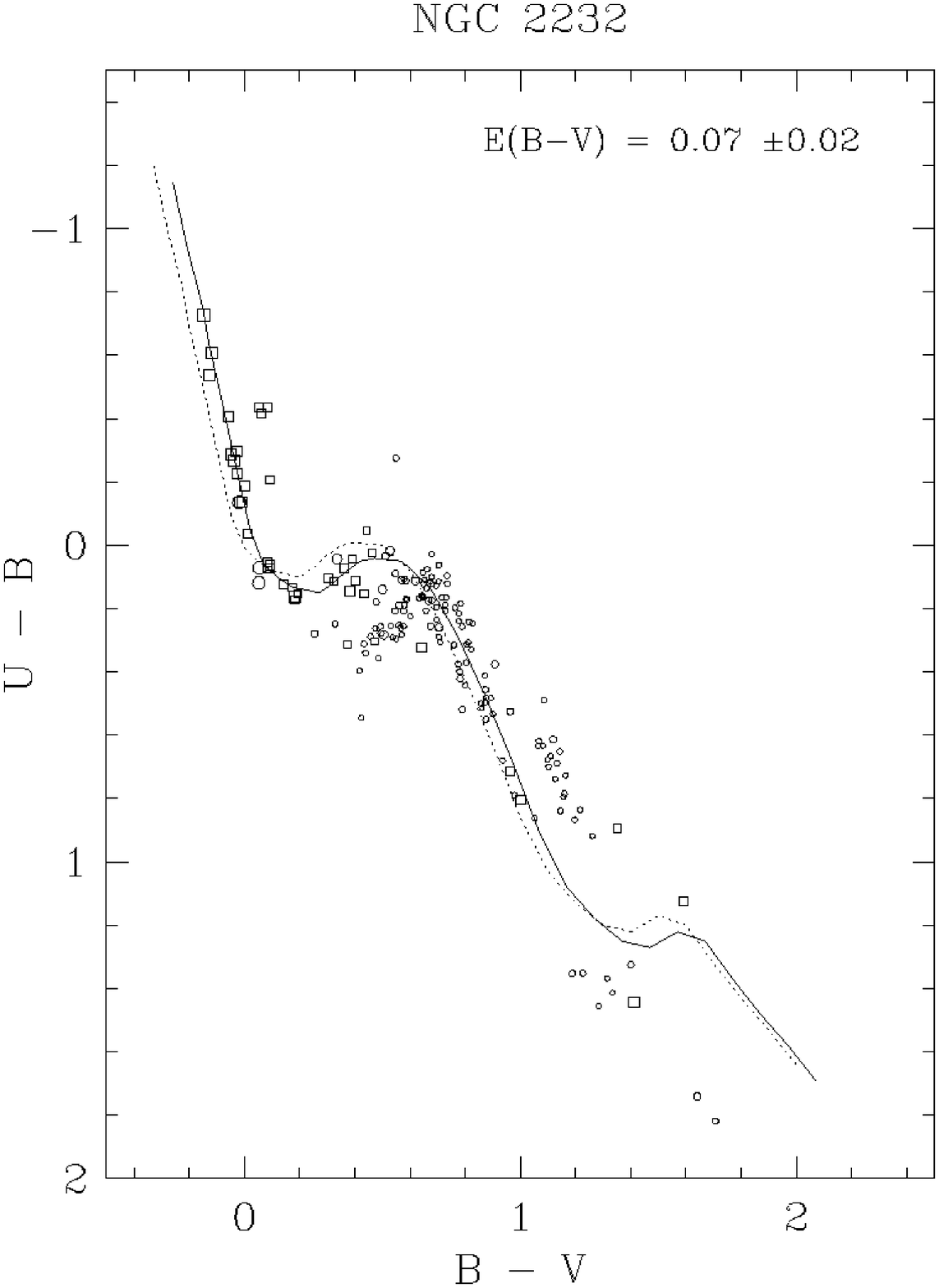}}
\resizebox{8cm}{!}{\includegraphics{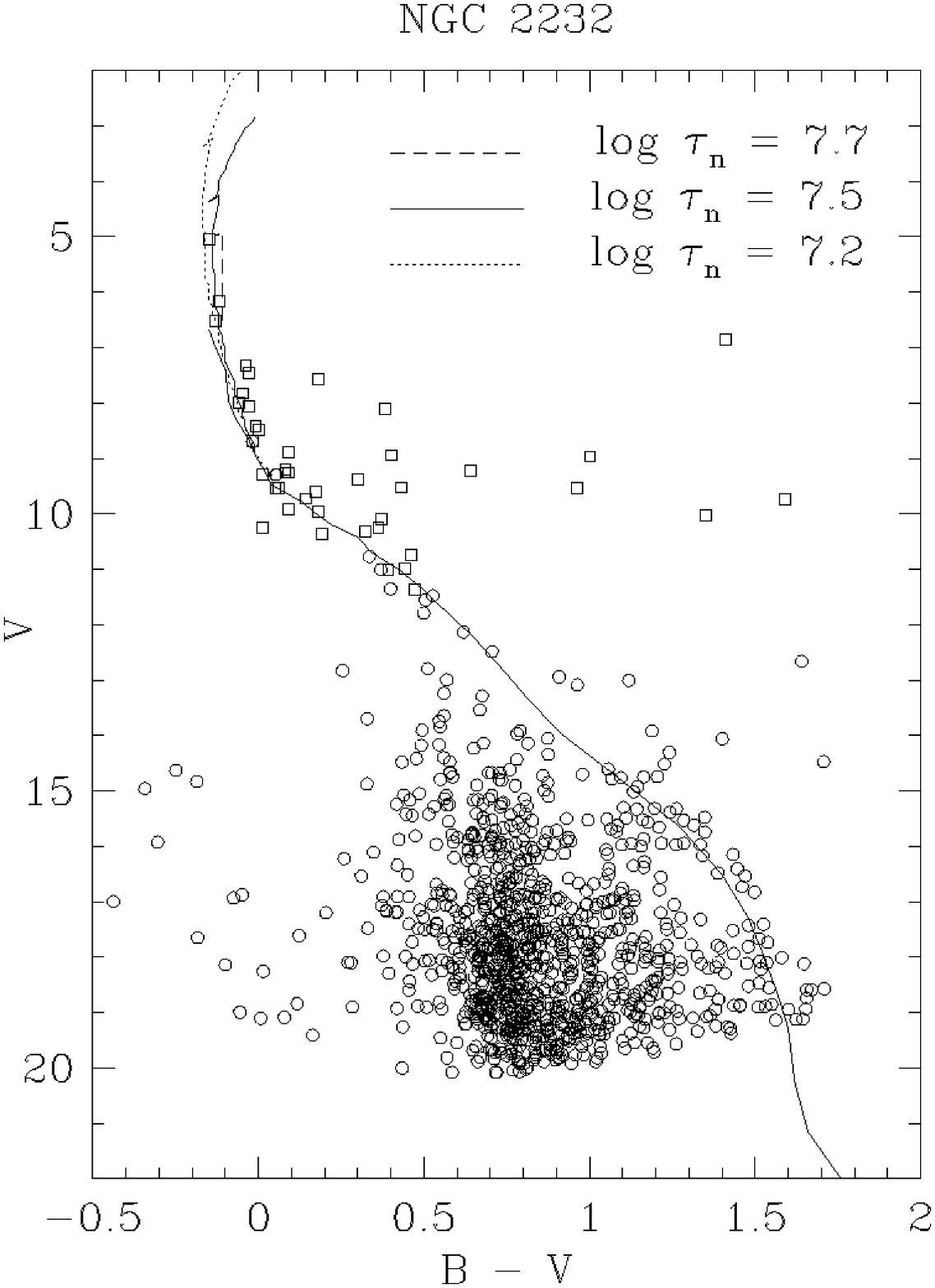}}
\resizebox{8cm}{!}{\includegraphics{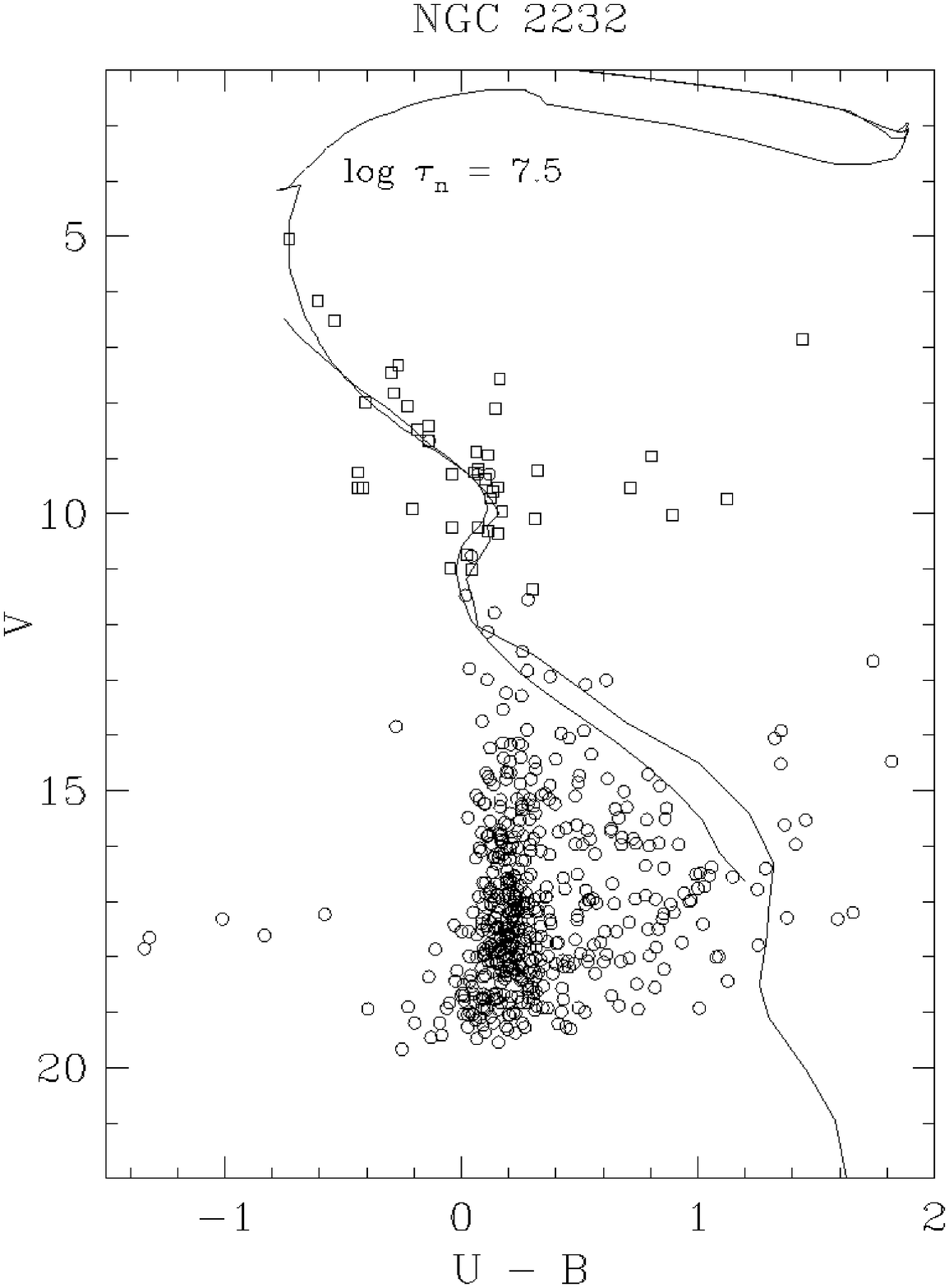}}
\end{center}
\caption[]{Diagrams for NGC~2232. A sparsely populated main sequence can be seen rising above the dense region of field stars. Reddening is fitted in the color-color diagram, with a best fit of $E(B-V)=0.07 \pm 0.02$. The size of the symbols is proportional to the $V$ magnitude. The dotted line is a SK82 ZAMS and the solid line is the adopted, reddened SK82 ZAMS which fits the observations. Distance is fitted in the $U-B$ vs. $V$ CMD, where the stars immediately brighter than the Balmer Jump show greater sensitivity than the sparse populated main sequence in the $B-V$ vs. $V$ CMD. Its value is $V-M_V=7.60\pm0.15$, or 300$\pm$20 pc. Post main sequence stars are fitted by a range of isochrones from 7.2 to 7.7 in log age (dashed lines), being 7.5 (solid line) the best fit in the $B-V$ CMD. In the $V$ vs. $U-B$ diagram, age is constrained by the position of the brightest star, HD\,45546 (see text). Its value is $\log \tau_n = 7.50$. Empty circles refer to our photometry whilst empty squares refer to corrected literature photometry.}
\label{isos2232}
\end{figure*}

\begin{figure*}
\begin{center}
\resizebox{16cm}{!}{\includegraphics[angle=270]{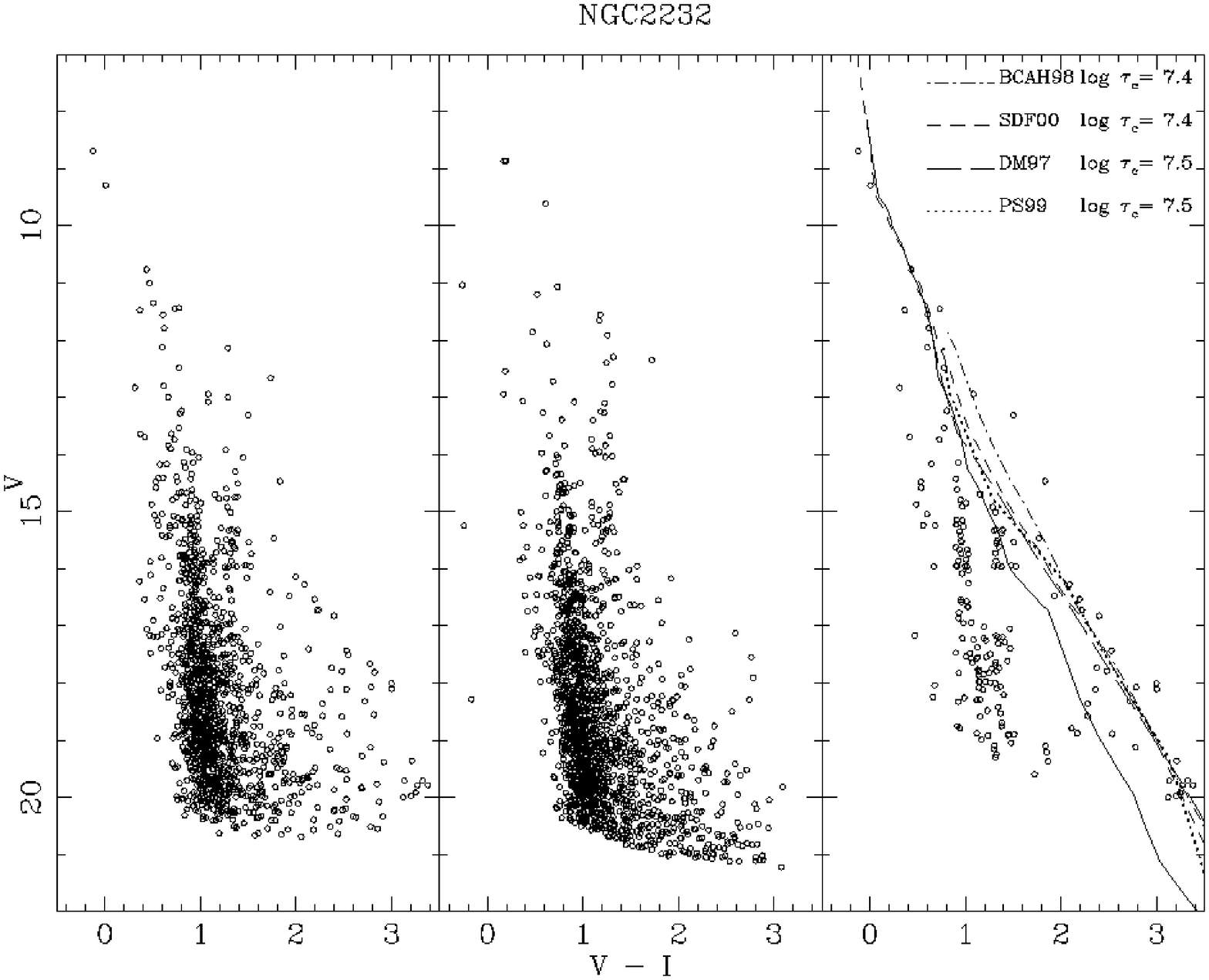}}
\resizebox{8cm}{!}{\includegraphics{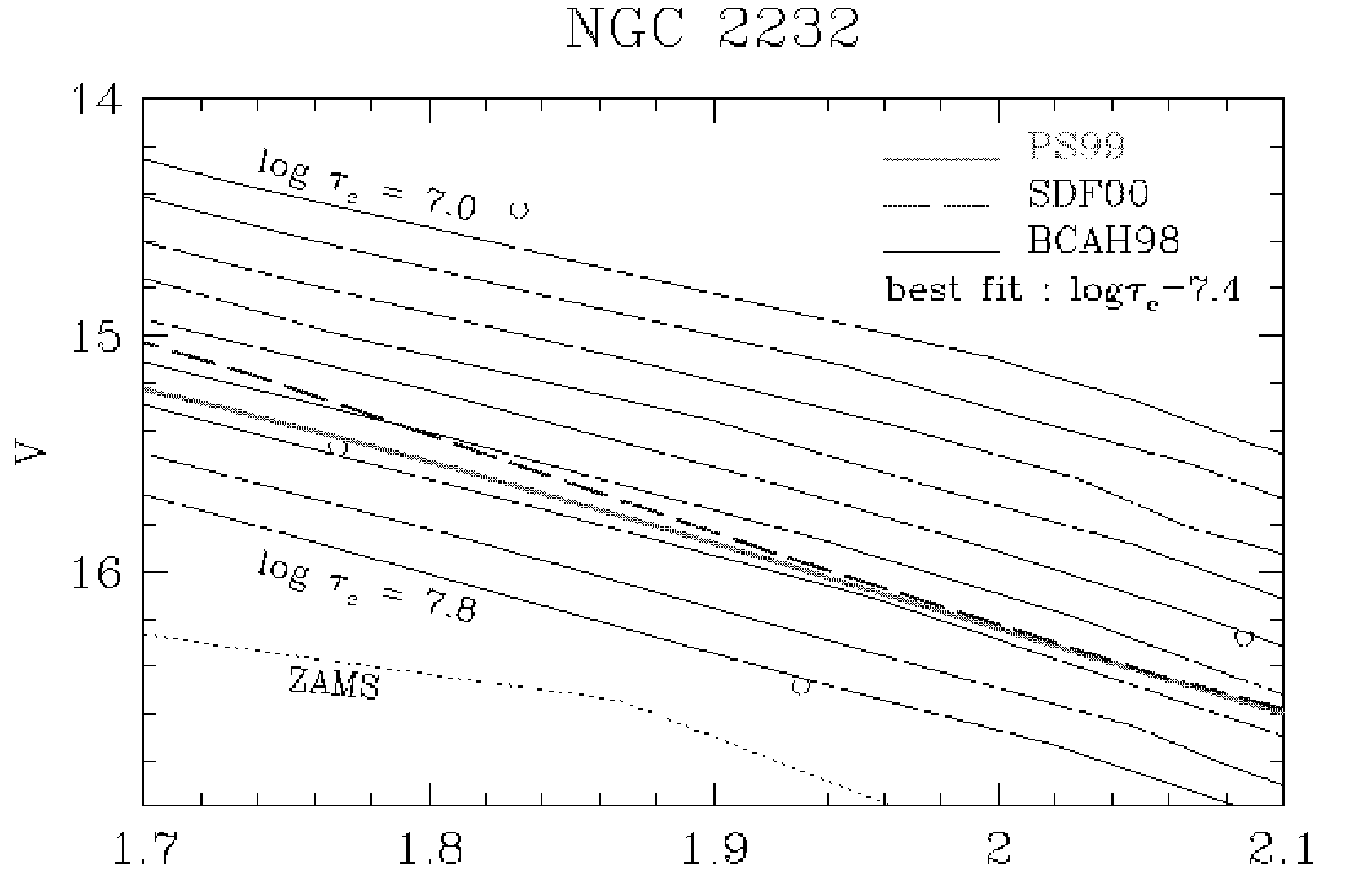}}
\end{center}
\caption[]{{\it Upper.} $V$ vs. $V-I$ color magnitude diagrams of NGC~2232. The diagram at the right is the result of statistical subtraction of the cluster diagram ({\it left}) by a nearby control field ({\it middle}). The stripe of red faint stars not erased by statistical subtraction is matched by the PMS models, an evidence that they belong to the cluster. BCAH98 and SDF00 assert contraction ages of 25~Myr, while DM97 and PS99 yield 32\,Myr. The SK82 ZAMS is plotted as solid line.\\
{\it Lower.} Determination of contraction age for NGG~2232 using the uncalibrated BCAH98 isochrones. In the range $1.6 < V-I < 2.0$ we search for the BCAH98 isochrone whose location in the CMD best agrees with the iscohrones of the models just fitted before. As seen from the figure, this corresponds to $\log\tau_c=7.4$. The SK82 ZAMS is shown as a dotted line for comparison.
}
\label{isoVI2232}
\end{figure*}

	Some care must be taken when including literature photometry for the brightest stars. Different instrumental setups and standard stars can introduce important differences between studies. So photometry of stars in common with other studies should be compared and correction terms should be derived and applied to those other data. Without this care, the simple insertion of data from the literature would generate distorted turnoff points. 

\subsection{Astrometry} 

	As said, to transform the literature photometry to our system, we must recognize stars in common in the two databases. The cross-identification of our stars and those found in the literature is done by positional coincidence. Therefore, first of all, we needed to do the astrometry in our frames, computing the world coordinates solution to transform the instrumental (x,y) coordinates to equatorial ($\alpha$,$\delta$). 

	Our astrometric solution was tied to the system defined by the US Naval Observatory Guide Star Catalog (Monet et al. 2003) using ESO's (European Southern Observatory) SKYCAT tool. An average of 40 stars were matched in each frame, which allowed for solutions with rms up to 0.5 pixel (0.2''). All solutions were computed making use of the tasks CCMAP, CCSETWCS and CCTRAN, within the IRAF package images/imcoords. The astrometric and photometric data are shown in Tables 6, 7 , 8 and 9 (corresponding to NGC~2232, NGC~2516, NGC~2547 and NGC~4755, respectively) in the format $\alpha^h$, $\alpha^m$, $\alpha^s$, $\delta^o$, $\delta^\prime$, $\delta^{\prime\prime}$, $U$, $B$, $V$, $R$, $I$, $\sigma_U$, $\sigma_B$, $\sigma_V$, $\sigma_R$, $\sigma_I$, meaning right ascension (hours, minutes, seconds), declination (degrees,arcminutes,arcseconds), $UBVRI$ magnitudes and their respective photometric errors.

\subsection{Literature Sources}

	Coordinates for a number of stars in each cluster were taken from WEBDA{\footnote {http://obswww.unige.ch/webda/}} and cross-identified with the coordinates from our astrometry. The recognition is done by positional coincidence within a 1'' radius tolerance, taking the closest stars within this distance. Magnitudes are then compared in order to determine if correction terms are necessary. 
	
	The mean of the differences between our photometry and literature photometry (left by a 2$\sigma$ clipping algorithm) is the correction term to be summed. These corrections are summarized in Table~4. Also indicated are the references where the photometric data were taken from. 

	For NGC~2232, a single star in common was found to correct the literature photometry. This is not surprising, as the cluster is poor in number of stars and extends over an angular diameter of $\sim$ 53' (Dias et al., 2002). With one single 13.8' x 13.8' frame, we are not observing the whole extent of the cluster. 

\section{Analysis and Results}

\subsection {Reddening and Distances}

	To relate the excess in each color to the excess in the color ($B-V$), $E(B-V)$, we used the standard value for the ratio of total-to-selective absorption, $R_V = 3.1$ and for the reddening slope, $E(U-B)/E(B-V)=0.72$ (e.g., Turner 1989). For other color-color combinations, we use the values given by Strai\v{z}ys (1992):

\begin{equation}
\label{slopes}
\left[
\begin{array}{c}
E(V-R)\\
E(V-I)\\
E(R-I)
\end{array}
\right]=E(B-V)\left[
\begin{array}{c}
0.56\\
1.25\\
0.69
\end{array}
\right].
\end{equation}

The adopted method for deriving reddening is the standard method of shifting a Schmidt-Kaler (1982, hereafter SK82) ZAMS in the direction defined by the reddening slope in a color-color diagram ($U-B$ vs. $B-V$ or $R-I$ vs. $B-V$) until a match with the lower envelope of the cluster sequence is achieved. By fitting the ZAMS only to stars with spectral type earlier than A0, we expect to minimize effects of metallicity, binarity, rotation or photometric errors (Phelps and Janes, 1994 and references therein).

Once the reddening has been fixed, we can proceed to fit distances. This is done in the color-magnitude diagrams, by shifting the fiducial reddened ZAMS by different distance moduli until a match with the lower envelope of the cluster's Main Sequence is achieved. Due to binary population and rotation, the main sequence is spread toward brighter magnitudes and redder colors. To avoid this bias, we attempt to fit the ZAMS to the lower envelope of the stellar distribution.

\subsection {Isochrone Fitting}

      Nuclear ages are derived using isochrones of the Padova group (Girardi et al. 2000), with core overshooting and for solar metallicity. Four different models of low mass stars are used to fit the pre main sequence, namely Siess et al. (2000, hereafter SDF00), Palla \& Stahler (1998, herafter PS99), Baraffe et al. (1998, hereafter BCAH98) and D'Antona \& Mazzitelli (1997, hereafter DM97). 

      Having distance and reddening fixed, we survey several ages, until a match between the isochrone and the cluster sequence is achieved. The accuracy of this fit for nuclear ages depends strongly in the number of evolved stars present in the cluster, since all isochrones fit well the part of the sequence near the turnoff point. The same process holds for contraction ages. In this case, a Pre Main Sequence (PMS) must be spotted, with no counterpart in the control field. As the PMS extends through the red tail of a color-magnitude diagram, it is better seen in the $V-I$ color.

      The need for several pre main sequence models while just one is used for post main sequence fitting is justified by the uncertainty still present in our theoretical understanding of young cool stars. It is a known fact that when deriving pre main sequences ages, different tracks lead to different values, which can diverge by several tens of millions of years in the worst cases. 

      The tracks of PS99{\footnote {available upon request}} were calculated to model the specific problem of the star formation history of the Orion Nebula, using the initial properties of the protostellar environment to create an empirical birthline used as initial condition in their evolution code. DM97 use a somewhat more arbitrary initial conditions and treat convection not with the standard mixing length theory, but instead with a model called the Full Spectrum of Turbulence (FST), suitable for low viscosity fluids (Canuto et al. 1996). Both use a grey atmosphere approximation, which in turn was abandoned by SDF00 in favor of more realistic atmospheres.

      A common feature of these three models is that they solve for luminosity and $T_{\rm{eff}}$, requiring empirical color-$T_{\rm_{eff}}$ calibrations along with bolometric corrections to match observational photometric data. The authors of the PS99 model provide their isochrones already tranformed to observables using the procedure described in Testi et al. (1998). For the SDF00 models, the user{\footnote {The files with isochrones were downloaded using the form that the authors made available at their internet server {\tt http://www-astro.ulb.ac.be/~siess/server/iso.html}}} can choose whether to use the color-temperature calibration of Siess et al. (1997) or Kenyon and Hartmann (1995). We chose the ones of Siess et al (1997), which for $V-I$ are based on the blackbody temperatures of Bessel (1979, 1991). We applied the same table for the DM97 models, since these isochrones were published as Luminosity-Temperature only. In all cases, the models used are those of solar composition.

      BCAH98 colors are instead derived theoretically from the models, which include the sophisticated non-grey atmospheres of Allard et al. (1998). It is important to note, as stressed by BCAH98 themselves, that their models diverge from observations for colors redward of $V-I \sim 2$. We use the BCAH98 models with ratio of mixing length to scale height $l/h$ = 1.9 for stellar masses $M>0.7M\odot$. For lower masses, where overshooting is not important, we use the models with $l/h$ = 1.0. 

     Steffen et al. (2001) tested several PMS models upon the known parameters of the pre main sequence binary NTT 045251+3016, concluding that BCAH98 and PS99 provided masses that best matched the observations, while DM97 was inconsistent at significant confidence levels. Yet, they add, all PMS tracks underestimated the masses of both stars. White et al. (1999) studying the young quadruple system GG Tauri also conclude that the BCAH98 isochrones yield the most coeval ages for the components of the system. On the other hand, a similar study carried by Ammler et al. (2005) using the young eclipsing binaries RX J0529.4+0041 and V1174\,Ori, concluded that none of the evolutionary models gives coeval solutions simultaneously in mass, radius and effective temperature for the former, while the coevality of the latter was consistent with the model of PS99 and a BCAH98 model of [M/H]=-0.5. However, the authors note that it is unlikely that the system is that metal-poor.

If doubt can be casted on the reliability of the PMS models on providing coeval ages, the problem gets worse when one is insterested on absolute age determination. In Fig.~\ref{gliese}, we provide a test of these theoretical models by comparing them in the $M_{\rm V}$ vs. $V-I$ plane to K and M dwarfs from the Gliese catalogue (Gliese, 1969) with photometry from Bessell (1990). These field stars are presumably old and should be located along the ZAMS. We take only stars with good {\tt HIPPARCOS} parallaxes (error smaller than 5\%). 

The plot shows that the SK82 ZAMS is a good fit in the whole range of colors. The SDF00 ZAMS is a fairly good fit to the K dwarfs ($1.0 < V-I < 1.60$), but falls below the true ZAMS as defined by the majority of the M dwarfs ($V-I > 1.6$). Even the 100 Myr isochrone is considerably below the main locus of points for $V-I > 1.8$ while the 30 Myr isochrone is a reasonably good match to the M dwarfs. 

The oldest isochrone we have from the PS99 models is that of 100 Myr. It is seen that this model shows similar behaviour to what was seen in SDF00: the field M dwarfs fall well above the 100 Myr isochrone, while the track of 50 Myr and 30 Myr match the sequence, 30 Myr being slightly better. 

As for the model of DM97, we also see that the 100 Myr isochrone is below the majority of stars. Isochrones of 50 and 30 Myr achieve better agreement with the sequence.

As these M dwarfs are presumably old field stars, it is highly unlikely that these young ages are real, and we assume that the match occurs by accident: the ages provided by SDF00, PS99 and DM97 are unreliable in this regime of low mass young stars.

In the lower right plot of Fig.~\ref{gliese} we show isochrones of the BCAH98 model. It is clear that the isochrones diverge from the observations for redder colors. It is interesting to note that in the range $1.6 < V-I < 2.0$ the BCAH98 8 Gyr isochrone (the oldest of the set) coincides with the SK82 ZAMS and is a good fit to the sequence of stars, indicating that BCAH98 might provide, among the four models used, the only consistency with the presumably very old age of the field M dwarfs. However, the 0.4 mag color range will in many cases be too small to allow for accurate age determination.

From these considerations, it is clear that some work is needed on the isochrones before they can be used to fit ages. In order to understand how the models differ, we plot them in theoretical spaces in fig~\ref{theo-plane}, taking the isochrone of 100 Myr as proxy of each model. We plot absolute bolometric magnitude ($M_{\rm bol}$) upon mass on the upper left diagram; ($M_{\rm bol}$) upon effective temperature ($T_{\rm eff}$) on the upper right; $T_{\rm eff}$ upon mass on the lower left and finally color upon $T_{\rm eff}$ on the lower right. The model of PS99 is not included in this analysis because the authors did not provide luminosities or temperatures but only masses.

From the $M_{\rm bol}$-mass relation, one sees that BCAH98 and DM97 meet good agreement, while SDF00 differs by $\sim$0.2 mag in the mass range of 0.2-0.6 $M_{\odot}$. We conclude that the predicted mass-luminosity relations of the three models are consistent with each other.

As for temperatures, the $M_{\rm bol}$-$T_{\rm eff}$ relation shows that the models do not differ much from each other for $T_{eff} <$ 4000, while DM97 is $\sim$0.5 mag fainter for hotter temperatures. The $T_{\rm eff}$-mass relation shows that DM97 gives temperatures approximately 400~K hotter than SDF00 and BCAH98 for $M/M_{\odot} >$ 0.6. This difference in temperature changes the bolometric correction and accounts for the difference in visual magnitude seen between DM97 and SDF00 in fig~3. For lower masses, the models behave very similarly, except for a small variation of $\sim$100~K. 

The color-$T_{\rm eff}$ relation, however, seems to be dominant effect. BCAH98 and SDF00 diverge in the whole interval, reaching 0.4 mag of difference in $V-I$ for $T_{\rm eff} =$ 3000~K. Moreover, none of them yield a precise color-$T_{\rm eff}$ conversion since, as noted by van Leeuwen et al. (1987) and Stauffer et al. (2003), young K and M dwarfs have different spectral energy distributions than old K and M dwarfs and thus different color-$T_{\rm eff}$ relations. Based on this, the placement of the theoretical isochrones onto an $V$ vs. $V-I$ CMD depends on knowing this conversion for young stars. 

A workaround to this problem was outlined by Jeffries et al. (2001), consisting of generating an empirical color-Teff relation using the well-observed Pleiades cluster as a calibrator. We assumed an age of 100 Myr and for a given PMS model the isochrone for this age is forced to follow the photometry of low mass stars. 

The procedure is shown in fig~\ref{calib-models}. We took $(VI)_K$ Pleiades data from Stauffer (1982a), Stauffer (1982b), Stauffer (1984), Stauffer \& Hartmann (1987) and Prosser et al. (1991) with the Bessell \& Weis (1987) $(V-I)_K$ - $(V-I)_C$ transformation. We assume that the Pleiades' distance modulus and reddening are $(V-M_V)_0 =$ 5.53 and $E(B-V) =$ 0.04 (Crawford \& Perry 1976; Pinfield 2000). 

The same empirical color-$T_{\rm eff}$ relation is then used for other clusters. Here we assume that for low mass stars the relation does not significanly change with age. Such an assumption is not contradicted by theoretical atmosphere models and is supported to some extent by observational evidence that various young low mass stars seem to be on a saturated level of activity therefore having similar spottedness. This assumption also validates the fact that we calibrated the empirical color-$T_{\rm eff}$ relation with 100~Myr isochrones although the best age estimates for the Pleiades cluster are around 120-130~Myr.

As a final remark on the isochrone fitting, we also want to derive ages given by the BCAH98 models with $V$ and $V-I$ exactly as provided by the authors. The {\it ad hoc} modification on the color-$T_{\rm eff}$ relationship can be justified in the case of DM97, PS99 and SDF00, where such a conversion is already applied. But the case of BCAH98 is different because the author do not simply adopt a color-$T_{\rm eff}$ conversion but fully integrate their isochrones with model atmospheres. Thus, it is not justified to dismiss these state-of-the-art calculations in favor of empirical calibrations. We therefore used both uncalibrated and the empirically calibrated version of BCAH98 isochrones for the fit.

\begin{figure*}
\begin{center}
\resizebox{8cm}{!}{\includegraphics{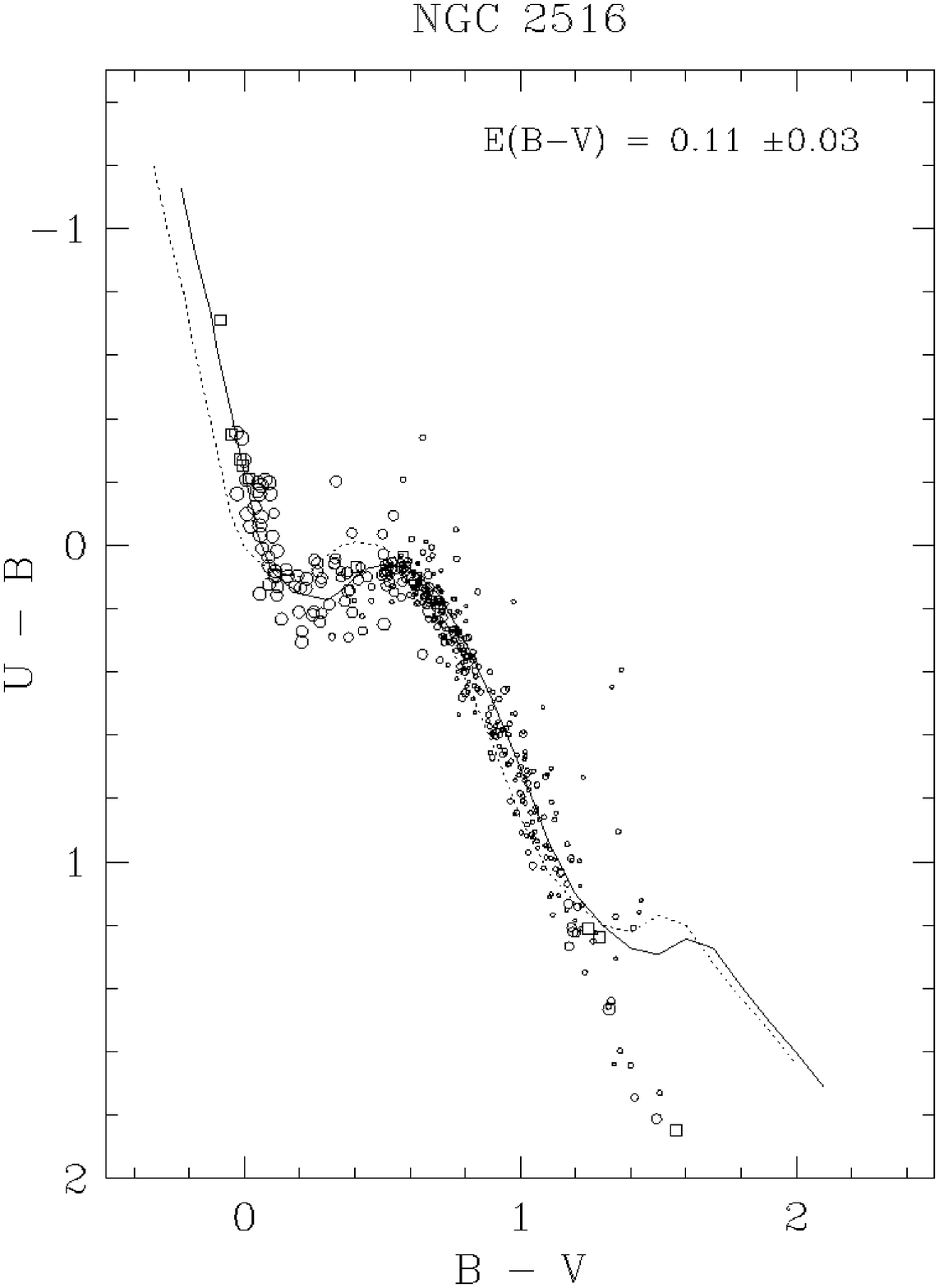}}
\resizebox{8cm}{!}{\includegraphics{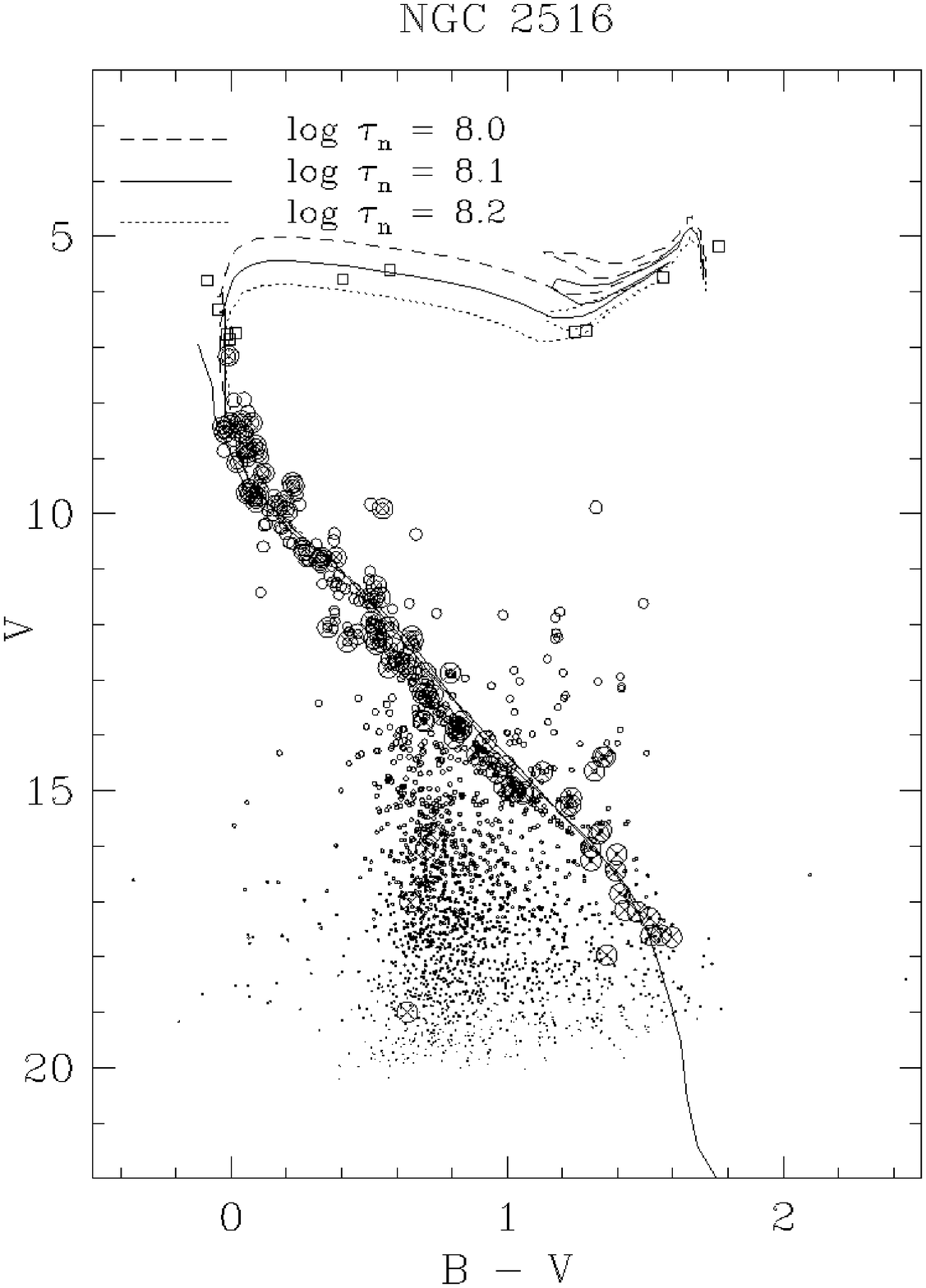}}
\resizebox{8cm}{!}{\includegraphics{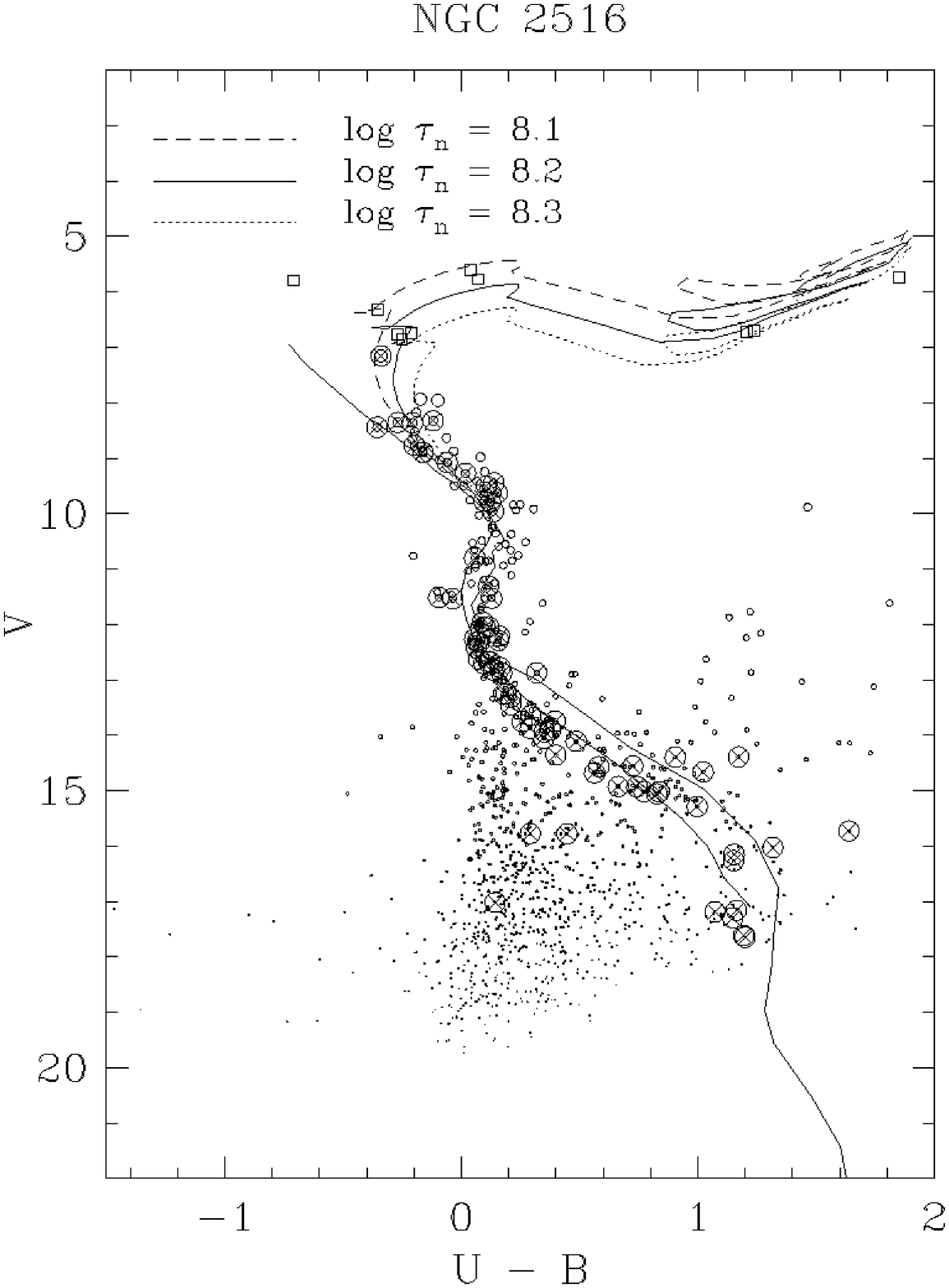}}
\end{center}
\caption[]{
         {\it Upper Left.} A good number of stars define the reddening in the direction of NGC~2516 as $E(B-V)=0.11\pm0.03$. For better visualization, only stars brighter than $V=16$ are plotted and the size of the dot is proportional to the $V$ magnitude. The non-reddened SK82 ZAMS (dotted line) is shown for comparison.\\
\\
         {\it Upper Right.} The state of a cluster after its first hundred million years of evolution is shown in this $V$ versus $B-V$ CMD of  NGC~2516. The best fit for nuclear age is achieved at $\log \tau_n = 8.10$, or 125 Myr. The uncertainty is very small, because the cluster has a very populated red giant branch. It is seen that most X-rays sources (crossed circles) follow the cluster sequence. Empty circles refer to our photometry whilst empty squares refer to corrected literature photometry.\\
\\
          {\it Lower.} Fit using $U-B$ as color index. The isochrones shown are $\log \tau_n =  8.1$ and $8.3$ (dashed and dotted lines, respectively), and the best fit at $\log \tau_n = 8.2$.  No pre main sequence structure is seen at all at these blue colors. True distance modulus is 8.15, equivalent to a distance of 430 pc. The ZAMS is plotted as solid line. Empty circles refer to our photometry whilst empty squares refer to corrected literature photometry.
}
\label{2516UBVIupper}
\end{figure*}

\subsection{NGC~2232}

\subsubsection {Historical Review}

	NGC~2232 is an open cluster in the constellation of Monoceros, which was catalogued by Herschel (1864) and Dreyer (1888), and almost ignored thereafter. Collinder (1931) estimated its distance (425 pc), followed by only a handful of other studies, which provided photometry for some of its stars (e.g., Cousins 1962). 

	The first -- and to date the only -- extensive photometric study in this cluster was done by Clari\'a (1972), who performed UBV and H$\beta$ photometry on 43 stars in its vicinity. This study asserted a nuclear age of 20 Myr, color excess of 0.01 mag and distance of 360 pc. A spectroscopic study of 16 members carried by Levato \& Malaroda (1974) found a color excess of 0.06 {$\pm$} 0.03 and placed the cluster at 375 pc of distance. NGC~2232 was further investigated by Pastoriza \& Ropke (1983), who performed DDO photometry on five red giants supposed to be contained in the cluster. Their values for distance and reddening, 316 and 0.02 $\pm$ 0.03, respectively, reasonably agree with the values of Clari\'a (1972) and Levato \& Malaroda (1974). However, the age estimate of $>$ 1Gyr, which was derived by extrapolating a calibration by Mermilliod (1981) relating the age with the mean magnitude of the red giants, seems quite unrealistic. Indeed, Clari\'a (1985) later published evidence that these five red giants do not match the criteria for membership, concluding that NGC~2232 does not contain red giant members. 

	To our knowledge, no other photometric studies exist in the literature. This is not only the first UBVRI work done on the cluster, but also the first one to do photometry in NGC~2232 using a CCD camera.   

\subsubsection{Analysis}

        Of all our clusters, NGC~2232 proves to be the poorest one in number of stars, as its CMDs show (Fig.~\ref{isos2232}). In these diagrams, one can devise an obvious although sparsely populated sequence. The background contamination is high and, due perhaps to statistical fluctuation, the total star counts in the control exceeds that in the cluster field (see Fig.~\ref{isos2232} and ~\ref{isoVI2232}). 

	The reddening value is constrained by eight evolved B stars (Fig.~\ref{isos2232}a), with good agreement on $E(B-V) = 0.07$. Reddening values of 0.05 and 0.09 mark the limits of good fits, thus constraining the error in 0.02. On estimating distance, in the $V$ vs. $B-V$ CMD the best fit to the A stars is achieved with a ZAMS shifted by a distance modulus of $V-M_V = 7.94$, corresponding to 350pc. 

We note that fits up to 380pc and down to 320pc also meet good agreement, defining the uncertainty in 30pc. Using $U-B$ as color index, a range of distance moduli from $V-M_V=7.45$ to $V-M_V=7.75$ (280 to 320 pc) also fits the cluster sequence. The agreement is therefore marginal, with the value of 320pc representing the best compromise between the two determinations.  

	On dating the cluster, the $V$ versus $B-V$ CMD is well fitted for a large range of post-main sequence isochrones, from 7.2 to 7.7 dex in $\log \tau_n$ (Fig.~\ref{isos2232}b). In the $V$ versus $U-B$ CMD (Fig.~\ref{isos2232}c), the age is constrained based on the position of a single object, HD\,45546, the brighest star belonging to the cluster. Although constraining nuclear age by just one star poses the problem of poor statistics, we believe that HD\,45546 can be used for this purpose without major concerns. As seen from the color-color diagram (Fig.~\ref{isos2232}a), the star presents typical redenning. Its rotational velocity of 70 km\,s$^{-1}$ is lower than the average of 127$\pm$8 for B0V-B2V stars (Abt et al. 2002). Indeed, the only problem affecting the usefulness of HD\,45546 is that it is in fact a triple system: the primary is accompanied by two other stars. However, we notice that the companions have magnitude V$\sim$9.2, i.e, $\sim$38 times fainter. Thus, the multiple nature of HD\,45546 does not affect the fit, since the enhancement in brightness due to these two faint companions is not considerable ($\sim$ 0.05 mag). Moreover, the isochrone curve gets practically vertical in the region of the CMD where HD\,45546 lies, as shown in Fig.~\ref{isos2232}c, so the small shift in magnitude does not affect the determination.

An isochrone fit of $\log\tau_n = 7.6$ is slightly better than 7.4, which leads us to consider that the best value is around 7.5 and 7.6. Post main-sequence nuclear age is then placed at $\log \tau_n=7.55 \pm 0.10$, linear scale equivalent = 35 Myr. 

	In $V-I$ (Fig.~{\ref{isoVI2232}), we performed statistical subtraction of the control field (middle diagram) population from the cluster field population (left diagram). We subdivide the CMDs in bins of 0.1 in color and 0.5 in magnitude, counting the number of stars in the control and randomly deleting the same number of stars in the corresponding bin of the cluster's CMD.  

As aforementioned, the control field has more stars than the cluster field, so we do not expect the statistical subtraction to behave perfectly. As seen in the upper right panel of Fig.~\ref{isoVI2232}, several bins of the cluster CMD were completely deleted, but a number of faint blue stars, clearly pertaining to the background, remain. There is also an edge of faint red stars left by statistical subtraction of the control field. Although the flawed subtraction of faint blue objects could indicate that these red objects could be a problem with the control field as well, this is not likely since this stripe of red objects matches the shape of the PMS models of SDF00, PS99 and DM97 reasonably well. We therefore take it an indication that it is composed mainly by pre main sequence stars belonging to the cluster.

The fitted contraction ages $\log\tau_c$ are shown in Fig.~\ref{isoVI2232}. The empirically calibrated models of BCAH98 and SDF00 yield $\log\tau_c$=7.4. DM97 and PS99 read $\log\tau_c$=7.5. The limits of the fit ($0.1$ dex) are very tight.
The uncalibrated isochonres of BCAH98 do not provide a direct age measurement of NGC~2232, since its PMS stars lie in the red edge of the sequence, where the model diverges, as mentioned before (see Sec.\,6.2). Indeed, there are only about 5 PMS stars in the range of validity of the uncalibrated BCAH98 model ($1.6<V-I<2.0$), two of which appear to be photometric binaries, since they are much brighter than the isochrones of SDF97, PS99 and DM97.

        To solve this problem, we perform an indirect fit with the uncalibrated isochrones of BCAH98, fitting just the range between $1.6 < V-I < 2.0$  to the range of magnitudes covered by the SDF00, DM97 and PS99 fit. These steps are shown in the lower panel of Fig.~\ref{isoVI2232}. By doing this, we are in principle able to perform a BCAH98 fit to a cluster devoid of PMS stars in the color range $1.6 < V-I < 2.0$. 

As shown in Fig.~\ref{isoVI2232} (lower), the BCAH98 track of $\log\tau_c = 7.4$ is the one whose location in the CMD best agrees with the location of the best fitted isochrones of SDF00 and PS99. We note that the value of $\tau_c=7.4$ yielded by the indirect fit of uncalibrated BCAH98 isochrones agrees with the values yielded by the empirically calibrated models and is very close to the the nuclear age $\tau_n=7.5$. 

\begin{figure*}
\begin{center}
\resizebox{16cm}{!}{\includegraphics[angle=270]{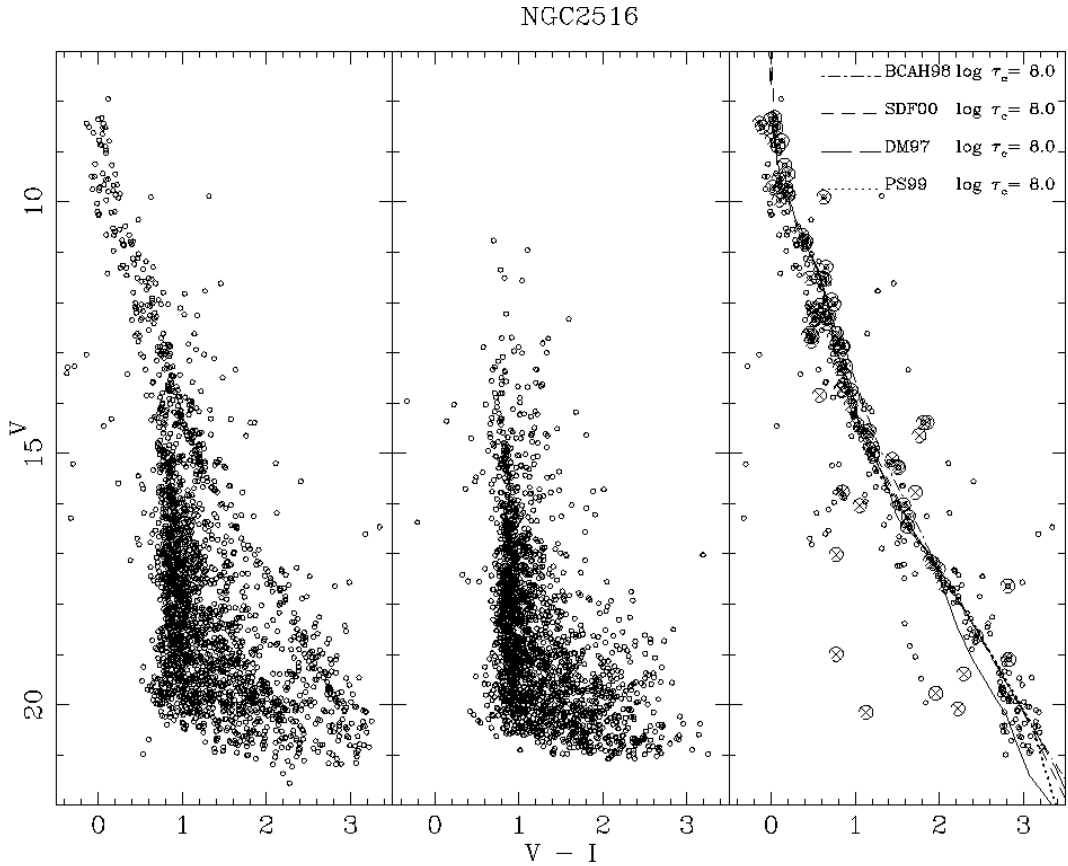}}
\end{center}
\caption[]{Using $V-I$ color as color index, a pre main sequence structure not visible in the $B-V$ plane arises. After statistical subtraction by the control field ({\it middle}) and cross-identification with the X-ray sources, we proceed to isochrone fitting, as shown above. The oldest isochrone on the set of DM97 and PS99 is that of 100~Myr, shown in the figure. As for the sets of BCAH98 and SDF00, any isochrone older than 100~Myr provides a good fit. This lower limit of $\tau_c=$100~Myr is consistent with the nuclear age derive before.
} 
\label{2516VI}
\end{figure*}

\subsection{NGC~2516}

\begin{figure*}
\begin{center}
\resizebox{8cm}{!}{\includegraphics{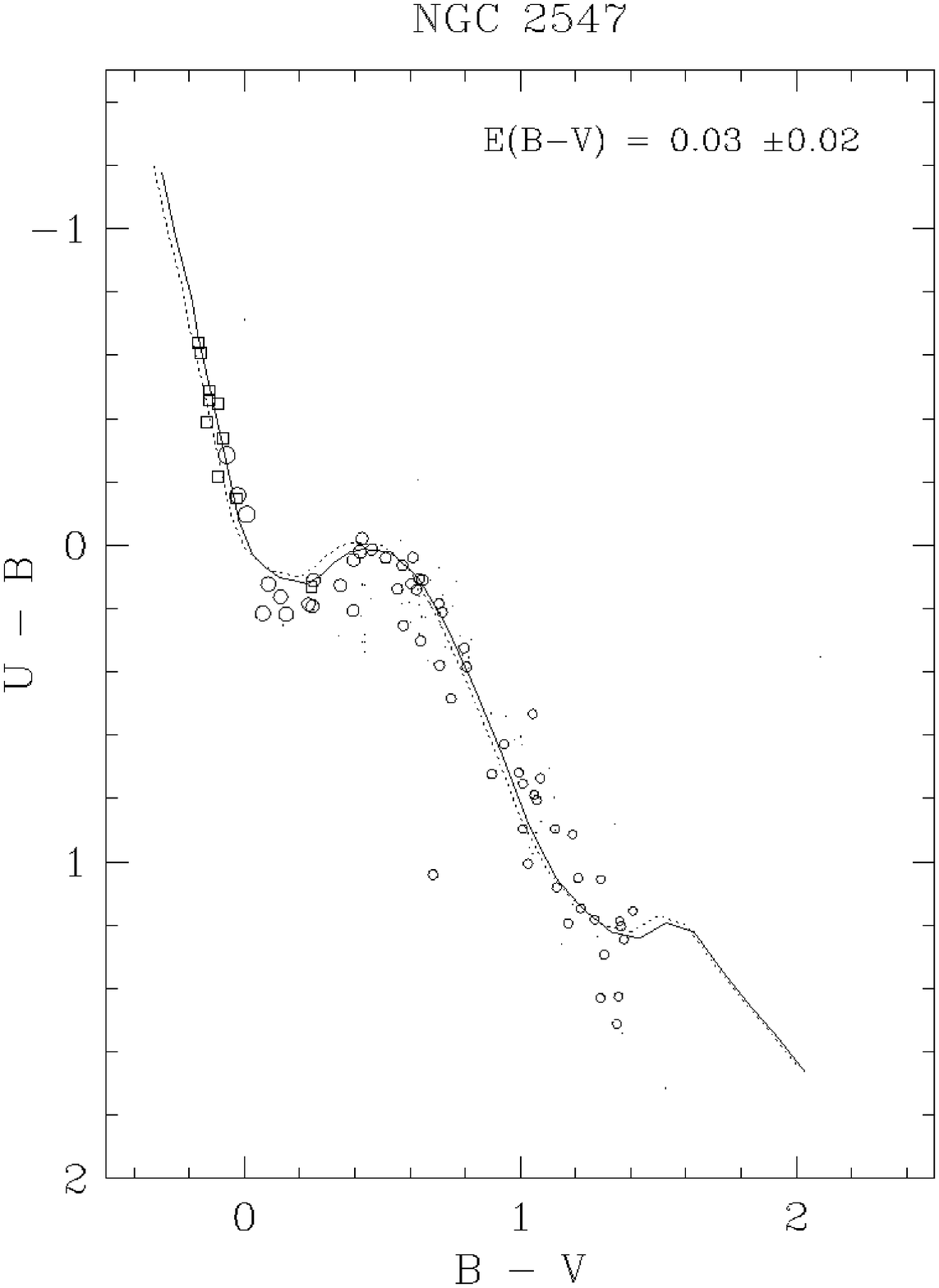}}
\resizebox{8cm}{!}{\includegraphics{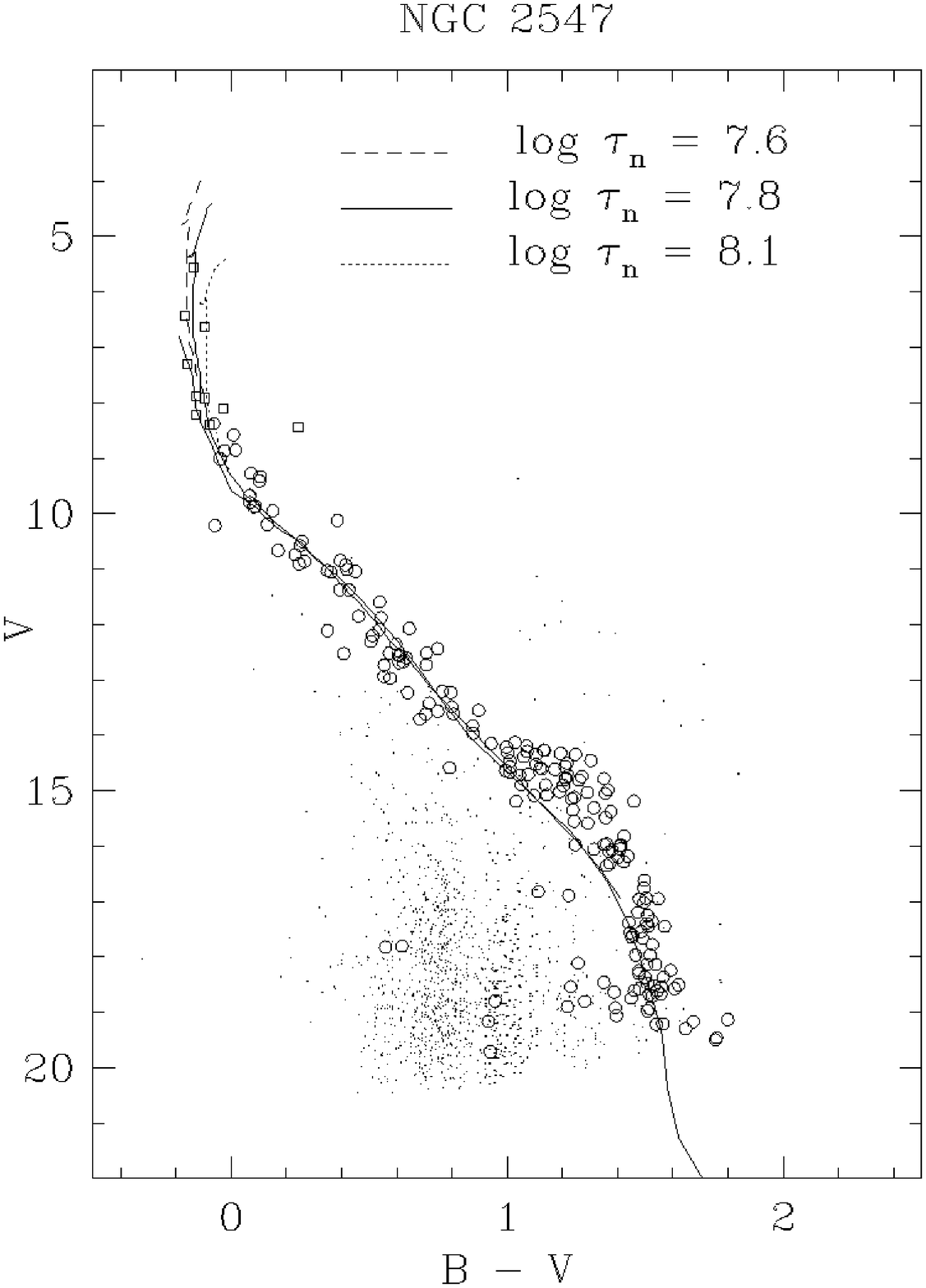}}
\resizebox{8cm}{!}{\includegraphics{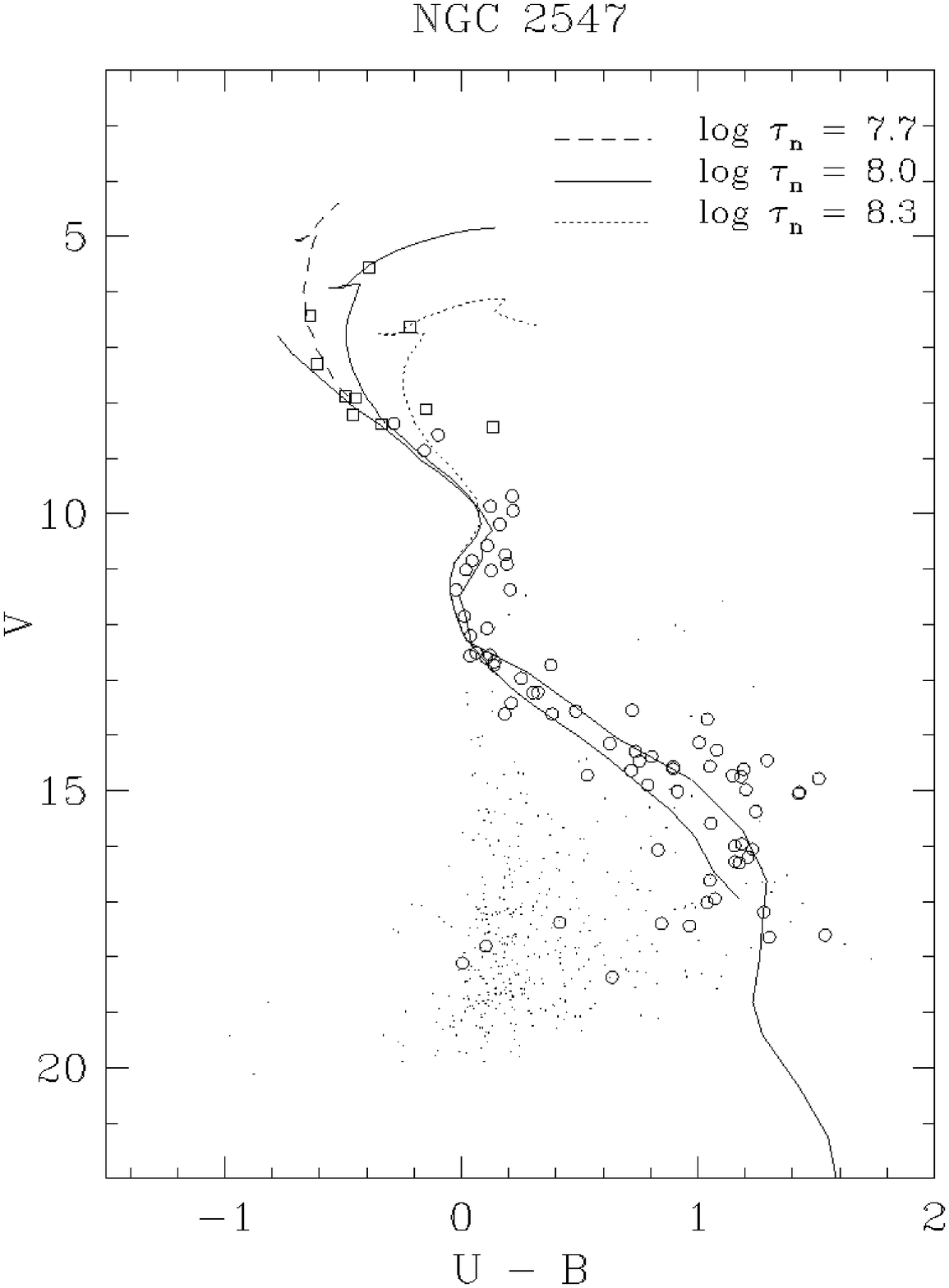}}
\end{center}
 \caption[]{Diagrams for NGC~2547. It is a well populated cluster with a main sequence clearly above the field star contamination. The reddening value yielded by the fitting on the Color-color diagram ({\it upper left}) is $E(B-V) = 0.03 \pm 0.02$. The size of the dot is proportional to the $V$ magnitude. The non-reddened SK82 ZAMS (dotted line) is shown for comparison.\\
   The best value for distance modulus is $V-M_V=8.05$, i.e., distance of 390 pc. In the $U-B$ diagram ({\it lower}), isochrones of 50, 100 and 200 Myr fit the position of three brightest stars. We disregard the oldest since it does not fit the blue ZAMS stars. \\
 	Using $B-V$ as color index ({\it upper right}), the isochrone of $\log \tau_n$ = 7.8 (solid line) fits the position of the brightest star. Isochrones of $\log \tau_n$ = 7.6 and 8.1 (dashed lines) provide the limit of good fit. As pointed out by Jeffries \& Tolley (1998), these nuclear ages should be taken with care, since they are defined by only three stars, which are not evolved enough to undoubtedly constrain a post main sequence fit. Nevertheless, an estimate is possible based on the lower limit they yield and on the upper limit set by the non-evolved stars in the ZAMS (see text). Empty circles refer to our photometry whilst empty squares refer to corrected literature photometry.
}
\label{2547blue}
\end{figure*}

\subsubsection{Historical Review}

        NGC~2516 is a populated cluster in the age range of 100 to 200 Myr. The presence of many bright stars in NGC~2516, alongside with the apparent spatial extent due to its proximity, rendered this cluster the nickname of {\it Southern Pleiades}. Early photometric works were presented by Cox (1955), Eggen (1972) and Dachs \& Kabus (1989), yeilding a mean reddening estimate of $E(B-V)=0.12$ and distance of 400 pc. Meynet et al. (1993) derived an age of 140 Myr.{\footnote {For a complete historical review of earlier works, the reader is addressed to the introduction of Sung et al. (2002).}}
	
	NGC~2516 underwent a burst of popularity after 159 X-ray sources were identified within its central 20-arcmin radius using the ROSAT satellite (Jeffries et al. 1997), 65 of these being identified as cluster members. Following this work, the cluster was later chosen as a target for calibration purposes on recent X-ray observatories, {\it Chandra} (Harnden et al. 2001) and {\it XMM-Newton} (Sciortino et al. 2001), which made NGC~2516 unique in volume of X-ray data and conclusively established the presence of UV excess in its F stars. 
	
	The cluster was recently the target of a deep $UBVI$ photometric study conducted by Sung et al. (2002), which re-derived the main parameters of the cluster, $E(B-V)=0.11 \pm 0.02$, distance of 360 pc and $\log \tau_n = 8.2\pm 0.1$. One interesting result of their work is that the cluster is found to present only a slight sub-solar metallicity, [Fe/H]=-0.10 $\pm$ 0.04, diverging from the metal poor value of -0.32 $\pm$ 0.06 determined by Jeffries et al. (1997). Thus, while the latter explained the UV excess of F stars as due to the metallicity deficit, Sung et al. (2002) invoke binarity and a subsequent enhancement in activity due to tidal locking as a more likely explanation in this metal richer scenario. Adding to the debate, Terndrup et al. (2002) identified possible flaws in the photometric analysis of Sung et al. (2002) and derived an even greater metallicity ([Fe/H]=+0.01 $\pm$ 0.07) based on the analysis of moderately high signal-to-noise, high resolution spectra of two cluster members.

\subsubsection{Analysis}

        As previously said, NGC~2516 is a well populated cluster. A well defined sequence can be seen in all color-magnitude diagrams, particularly reaching magnitude 21 in $V-I$. It is seen in the color-color diagram that the cluster has a good number of $B$ type stars among its members, thus allowing for a realiable reddening solution. A ZAMS shifted by $E(B-V) = 0.11 \pm 0.03$ provides the best fit (Fig.~\ref{2516UBVIupper}a). Stars fainter than V = 16 are not shown, clipping the background contamination.

	X-ray data, as said before, is available in great amount for NGC~2516. We cross-identify the positions of the catalogued X-ray sources with our astrometry, selecting the closest star in a 5'' radius. This higher tolerance is due to greater uncertainty in the position of X-ray sources. In all CMDs of NGC~2516, the matching X-ray sources are plotted as crossed circles, and help in better determining the cluster sequence.

	The cluster is evolved, given the presence of a populated red giant branch. Devoid of blue objects, its early A-type stars ($B-V$ $\sim$ 0) already show signs of evolution, starting to leave the main sequence. In the $U-B$ CMD (Fig.~\ref{2516UBVIupper}a), the ZAMS is shifted by a distance modulus of $V-M_V = 8.20 \pm 0.2$, placing the cluster, after reddening correction, at $380 \pm 35$ pc of distance.

	We note that in the CMDs using $U-B$ and $B-V$ as the color index, the ZAMS fitting is accurate for the lower envelope of A stars, but turns slightly redder than the cluster sequence for later types. This is apparently in agreement with the aforementioned UV excess found for F type stars in NGC~2516. The trend seen here, however, extends down to $B-V \sim 1.4$, well in the K spectral type. Stauffer et al. (2003) noticed that the K dwarfs in the Pleiades cluster show $B-V$ colors systematically bluer than the ZAMS, an effect that they attribute to spottedeness. It is possible that we are seeing the same effect taking place in NGC~2516 as well.

	The evolved stars of the cluster are best fitted by an isochrone of $\log \tau_n = 8.10$ (Fig.~\ref{2516UBVIupper}b), with small scatter. It is noticed that the reddest giant is slightly too red to be fitted by the isochrones. Also, the bluest star is too blue to be fitted by the young isochrones, and could be a blue straggler.

	In the $U-B$ vs. $V$ CMD, the isochrone that best fits the cluster is of $\log \tau_n = 8.20$, but tracks of 8.1 and 8.3 are needed to fit the other evolved stars. Nuclear age in then constrained at $\log \tau_n = 8.15 \pm 0.15$, or $140$ Myr.

	In the $V-I$ plane (Fig.~\ref{2516VI}), one can see a feature rising above the best fitted ZAMS, with no counterpart in the control field. As the control field for this cluster presents no obvious flaw, statistical subtraction of the background could be performed, confirming that the feature is composed of cluster stars. The oldest isochrone on the sets of DM97 and PS99 is that of 100 Myr, which is a good fit to the cluster PMS. Any SDF00 or BCAH98 isochrone older than 100 Myr meets good agreement with the cluster sequence. Although an accurate measurement is not possible, this imposed lower limit of 100 Myr is within one standard deviation of the nuclear age. The uncalibrated BCAH98 model isochrones behave similarly.

\subsubsection {Historical Review}

	We follow our study with NGC~2547, a well-studied cluster. Fernie (1959) concluded from UBV photometry that the cluster presents a uniform reddening of $E(B-V) = 0.03$ and a distance of 380 pc. Being a close cluster, the cluster sequence lies above the field star contamination, so he could immediately define its PMS, concluding that the cluster is young. In a subsequent work, Lindoff (1968) gave an age of 75 Myr. Clari\'a (1982) conducted an extensive photoelectric study of NGC~2547, placing it at a distance of 450 pc, being extincted by 0.06 of a magnitude, and dated it on 57 Myr. 

	Jeffries \& Tolley (1998) conducted the first CCD study on NGC~2547, also introducing X-ray observations to select cluster members. As said before (Sec. 1), a curious result of their work is that the nuclear and contraction ages greatly disagree. Fitting PMS models to the $V$ - $V-I$ sequence, they found an age of 14 $\pm$ 4 Myr, whereas the post main sequence stars yielded the same age derived by Clari\'a (1982). 

	On an updated version of the same work, Naylor et al. (2002) presented ``optimal photometry'' of the cluster, in which they rederive $\tau_c$, in both $B-V$ and $V-I$ vs. $V$ color-magnitude diagrams. They find a somewhat older value than Jeffries \& Tolley (1998), between 20 and 35 Myr, but still yonger than the nuclear age. In this study, we use different PMS models to yield another value to the contraction age of the cluster.

\subsubsection {Analysis}

        Having a well populated main sequence, the fitting for reddening and distance of NGC~2547 is straighforward. In the color-color diagram (Fig.~\ref{2547blue}a), after clipping the background contamination below $V \sim 16$, the cluster stars are seen to tightly follow a ZAMS reddened by $E(B-V) = 0.03 \pm 0.02$. CMDs using $B-V$ (Fig.~\ref{2547blue}b), $U-B$ (Fig.~\ref{2547blue}c) and $V-I$ (Fig.~\ref{2547red}) all agree that a distance modulus of $V-M_V=8.05$ (distance = 390 pc) fit the cluster sequence above the field star contamination, with small dispersion. 

	Devoid of red giants and with a small number of stars in the upper main sequence, the cluster does not allow for an accurate measure of nuclear age, which is indeed defined only by a few stars as noticed by Jeffries \& Tolley (1998). In the $B-V$ color, the three brightest stars are the only ones showing any evolutionary effect. Therefore, a broad range in log $\tau_n$, from 7.6 to 8.1, seem to fit the blue region, 7.8 being the best value (Fig.~\ref{2547blue}b). In the $U-B$ diagram (Fig.~\ref{2547blue}c), isochrones of 7.7 (50 Myr),  8.0 (100 Myr) and 8.3 (200 Myr) fit the position of the three brightest stars. It is seen, however, that the track of 200 Myr does not fit the non-evolved blue ZAMS stars, and therefore should be disregarded. The same thing can be said of the track of $log \tau_n = 8.0$, which lead us to an upper limit of 100 Myr to the nuclear age of NGC~2547. 

We call the attention of the reader to clearly stress two important points often understated: 

\begin{enumerate} 

\item{Although an accurate measurement is not allowed by the 3 evolved high-mass stars, the non-evolved stars in the upper main sequence yield an upper limit to the nuclear age.}

\item{The 3 evolved high mass stars allow for at least a lower limit to the nuclear age}
\end{enumerate}

These obvious statements were not touched by Jeffries and Tolley (1998) or other studies who attempted to derive a nuclear age for NGC~2547. In our case, we have an upper limit of $\tau_n$=100\,Myr, and an lower limit of $\tau_n$=50\,Myr. We take the average of this interval as the most likely value, thus determining at $75 \pm 25$ Myr the age as yielded from the $U-B$ CMD. 

	On deriving contraction ages, we faced a small problem with the background contamination. It is seen in Fig.~\ref{2547red} that the stars of the control field CMD are systematically redder than the background stars on the cluster CMD. This could be due to the fact that our control fields for NGC~2547 were taken in a region of the sky with a somewhat higher extinction. Not being a good representation of the background, we could not use this control field to perform statistical subtraction as done in NGC~2516. 

	Fortunately, a solution is straighforward, since the whole cluster sequence arises above the background contamination in $V-I$ (Fig.~\ref{2547red}a). The red tail of the sequence is fitted by a PS99 isochrone of $\log tau_c$ = 7.0, with tracks of 6.5 and 7.2 marking the limits of good fit (Fig.~\ref{2547red}c). The stars between these limits were flagged in the photometry list, revealing the location of the PMS in CMDs using other colors (Fig.~\ref{2547blue} and Fig.~\ref{2547VIfull}). 

\begin{figure*}[tb]
\begin{center}
\resizebox{16cm}{!}{\includegraphics{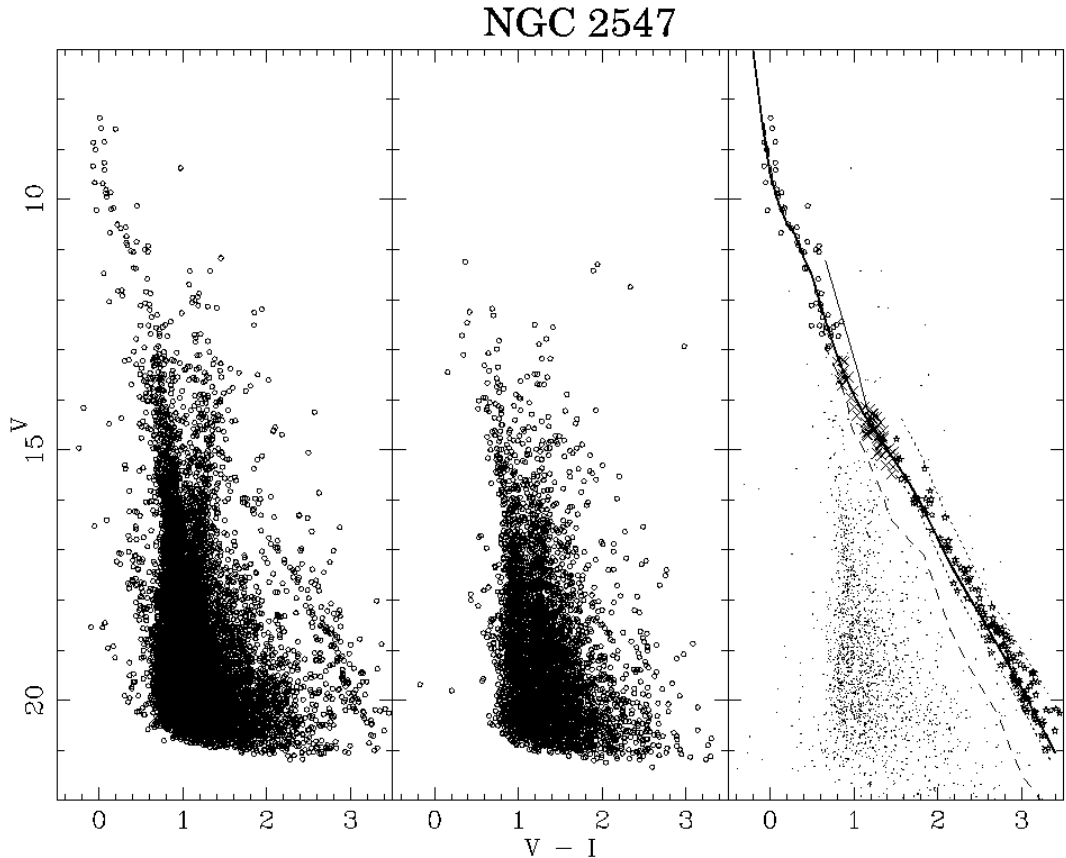}}
\end{center}
\caption[]{The control field for NGC~2547 could not be used to statistically subtract the background stars, because of the presence of an interstellar cloud. As seen in the figure, stars in the control field CMD ({\it middle})are systematically more extincted than in the cluster field ({\it left}). Nevertheless, the PMS is easily seen completely above the background contamination when $V-I$ is used as color index. \\

  In the {\it right} diagram, we show the best fits using PS99 tracks (solid line) and the limit of fit for the red PMS stars (short-dashed lines). Stars between these limits and redder than $V-I = 1.5$ are flagged as belonging to the PMS (plotted as {\it stars}). Stars brighter than $V = 13$ and not farther than $1.5^{mag}$ from the ZAMS (long-dashed line) are marked as {\it empty circles}. Background stars are marked as {\it small dots} and stars with dubious classification are marked as {\it crosses}. The solid line marks the semi-empirical cluster sequence, achieved by joining the bright part of the ZAMS with the best PS99 isochrone, with a spline interpolation between them. The fit to the low-mass stars yields a lower limit $\log \tau_c = 7.6$ (see text). 
}
\label{2547VIfull}
\end{figure*}

	The flagging showed in Fig.~\ref{2547red}c follows the criteria: the {\it star} symbol stands for stars redder than $V-I = 1.5$ that fall between the limits of good fit; {\it open circles} mark stars brighter than $V = 13$ that does not deviate more than 1.5 magnitude from the best fitted ZAMS, the high deviation occurring due to verticalization of the ZAMS in this range. These criteria define the fiducial PMS and ZAMS stars free from background contamination. The two tracks are then joined by a spline interpolation to produce a semi-empirical cluster sequence (solid line). The stars that fall in this interpolation range should be mainly cluster members, although some few field interlopers will also be included, and are marked as {\it crosses}. All others are taken as background stars and plotted as {\it small dots}.

\begin{figure}
\begin{center}
\resizebox{8cm}{!}{\includegraphics{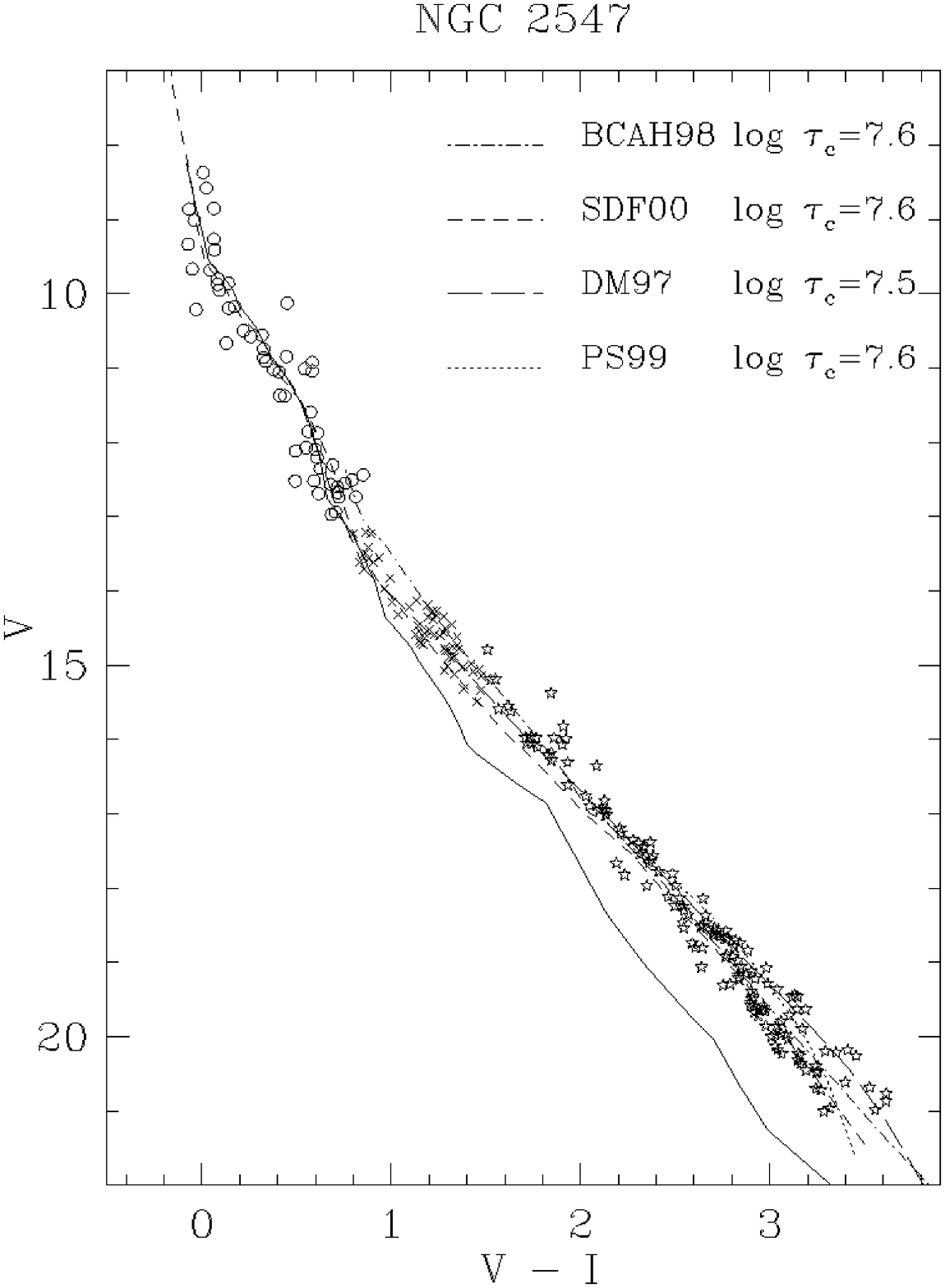}}
\resizebox{8cm}{!}{\includegraphics{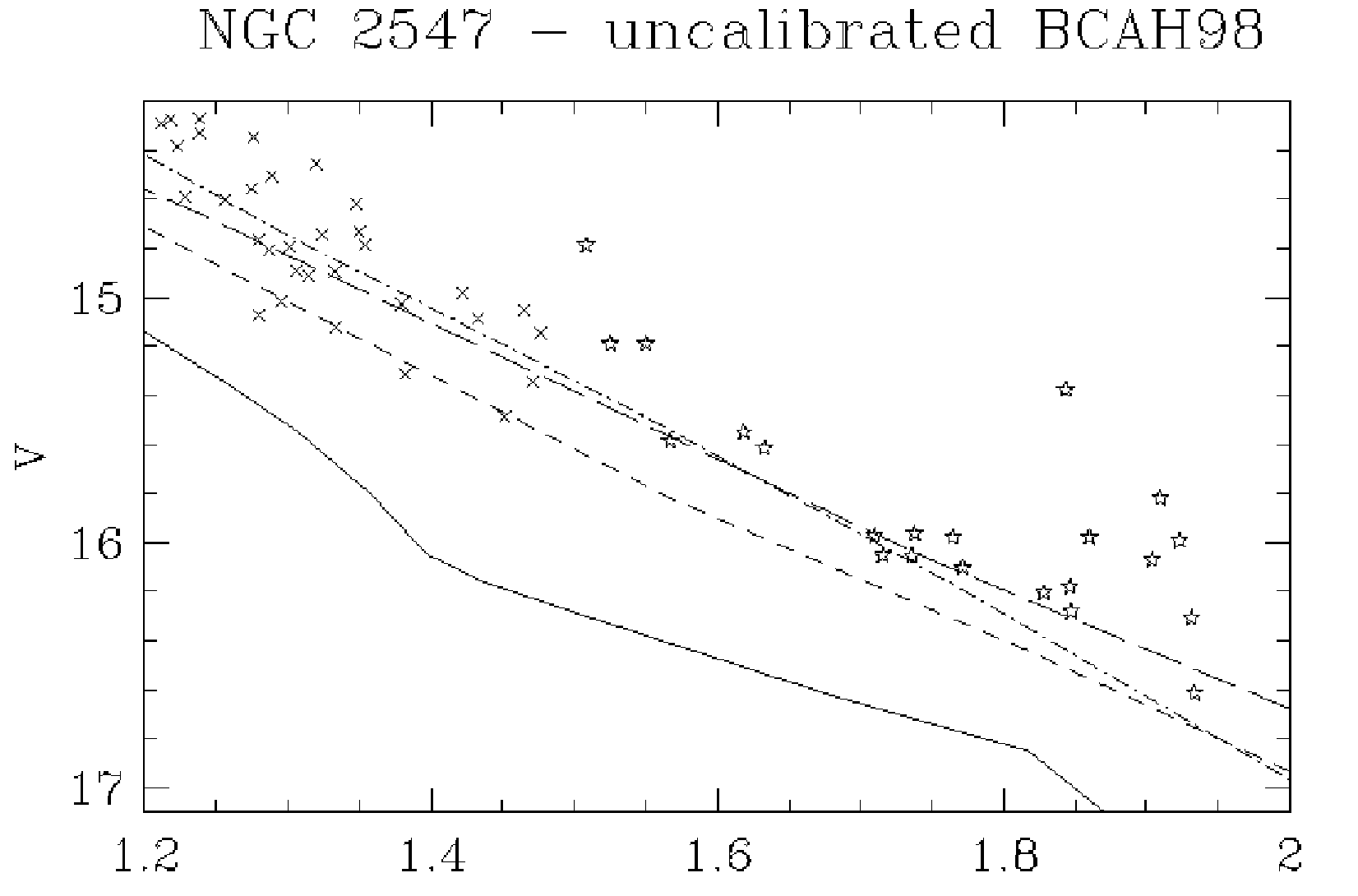}}
\end{center}
 \caption[]{
   {\it Upper}.Isochrone fitting in the $V$ vs. $V-I$ CMD for NGC~2547 with the empirically calibrated PMS models. The SK82 ZAMS is plotted as a solid line.\\

   {\it Lower}. Fit for the uncalibrated BCAH98 models (see text). We zoom in the color range of interest. The age as yielded by this fit is $\log \tau_c=7.7$. The same line style is used.
}

\label{2547red}
\end{figure}

We shown in fig.~\ref{2547red} the PMS isochrone fit for NGC~2547. The tracks of PS99, SDF00 and BCAH98 meet very good agreement, indicating $\log \tau_c = $7.6. DM97 yields a slightly younger age of $\log \tau_c = $7.5.

\subsection{NGC~4755}

\begin{figure*}
\begin{center}
\resizebox{8cm}{!}{\includegraphics{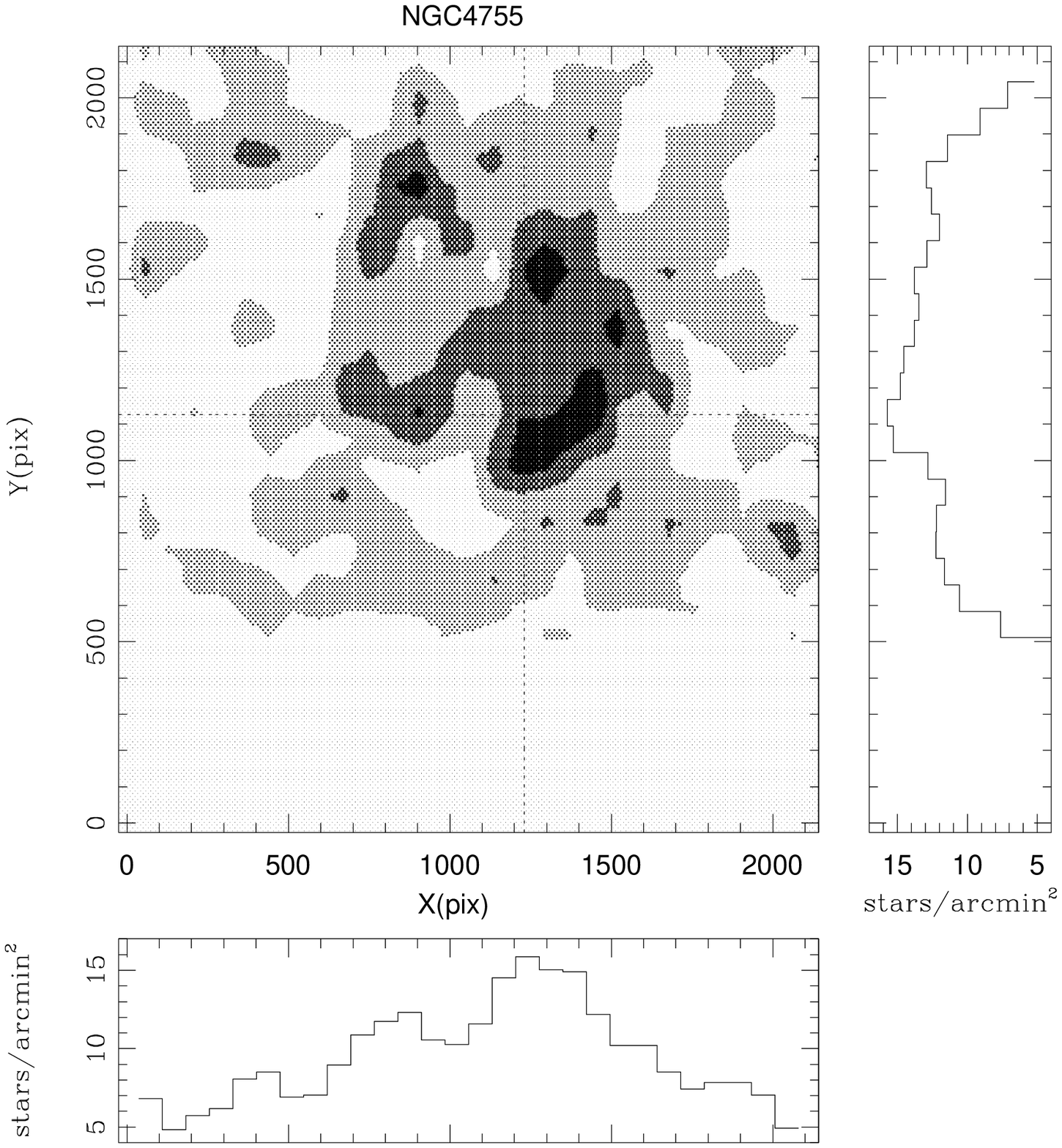}}
\resizebox{8cm}{!}{\includegraphics{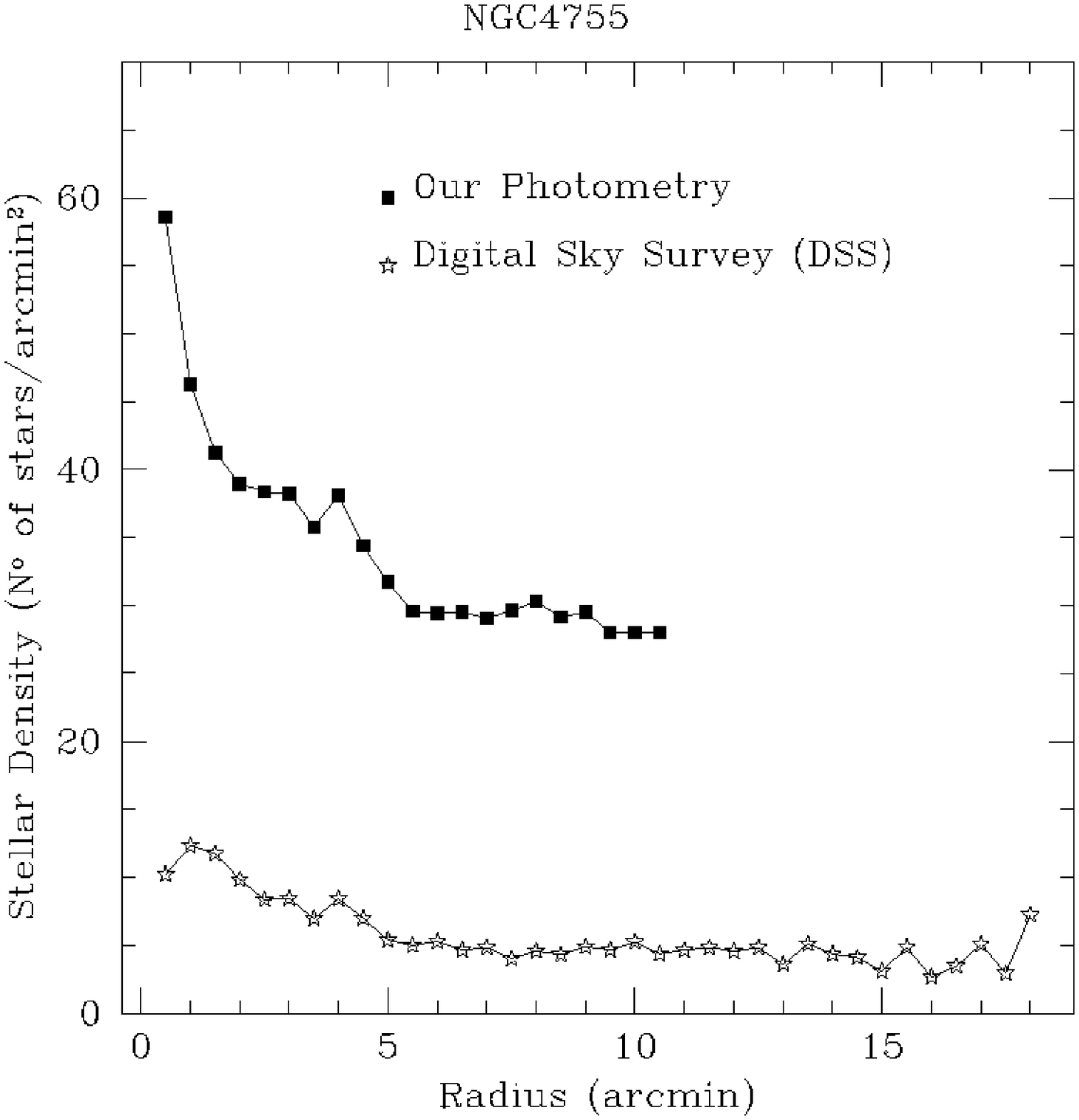}}
\end{center}
\caption[]{Determination of the center and radius of NGC~4755. Stars are counted in {\it x} and {\it y} in bins of 0.5'. The center is defined when the two distributions meet a common enhancement. From this center, the stars are counted concentrically, yielding the cluster radius when the distribution reaches a steady value. Data from DSS is shown as comparison.}
\label{radius}
\end{figure*}

\subsubsection{Historical Review}

NGC~4755, called $\kappa$ Crucis because of its starlike appearence in
the vicinity of the southern cross, was also dubbed {\it Jewel Box} by
John Herschel. The charming name was given because one of the bright
stars appears extremely red ($B-V$ = 2.2), whereas all others are
conspicuously blue ($B-V < 0.3$).

The cluster was the object of at least three CCD photometric studies in
the past ten years, by Sagar \& Cannon (1995), Sanner et al. (2001)
and Piskunov et al. (2004) {\footnote {the introduction of Sagar \&
    Cannon provides a complete historical review of earlier
    works}}. The determination of nuclear ages for the cluster do not
find large differences among the studies. Sanner et al. (2001) and
Sagar \& Cannon (1995) agree with an age of 10 $\pm$ 5 Myr. Contraction ages were
derived by Sagar \& Cannon (1995) using $B-V$ as color index, yielding
ages between 3 and 10 Myr.  Piskunov et al. (2004) estimate the age of
NGC~4755 as 16\,Myr, based on the location of the so-called H-feature
in the luminosity function, which arises from the transition from the
PMS to the main-sequence. The published heliocentric distances place
the cluster at $\sim$ 2000 parsecs, being variably extincted by
0.30-0.55 of a magnitude, with a mean $E(B-V) = 0.41$ (Sagar 1987).

\subsubsection{Analysis}

	Being a distant cluster, NGC~4755 shows heavy field star contamination, as shown in the CMDs, masking the cluster sequence in the range of low mass stars. Nevertheless, part of the cluster sequence is clearly visible at blue colors. An analysis to determine the cluster center and radius will therefore help in diminishing the strong contamination.

	We construct distributions of stars in the cluster in {\it x} and {\it y}, at 0.5' interval (75 pixels). The cluster center is determined to be where the two distributions peak. The result is shown in Fig.~\ref{radius}a, alongside with the contour plot of the stellar density of the cluster. The center is determined at ($x,y$)$=$($1153,1117$), or  ($\alpha , \delta$)$=$($12^h 53^m40^s_.04 , -60^o22'04.9''$).

	Having the coordinates of the center, we perform star counting in concentric rings, transformed into stellar surface densities by dividing by the area of the rings. One must pay attention to the fact that as the rings become too big in radius, they will eventually emcompass an area not observed in the frame. If not subtracted, this void area would lead us to underestimate the true stellar density. 

	To solve this problem, we computed the area by a Monte Carlo method, constructing 2 sets of 10$^{\rm 6}$ random numbers in the pixel range of our frame. The area of the fraction of the ring that falls on the frame is thus taken as the hit/(hit+miss) ratio of the random numbers that ``hit'' the ring area. The resulting densities are shown in Fig.~\ref{radius}b, in which we see that the cluster extends itself over an angular area of $\sim$5'. Data from the Digital Sky Survey (DSS) are shown for comparison.

	Reddening is fit on the $U-B$ versus $B-V$ color-color diagram (Fig.~\ref{red4755}a), yielding $E(B-V) = 0.33 \pm 0.03$. Similar fit in the $R-I$ versus $B-V$ CCD (Fig.~\ref{red4755}b) lead us to $E(B-V) = 0.35 \pm 0.07$, the greater uncertainty being due to the lesser sensitivity of $R-I$ to variations in reddening when compared to $U-B$. 

	Distance modulus is fit in the $B-V$ color magnitude diagram (Fig.~\ref{ubv4755}a), $V-M_V = 11.60 \pm 0.30$, i.e, Dist = 2130 $\pm$ 250 pc. Such a high error is distance is due in part to the spread in the main sequence, which is not tight, because of variable reddening. We note that with $U-B$ as color index, a better fit to the B stars is achieved with a distance of 1950 $\pm$ 100 pc. We estimate the heliocentric distance of NGC~4755 as the average of these determinations.

	The small branch of evolved stars present in NGC~4755 is fitted by isochrones ranging 7.0 to 7.4 in $\log \tau_n$ when $U-B$ is used as color index; and ranging 6.9 to 7.3 with $B-V$. As noticed by Sanner et al. (2001), the single red giant star is too red to be fitted by the isochrone.

	As for contraction ages, the statistical subtraction of the field stars (Fig.~\ref{4755VI}) unveils a sequence of non-deleted stars in the color range 1.0 $< V-I <$ 2.6, magnitude range 15.5 $< V <$ 20. This sequence is likely to be the PMS of the cluster, because it is above the ZAMS and because it presents a slope in the CMD that is consistent with the slope of the PMS as modeled by PS99 and SDF00, at colors redder than $V-I = 1.3$. 

	  Having defined the location of the PMS, we perform the fit (Fig.~\ref{4755VI}, right panel). The PMS isochrones of PS99 and SDF00 read, respectively, $\log \tau_c = 6.9$ and $\log \tau_c = 7.0$. The model of DM97 yields $\log \tau_c = 6.9$. We notice the presence of a concentration of stars brighter than the isochrones at $V-I\sim2.2$, which probably consists of binary stars and/or leftovers from the statistical subtraction. 

We find that these contraction ages are within one standard deviation of nuclear age $\log \tau_n = 7.1 \pm 0.2$ (13 $\pm^7_5$). However, it should be noted that these contraction ages have greater uncertainty than the others derived because of the relatively poor statistical subtraction of the field population.

\begin{figure*}
\begin{center}
\resizebox{8cm}{!}{\includegraphics{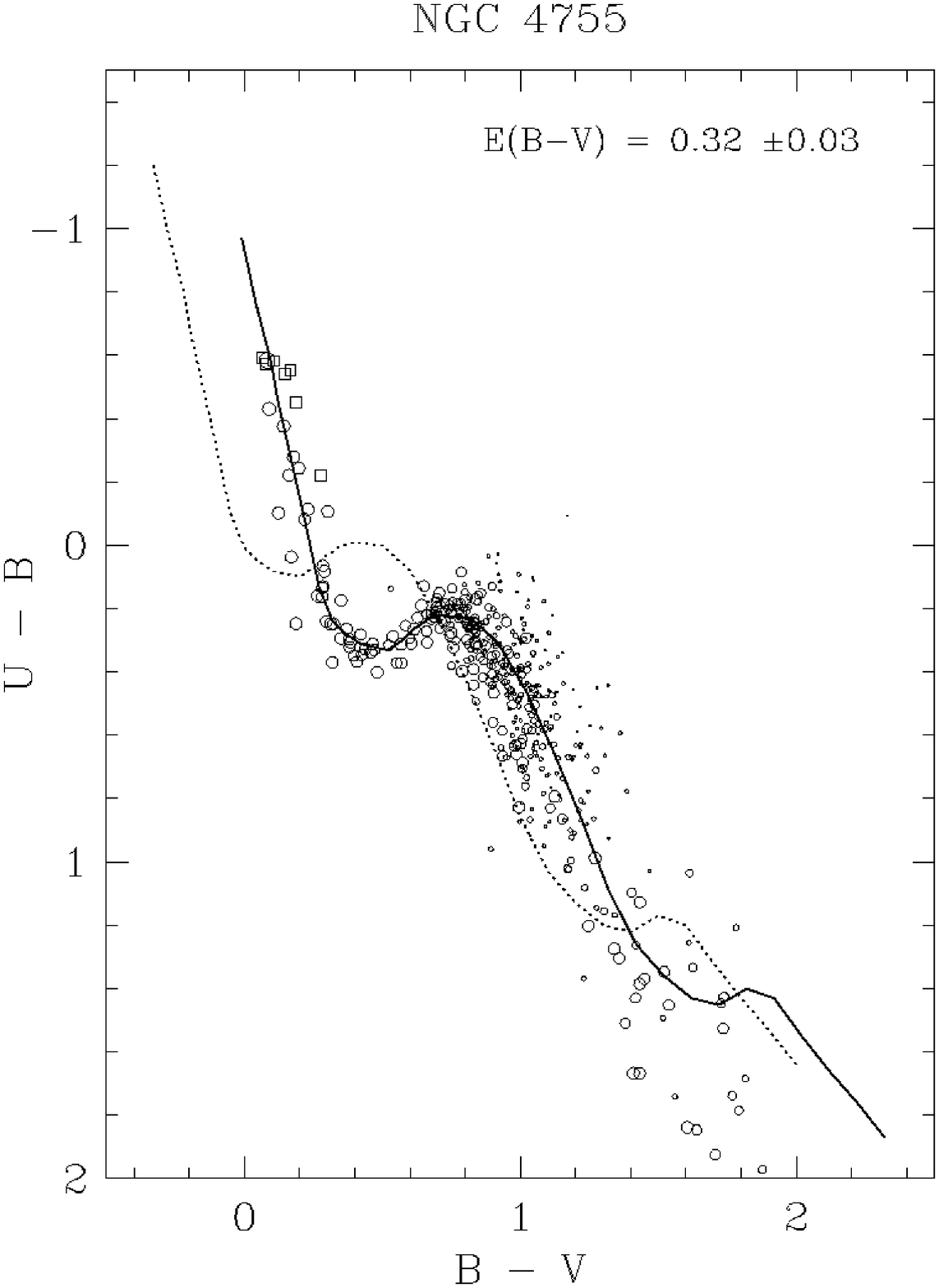}}
\resizebox{8cm}{!}{\includegraphics{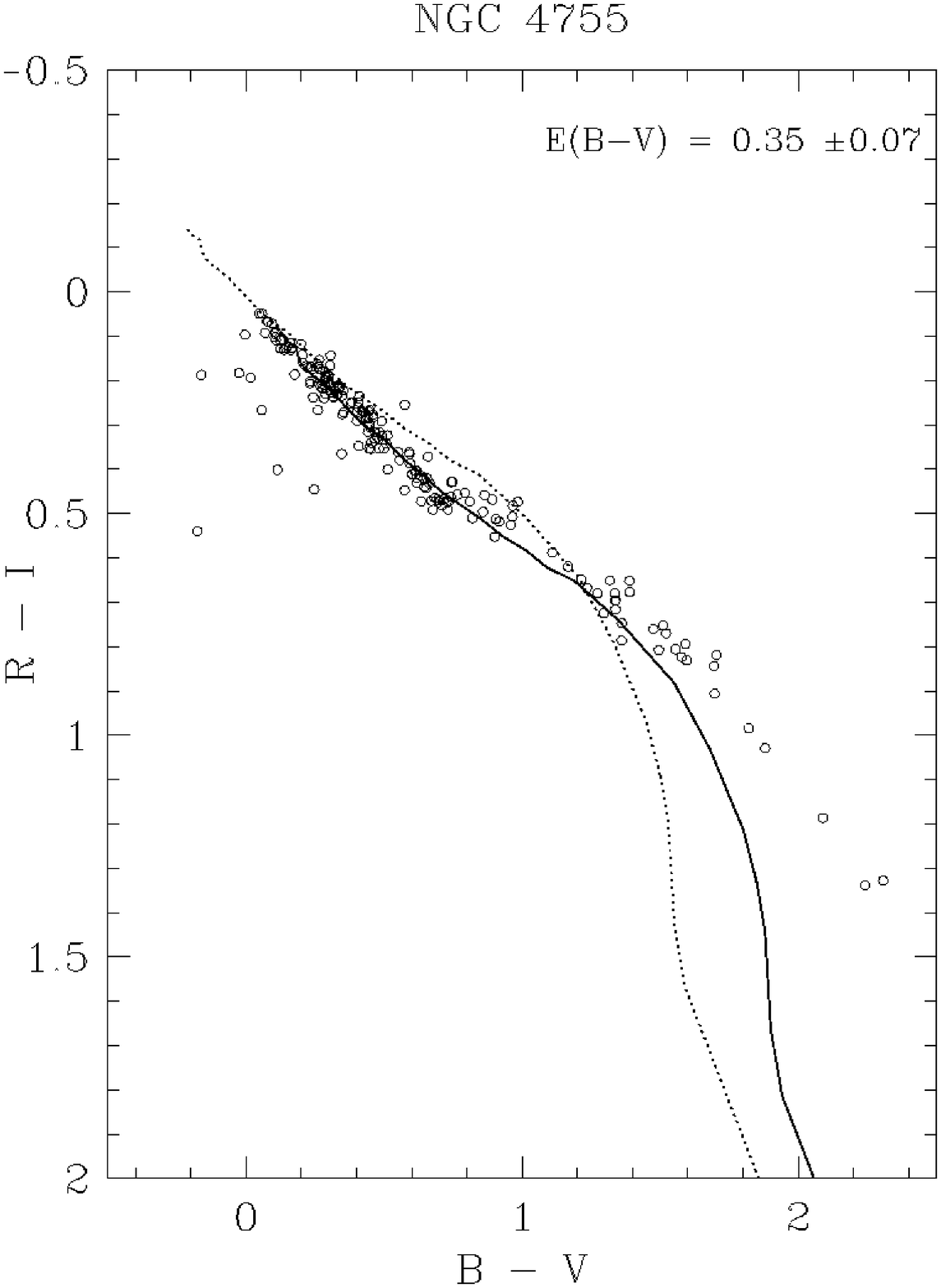}}
\end{center}
\caption[]{Color-color diagrams providing the reddening fit for NGC~4755. Only stars brighter than $V = 16$ are plotted, completely erasing the background contamination. The best fit in $U-B$ versus $B-V$ is $E(B-V) = 0.33$, with small dispersion. A CCD plotted using $R-I$ instead of $U-B$ shows less sensitivity, with a wider range of reddening values fitting the sequence. The best value in this diagram is $E(B-V) = 0.35$. In both diagrams the non-reddened ZAMS is plotted as a dotted line for comparison. Empty circles refer to our photometry whilst empty squares refer to corrected literature photometry.
}
\label{red4755}
\end{figure*}

\begin{figure*}
\begin{center}
\resizebox{8cm}{!}{\includegraphics{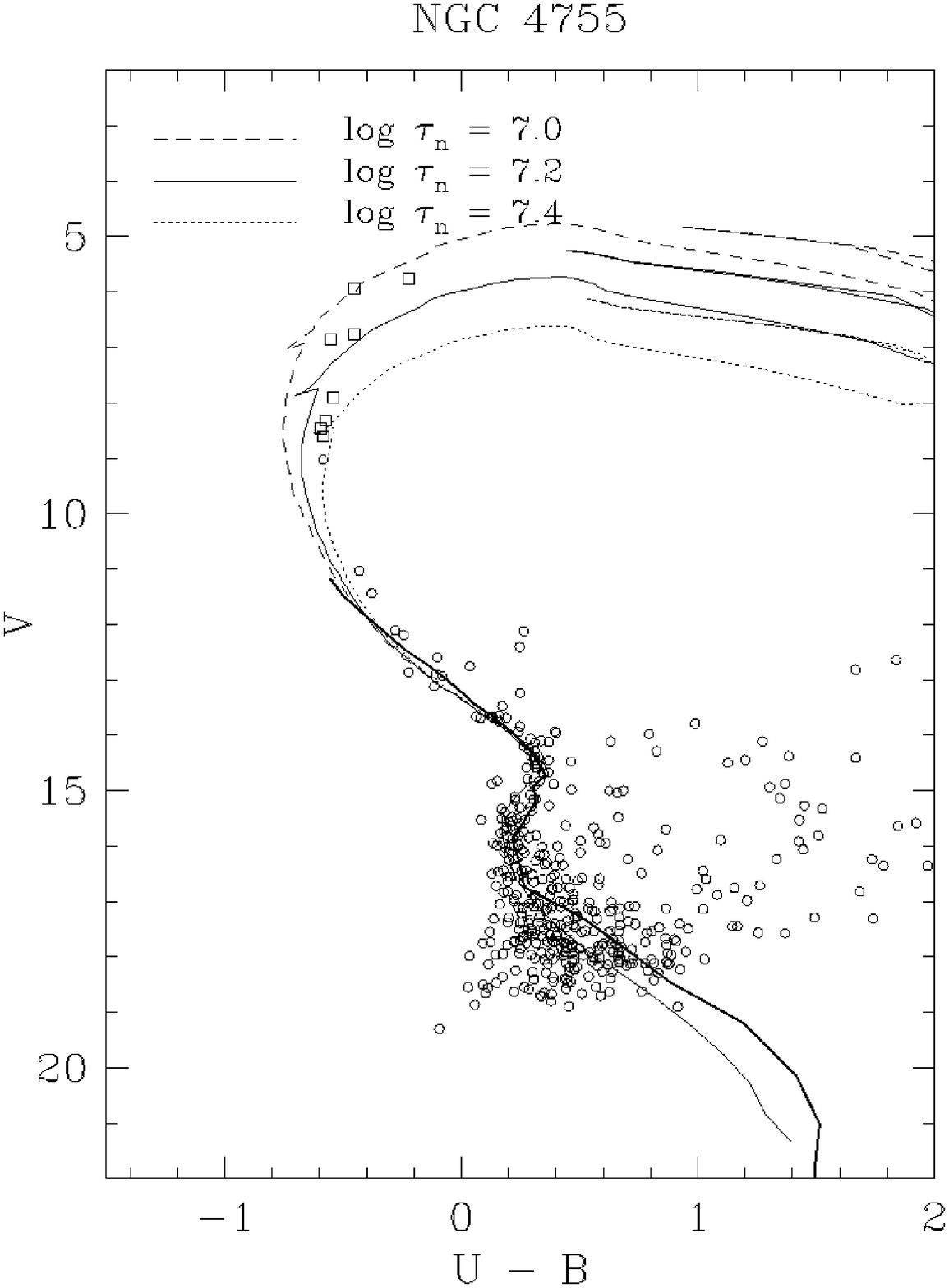}}
\resizebox{8cm}{!}{\includegraphics{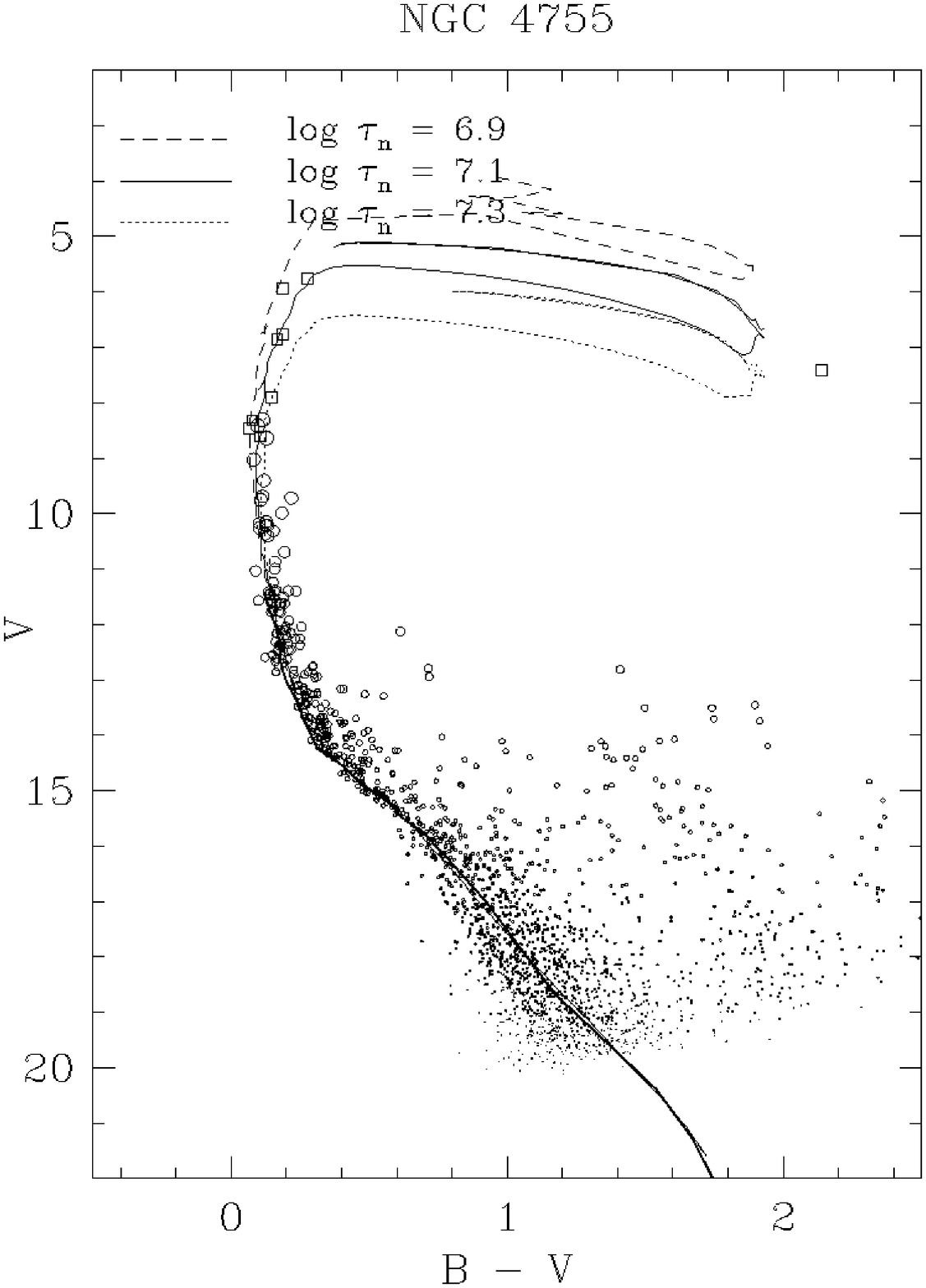}}
\end{center}
\caption[]{{\it Left}. $U$ vs. $U-B$ color magnitude diagram of NGC~4755. Ages between 10 and 25 Myr fit the bright sequence, the best value being $\log\tau_n=7.2$. A ZAMS (thin solid line) is shown for comparison.\\
\\
            {\it Right}. $V$ vs. $B-V$ color magnitude diagram of NGC~4755. The best fit occurs for $\log \tau_n$ = 7.1. Isochrones of $\log \tau_n$ =  7.0 and 7.2 mark the limit of good fits to the luminous branch of post main sequence stars. The red star is slightly too red to be fitted by the isochrones. Empty circles refer to our photometry whilst empty squares refer to corrected literature photometry.
 }
\label{ubv4755}
\end{figure*}

\begin{figure*}
\begin{center}
\resizebox{16cm}{!}{\includegraphics[angle=270]{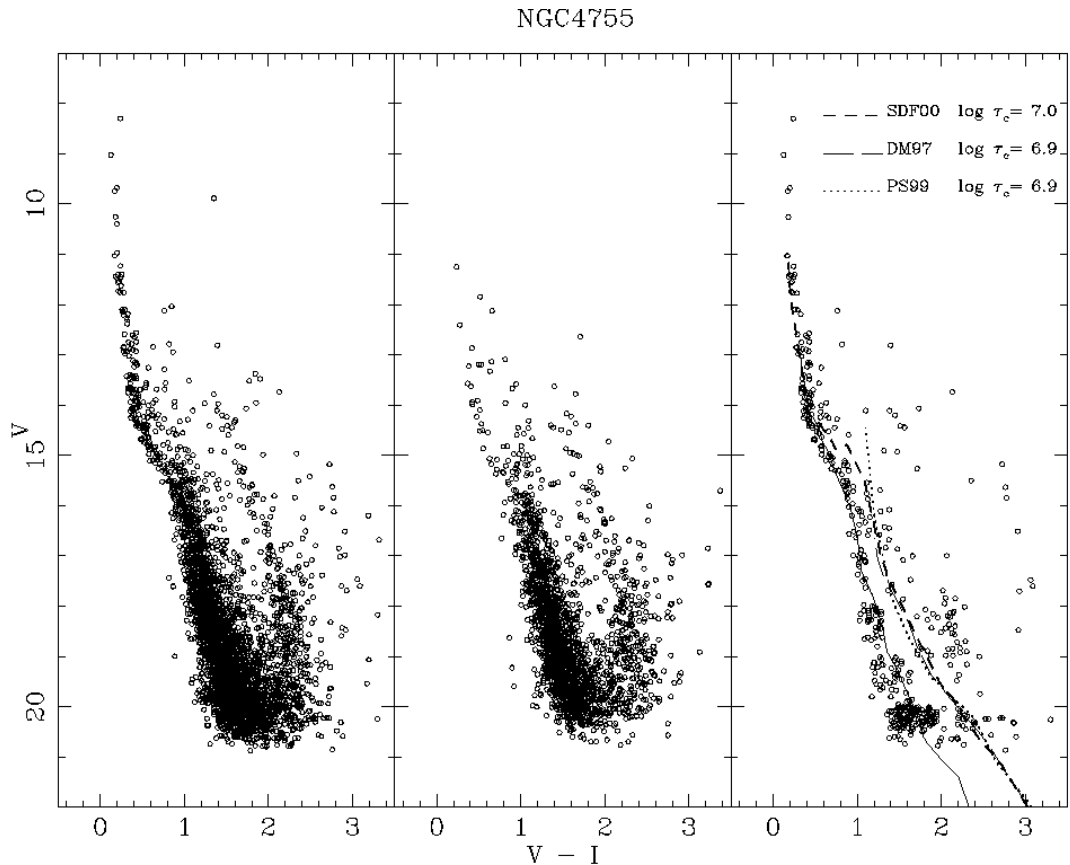}}
\end{center}
\caption[]{$V$ vs. $V-I$ color magnitude diagram of NGC~4755. The control field ({\it middle}) is statistically subtracted from the cluster field ({\it left}), leading to the rightmost diagram. The subtraction reveals a sequence of stars lying above the ZAMS (thin solid line) that is likely to constitute the PMS of the cluster. The fit to this sequence reads $\log \tau_c = 6.9$ for PS99, $\log \tau_c = 7.0$ for SDF00 and $\log \tau_c = 6.9$ for DM97.  The fit is done in the color range 1.3 $< V-I <$ 2.4. Although this is our formally best result, the uncertainties are too large to consider this result seriously.}
\label{4755VI}
\end{figure*}

\section{Conclusions}

\begin{table*}
\label{results}
\caption[]{Parameters of the four open clusters studied, derived from ZAMS and isochrones fitting to color-color and color-magnitudes diagrams. The contraction ages yielded by the four models are consistent with the nuclear ages. For BCAH98, we show ages derived with the calibrated and uncalibrated versions of the model.}

\begin{center}
\begin{tabular}{lccccccccccc} \hline
Cluster&$E(B-V)$&Distance (pc)&&\multicolumn{2}{c}{Nuclear Age (Myr)}&&\multicolumn{5}{c}{Contraction Age (Myr)}\\
\cline{5-6}\cline{8-12}      
	 &		         &	         	&&        $B-V$         &     $U-B$            &&
\multicolumn{2}{c}{BCAH98}& SDF00&  DM97&  PS99 \\
 &&&&&&&{\tiny{\it uncalib.}}&{\tiny{\it calib.}}&&&      \\
\hline
NGC~2232&0.07{\tiny$\pm$0.02}& 320{\tiny$\pm$30}  && 32{\tiny$\pm$15}  & 35{\tiny$\pm$5}    &&25    &    25&    25&    32&    32\\  
NGC~2516&0.11{\tiny$\pm$0.03}& 380{\tiny$\pm$35}  &&125{\tiny$\pm$25}  &160{\tiny$\pm$60}   &&$>$100&$>$100&$>$100&$>$100&$>$100\\  
NGC~2547&0.03{\tiny$\pm$0.02}& 390{\tiny$\pm$25}  && 63{\tiny$\pm$30}  & 75{\tiny$\pm$25}   &&50    &    40&    40&    32&    32\\  
NGC~4755&0.33{\tiny$\pm$0.03}&2040{\tiny$\pm$250} && 13{\tiny$\pm^7_5$}& 16{\tiny$\pm^9_6$} &&-     &     -&    10&     8&     8\\
\hline

\end{tabular}
\end{center}
\end{table*} 

The main results of this paper can be summarized as follows:

\begin{enumerate}
\item We performed UBVRI photometry on four young open clusters,
  spanning an age range of $\sim$10-150 Myr, using the same
  instruments and observational techniques, and favoring homogeneity
  whenever possible. The parameters derived are summarized in Table~5.

\item We find that when using exclusively the $V$ vs. $V-I$ CMD and 
 empirically calibrated PMS isochrones there is consistency between 
 nuclear ($\tau_n$) and contraction ($\tau_c$) ages. The uncalibrated BCAH98
 isochrones also read consistent ages, but the limited color range in which
 they are applicable hinders its utility. This has also been found for the 5 Myr 
 cluster NGC~2362 (Moitinho et al. 2001) using the $V-I$ vs. $V$ diagram and 
 uncalibrated BCAH98 isochrones. 

\item We performed the first contraction age fit using $V-I$ as color
  index for NGC~4755. Sagar \& Cannon (1995) used $B-V$, where color
  differences of red stars are suppressed, rendering the PMS harder to
  identify. The $B-V$ color index shows the disadvantage of showing bluer 
  color for young active stars and therefore leading to wrong determinations 
  of $\tau_c$.

\item We recall that Stauffer et al.(2003) could not decide whether
  NGC~2516 presented activity-related shifts in $B-V$, since they did not
  have $RI$ photometry to compare with $BV$. In our work, we see the
  cluster sequence of NGC~2516 above the best-fitted SK82
  ZAMS in $V-I$ whilst $B-V$ shows the K stars of NGC~2516 lying below
  it. We take it as an evidence that such shifts indeed occur in
  NGC~2516. 

\item We also note that the flagged PMS stars of NGC~2547 (Fig~11, right) are
  above the ZAMS in $B-V$ (Fig~10, upper right). It seems that activity-driven color
  shifts are seen in NGC~2516 as well as in the Pleiades but not in a
  younger cluster such as NGC~2547. This is not easily explained by
  the magnetic activity evolution paradigm in young late-type stars
  (e.g., Preibisch \& Feigelson 2005).

\item We present an estimate for the nuclear age of NGC~2547, 
  based on the lower limit given by 3 evolved stars and the upper limit 
  given by the non-evolved high mass stars still in the ZAMS. In our opinion, 
  there is no reason to dismiss these limits in favor of flawed contraction ages. 

\item We present the first CCD analysis of NGC~2232, allowing us to go
  9 magnitudes deeper than previous studies. This is also the first
  study to perform photometry in the $R$ and $I$ passbands on this
  cluster, thus unveiling this cluster PMS as well. The contraction age 
  derived for NGC~2232 is consistent with the nuclear age.

\item Although the contraction ages seem systematically
  underestimated, we notice that in none of cases they
  deviate by more than one standard deviation from the nuclear ages.
  However, D'Antona et al. (2000) argue that the effective
  temperatures predicted by current PMS evolutionary models are
  actually upper limits, because surface magnetic fields, which are
  not included in any of the models, act to reduce effective
  temperature.

\end{enumerate}

\begin{acknowledgements}
The authors thank the referee, Dr. John Stauffer, for his constructive comments that substantially improved the manuscript. We also thank David Gonzalez for performing the observations. W. Lyra wishes to thank the staff of ESO - Garching and of Lisbon Observatory for the pleasant atmosphere encountered during the visiting trips for this project and, in particular, to the staff of CTIO - La Serena, for the whole year of partnership. A. Moitinho acknowledges financial support from FCT (Portugal) through grant SFRH/BPD/19105/2005. This research has made use of the WEBDA database, operated at the Institute for Astronomy of the University of Vienna.
\end{acknowledgements}

\end{document}